\documentclass[11pt,a4paper]{article}
\input amssymb.sty
\usepackage{amssymb}
\usepackage{amscd}
\begin{document}


\newcounter{mo}
\newcommand{\mo}[1]
{\stepcounter{mo}$^{\bf MO\themo}$%
\footnotetext{\hspace{-3.7mm}$^{\blacksquare\!\blacksquare}$
{\bf MO\themo:~}#1}}

\newcounter{bk}
\newcommand{\bk}[1]
{\stepcounter{bk}$^{\bf BK\thebk}$%
\footnotetext{\hspace{-3.7mm}$^{\blacksquare\!\blacksquare}$
{\bf BK\thebk:~}#1}}


\newcommand{\Si}{\Sigma}
\newcommand{\tr}{{\rm tr}}
\newcommand{\ad}{{\rm ad}}
\newcommand{\Ad}{{\rm Ad}}
\newcommand{\ti}[1]{\tilde{#1}}
\newcommand{\om}{\omega}
\newcommand{\Om}{\Omega}
\newcommand{\de}{\delta}
\newcommand{\al}{\alpha}
\newcommand{\te}{\theta}
\newcommand{\vth}{\vartheta}
\newcommand{\be}{\beta}
\newcommand{\la}{\lambda}
\newcommand{\La}{\Lambda}
\newcommand{\D}{\Delta}
\newcommand{\ve}{\varepsilon}
\newcommand{\ep}{\epsilon}
\newcommand{\vf}{\varphi}
\newcommand{\vfh}{\varphi^\hbar}
\newcommand{\vfe}{\varphi^\eta}
\newcommand{\fh}{\phi^\hbar}
\newcommand{\fe}{\phi^\eta}
\newcommand{\G}{\Gamma}
\newcommand{\ka}{\kappa}
\newcommand{\ip}{\hat{\upsilon}}
\newcommand{\Ip}{\hat{\Upsilon}}
\newcommand{\ga}{\gamma}
\newcommand{\ze}{\zeta}
\newcommand{\si}{\sigma}

\def\hS{{\hat{S}}}

\newcommand{\li}{\lim_{n\rightarrow \infty}}
\def\mapright#1{\smash{
\mathop{\longrightarrow}\limits^{#1}}}

\newcommand{\mat}[4]{\left(\begin{array}{cc}{#1}&{#2}\\{#3}&{#4}
\end{array}\right)}
\newcommand{\thmat}[9]{\left(
\begin{array}{ccc}{#1}&{#2}&{#3}\\{#4}&{#5}&{#6}\\
{#7}&{#8}&{#9}
\end{array}\right)}
\newcommand{\beq}[1]{\begin{equation}\label{#1}}
\newcommand{\eq}{\end{equation}}
\newcommand{\beqn}[1]{\begin{small} \begin{eqnarray}\label{#1}}
\newcommand{\eqn}{\end{eqnarray} \end{small}}
\newcommand{\p}{\partial}
\def\sq2{\sqrt{2}}
\newcommand{\di}{{\rm diag}}
\newcommand{\oh}{\frac{1}{2}}
\newcommand{\su}{{\bf su_2}}
\newcommand{\uo}{{\bf u_1}}
\newcommand{\SL}{{\rm SL}(2,{\mathbb C})}
\newcommand{\GLN}{{\rm GL}(N,{\mathbb C})}
\def\sln{{\rm sl}(N, {\mathbb C})}
\def\sl2{{\rm sl}(2, {\mathbb C})}
\def\SLN{{\rm SL}(N, {\mathbb C})}
\def\SLT{{\rm SL}(2, {\mathbb C})}
\def\PSLN{{\rm PSL}(N, {\mathbb C})}
\newcommand{\PGLN}{{\rm PGL}(N,{\mathbb C})}
\newcommand{\gln}{{\rm gl}(N, {\mathbb C})}
\newcommand{\PSL}{{\rm PSL}_2( {\mathbb Z})}
\def\f1#1{\frac{1}{#1}}
\def\lb{\lfloor}
\def\rb{\rfloor}
\def\sn{{\rm sn}}
\def\cn{{\rm cn}}
\def\dn{{\rm dn}}
\newcommand{\rar}{\rightarrow}
\newcommand{\upar}{\uparrow}
\newcommand{\sm}{\setminus}
\newcommand{\ms}{\mapsto}
\newcommand{\bp}{\bar{\partial}}
\newcommand{\bz}{\bar{z}}
\newcommand{\bw}{\bar{w}}
\newcommand{\bA}{\bar{A}}
\newcommand{\bG}{\bar{G}}
\newcommand{\bL}{\bar{L}}
\newcommand{\btau}{\bar{\tau}}

\newcommand{\tie}{\tilde{e}}
\newcommand{\tial}{\tilde{\alpha}}

\newcommand{\Sh}{\hat{S}}
\newcommand{\vtb}{\theta_{2}}
\newcommand{\vtc}{\theta_{3}}
\newcommand{\vtd}{\theta_{4}}

\def\mC{{\mathbb C}}
\def\mZ{{\mathbb Z}}
\def\mR{{\mathbb R}}
\def\mN{{\mathbb N}}

\def\frak{\mathfrak}
\def\gg{{\frak g}}
\def\gJ{{\frak J}}
\def\gS{{\frak S}}
\def\gL{{\frak L}}
\def\gG{{\frak G}}
\def\gE{{\frak E}}
\def\gk{{\frak k}}
\def\gK{{\frak K}}
\def\gl{{\frak l}}
\def\gh{{\frak h}}
\def\gH{{\frak H}}
\def\gt{{\frak t}}
\def\gT{{\frak T}}
\def\gR{{\frak R}}

\def\baal{\bar{\al}}
\def\babe{\bar{\be}}

\def\bfa{{\bf a}}
\def\bfb{{\bf b}}
\def\bfc{{\bf c}}
\def\bfd{{\bf d}}
\def\bfe{{\bf e}}
\def\bff{{\bf f}}
\def\bfg{{\bf g}}
\def\bfm{{\bf m}}
\def\bfn{{\bf n}}
\def\bfp{{\bf p}}
\def\bfu{{\bf u}}
\def\bfv{{\bf v}}
\def\bfr{{\bf r}}
\def\bfs{{\bf s}}
\def\bft{{\bf t}}
\def\bfx{{\bf x}}
\def\bfy{{\bf y}}
\def\bfM{{\bf M}}
\def\bfR{{\bf R}}
\def\bfC{{\bf C}}
\def\bfP{{\bf P}}
\def\bfq{{\bf q}}
\def\bfS{{\bf S}}
\def\bfJ{{\bf J}}
\def\bfz{{\bf z}}
\def\bfnu{{\bf \nu}}
\def\bfsi{{\bf \sigma}}
\def\bfU{{\bf U}}
\def\bfso{{\bf so}}

\def\clA{\mathcal{A}}
\def\clC{\mathcal{C}}
\def\clD{\mathcal{D}}
\def\clE{\mathcal{E}}
\def\clG{\mathcal{G}}
\def\clR{\mathcal{R}}
\def\clU{\mathcal{U}}
\def\clT{\mathcal{T}}
\def\clO{\mathcal{O}}
\def\clH{\mathcal{H}}
\def\clK{\mathcal{K}}
\def\clJ{\mathcal{J}}
\def\clI{\mathcal{I}}
\def\clL{\mathcal{L}}
\def\clM{\mathcal{M}}
\def\clN{\mathcal{N}}
\def\clQ{\mathcal{Q}}
\def\clW{\mathcal{W}}
\def\clZ{\mathcal{Z}}

\def\baf{{\bf f_4}}
\def\bae{{\bf e_6}}
\def\ble{{\bf e_7}}
\def\bag2{{\bf g_2}}
\def\bas8{{\bf so(8)}}
\def\baso{{\bf so(n)}}

\def\sr2{\sqrt{2}}
\newcommand{\ran}{\rangle}
\newcommand{\lan}{\langle}
\def\f1#1{\frac{1}{#1}}
\def\lb{\lfloor}
\def\rb{\rfloor}
\newcommand{\slim}[2]{\sum\limits_{#1}^{#2}}

\newcommand{\sect}[1]{\setcounter{equation}{0}\section{#1}}
\renewcommand{\theequation}{\thesection.\arabic{equation}}
\newtheorem{predl}{Proposition}[section]
\newtheorem{defi}{Definition}[section]
\newtheorem{rem}{Remark}[section]
\newtheorem{cor}{Corollary}[section]
\newtheorem{lem}{Lemma}[section]
\newtheorem{theor}{Theorem}[section]

\begin{flushright}
 ITEP-TH-24/10\\
\end{flushright}
\vspace{10mm}
\begin{center}
{\Large{\bf Characteristic Classes and Integrable Systems for Simple Lie Groups.}
}\\
\vspace{5mm}

A.Levin \\
{\sf State University - Higher School of Economics, Department of Mathematics, } \\
{\sf 20 Myasnitskaya Ulitsa, Moscow, 101000, Russia } \\
{\em e-mail alevin@hse.ru}\\
M.Olshanetsky\\
{ \sf
Institute of Theoretical and Experimental Physics, Moscow, Russia}
\\
{\em e-mail olshanet@itep.ru}\\
A.Smirnov\\
{\sf Institute of Theoretical and Experimental Physics, Moscow, Russia,}\\
{\em e-mail asmirnov@itep.ru}\\
A.Zotov \\
{\sf Institute of Theoretical and Experimental Physics, Moscow, Russia,}\\
{\em e-mail zotov@itep.ru}\\
\vspace{5mm}
\end{center}

\begin{abstract}
This paper is a continuation of our previous paper \cite{LOSZ}.
For simple complex Lie groups with non-trivial center
 i.e. classical simply-connected groups,
$E_6$ and $E_7$ we consider elliptic Modified Calogero-Moser systems
corresponding to the Higgs bundles with an arbitrary characteristic
class. These systems
are generalization of  the classical Calogero-Moser (CM) systems
related to a simple Lie groups and contain CM systems related to some
(unbroken) subalgebras.
For all algebras we construct  a special basis, corresponding to
non-trivial
characteristic classes, the explicit forms of
Lax operators and  Hamiltonians.

\end{abstract}


\tableofcontents


\section{Introduction}
\setcounter{equation}{0}

This paper is a continuation of our previous paper \cite{LOSZ}.
 Here we consider concrete
implementation of the generic formulae for all simple groups with a non-trivial center.
In particular, we find  the structure of the unbroken Lie subalgebras
$\ti\gg_0$ (see Table I in \cite{LOSZ}).
We refer the formulae from \cite{LOSZ} with a number I. The information about roots,
weights and so on was taken from \cite{Bou,Jac}.

\vspace{0.3cm}
\bigskip
{\small {\bf Acknowledgments.}\\
 The work was supported by grants RFBR-09-02-00393, RFBR-09-01-92437-KEa,
NSh-3036.2008.2, RFBR-09-01-93106-NCNILa (A.Z. and A.S.),
RFBR-09-02-93105-NCNILa (M.O.)  and to the Federal Agency for
Science and Innovations of Russian Federation under contract
14.740.11.0347. The work of A.Z. was also supported by the Dynasty
fund and the President fund MK-1646.2011.1. A.L and M.O. are
grateful for hospitality to the Max Planck Institute of Mathematics,
Bonn, where the part of this work was done.}


\section{$\SLN$ - the root system $A_{N-1}$}

\setcounter{equation}{0}

\subsection*{Roots and weights.}

For the $A_n$, ($n=N-1$) root system we have two groups
$\bar G=\SLN$, $G^{ad}=\PSLN$.
Choose the Cartan subalgebra $\gH\subset\gg$ as an subalgebra of traceless
 diagonal matrices. Then $\gH$ can be identify with the hyperplane in $\mC^N$
$\gH=\{\,\bfx=(x_1,\ldots,x_N)\in\mC^N\,|\,\sum_{j=1}^Nx_j=0\,\}$.
The simple roots $\Pi=\{\al_k\}$
$=\{\al_1=e_1-e_2\,\ldots,\al_{N-1}=e_{N-1}-e_N\}$
form a basis in the dual space $\gH^*$.
 Here $\{e_j\}$ $j=1,\ldots,N$ is a canonical basis in $\mC^N$.
 The vectors $e_j$ generate the set of roots $R=\{\,(e_j-e_k)\,,~j\neq k\,\}$ of type $A_{N-1}$.
 The minimal root is
 \beq{mra}
 \al_0=-\sum_{\al\in\Pi}\al_k=e_N-e_1\,.
 \eq
It defines the extended Dynkin diagram
\vspace{0.5cm}
\begin{center}
\unitlength 1mm 
\linethickness{0.6pt}
\ifx\plotpoint\undefined\newsavebox{\plotpoint}\fi 
\begin{picture}(62.142,41.099)(0,0)
\put(26.087,9.881){\texttt{Fig.1:} \large A$_{n}$ and action of $\la_{N-1}$}
\put(31.008,21.968){\circle{2.531}}
\put(38.997,22.074){\circle{2.322}}
\put(61.07,22.074){\circle{2.144}}
\put(46.039,36.894){\circle*{2.765}}
\put(32.479,21.968){\line(1,0){5.3607}}
\put(40.363,22.073){\line(1,0){5.781}}
\put(55.814,21.968){\line(1,0){4.099}}
\multiput(31.744,23.23)(.052555788,.0504702408){252}{\line(1,0){.052555788}}
\multiput(47.3,36.263)(.0505344115,-.0513429621){260}{\line(0,-1){.0513429621}}
\put(31.744,27.539){\vector(-1,-1){.07}}
\multiput(40.994,36.474)(-.052258863,-.05047731){177}{\line(-1,0){.052258863}}
\put(40.048,18.92){\vector(1,0){4.625}}
\put(56.235,18.92){\vector(1,0){3.574}}
\put(51.189,35.948){\vector(-1,1){.07}}
\multiput(61.28,25.752)(-.050453556,.050979114){200}{\line(0,1){.050979114}}
\put(30.903,15.767){$\alpha_1$}
\put(38.996,15.977){$\alpha_2$}
\put(61.07,16.082){$\alpha_n$}
\put(46.039,41.099){$\alpha_0$}
\put(31.008,18.71){\vector(1,0){7.042}}
\end{picture}
\end{center}

 Since the half-sum of positive roots is
 $\rho=\oh(N-1,N-3,\ldots,3-N,1-N)$
and $h=N$, $\,\ka= \frac{\rho}h=\f1{2N}(N-1,N-3,\ldots,3-N,1-N)$.
For $\al=e_k-e_{k+a}$ the level (A.3.I) $f_\al=a$ and
\beq{rhoa}
\lan\ka,\al\ran=a/N
 \eq
 (see (A.14.I)).

We identify $\gH^*$ and $\gH$ by means of the standard metric on  $\mC^N$.
Then the coroot system
coincides with $R$, and the coroot lattice $Q^\vee$ coincides with $Q$
\beq{rl}
Q=\{\sum m_je_j\,|\,m_j\in \mZ\,,~\sum m_j=0\}\,.
\eq

The fundamental weights
$\varpi_k$, $(k=1,\ldots,N-1)$,  dual to the basis of simple
roots $\Pi^\vee\sim\Pi\,$ $\,(\varpi_k(\al^\vee_k)=\de_{kj})$, are
\beq{fw1}
\varpi_j=e_1+\ldots+e_j-\frac{j}{N}\sum_{l=1}^Ne_l\,,~~~
\left\{
\begin{array}{l}
 \varpi_1=(\frac{N-1}N,-\f1{N},\ldots,-\f1{N})\\
 \varpi_2=(\frac{N-2}N,\frac{N-2}N,\ldots,-\frac{2}{N})   \\
 \ldots \\
   \varpi_{N-1}=(\frac{1}N,\frac{1}N,\ldots,\frac{1-N}{N}) \,. \\
\end{array}
\right.
\eq
In the basis of simple roots the fundamental weights are
$$
\varpi_k=
\f1{N}[(N-k)\al_1+2(N-k)\al_2+\ldots(k-1)(N-k)\al_{k-1}
$$
$$
+k(N-k)\al_k+k(N-k-1)\al_{k+1}+\ldots+k\al_{N-1}]\,.
$$
Similar to the roots and coroots we identify the
fundamental weights and the fundamental coweights. They generate the weight
(coweight) lattice
 \beq{wl}
 P\subset\gH\,,~P=\{\sum_ln_l\varpi_l\,|\,n_l\in\mZ\}\,,~~{\rm or}~
  P=\{\sum_{j=1}^Nm_je_j\,,~m_j\in \f1{N}\mZ\,,~m_j-m_k\in\mZ\}\,.
  \eq
or
\beq{gv}
P=Q+\mZ\varpi_{N-1}\,,~~(\varpi_{N-1}=-e_N+\f1{N}\sum_{j=1}^Ne_j)\,.
\eq


\subsection*{Transition matrices.}

The factor-group  $P^\vee/Q^\vee\sim P/Q$ is isomorphic to the
center $\clZ(\SLN)\sim\mu_N$. It is generated by $\zeta=\exp 2\pi
i\varpi_{N-1}$. Following Proposition 3.1.I define $\la_{N-1}$ and its
action on the extended Dynkin diagram. Consider the fundamental alcove
(A.17.I). It follows from (\ref{mra}) that
$$
C_{alc}= \{0,\varpi_1,\ldots,\varpi_{N-1}\}\,.
$$
Thus, any fundamental weight generates a nontrivial $\La$.
For $\xi=\varpi_{N-1}$  in (3.10.I)  we find the corresponding transformation
\beq{gac1}
\la_{N-1}\,:\,e_j\to e_{j+1}\,,~(\al_k\to\al_{k+1})\,.
\eq
The action of $\la_{N-1}$ on $\Pi^{ext}$ is presented on
Fig.1.
It means that in the canonical basis in $\mC^N\,$ $\la_{N-1}$ acts
as the permutation matrix
\beq{lamb}
\La=c\left(
    \begin{array}{ccccc}
      0 & 1 & 0 & \ldots & 0 \\
      0 & 0 & 1 & \ldots & 0 \\
      \vdots & \ddots & \ddots & \ddots & \vdots \\
      0 & 0 &   & 0 & 1 \\
      1& 0 & \ldots & 0 & 0 \\
    \end{array}
  \right)\,,~~~c=\left\{
  \begin{array}{cc}
    1\,,& N=2m+1\\
    \exp\,\frac{\pi i}{2m} \,,& N=2m
  \end{array}\right. \,.
\eq
Evidently, that $\La\in\SLN$ and $\La^N\sim Id$.
On the other hand
\beq{clqu}
\clQ=\bfe\,(\ka)=\di
(\bfe\left(\frac{N-1}N\right),\bfe\left(\frac{N-3}N\right),
\dots,\bfe\left(\frac{1-N}N\right))\,,~~(\bfe(x)=\exp\,(2\pi ix))\,.
\eq
Thus,
$$
[\La,\clQ]=\zeta \,,~~\zeta=\om\cdot Id\,,~\om=\bfe\,(1/N)\,,
$$
and $\zeta$ is an obstruction to lift  a $\PSLN$-bundle
to a $\SLN$-bundle. Since $Ker\,(\la_{N-1}-1)|_\gH=0$
$\ti\gH=\empty$ (see Proposition 3.1.I)
the moduli space is empty set (see (3.29.I) and (3.32.I)).

Consider another extreme case when $\xi$ lies in the root lattice $Q$. Then $\La=Id$ and
$\gH\in Ker\La$. Since $[\La,\clQ]=Id$, the bundles have a trivial
characteristic class.

Consider an intermediate case and
assume that $N$ has nontrivial divisors $N=pl$, $(l\neq 1,N)$.
It can be found that
\beq{gac}
\varpi_{j}\,\to\,\la_j\,:\, e_k\to e_{k+N-j}\,.
\eq
Then there exists a sublattice $P_l= Q+\mZ\varpi_{N-p}\,$,
$\,Q\subset P_l\subset P\,$,  such that
 $P/P_l\sim\mu_l$. It follows from (\ref{gac}) that
 $\la_{N-p}\,:\,e_j\to e_{j+p}$.
Therefore, $\La_{N-p}$ is equal the $p$-th degree of (\ref{lamb})
($\La_{N-p}^l=Id$)
 For $\zeta_p=\bfe (\varpi_{N-p})=\om ^p Id\,$,
\,($\om^N=1$) (3.3.I) takes the form
\beq{co0}
[\La_{N-p},\clQ]=\om^p\cdot Id_N\,,
\eq
where $\clQ$ is (\ref{clqu}).

Define the factor group $G_l$
\beq{co1}
1\to\mu_l\to\SLN\to G_l\to 1\,.
\eq
It has the center $\clZ(G_l)\sim\mu_p=\mZ/p\mZ$. Therefore,
$$
1\to\mu_p\to G_l\to\PSLN\to 1\,.
$$
The sublattice $P_l$ is isomorphic the group of cocharcters $t(G_l)$ (A.43.I)0

For the dual group $^LG_l=G_p$ we have the similar exact sequences, where
the role of $l$ and $p$ are changed
$$
1\to\mu_p\to\SLN\to G_p\to 1\,,
$$
\beq{co2} 1\to\mu_l\to G_p\to\PSLN\to 1\,. \eq It follows from
(\ref{co1}) and (\ref{co2})  that the cocycle $H^2(\Si_\tau,\mu_l)$
is obstruction to lift  $G_l$-bundles to $\SLN$ bundles or to lift
$\PSLN$-bundles to $G_p=^LG_l$-bundles.


\subsection*{Bases}

\subsubsection*{Sin-basis.}

Consider the case $\zeta=\exp(2\pi i\varpi_{N-1})$. Therefore $\La$ is (\ref{lamb})
and all it orbits in $R$ have the same length $N$. The number of orbits is $N-1$. In other words $\sharp R=N(N-1)$.
Let us take  the root subspaces $E_{1,a}$ $(a\neq 1)$ as representatives of orbits
in the space  $\gL$.
Then the basis (5.4.I) in $\gL$ is
$$
\gt^c_a=\f1{\sqrt{N}}\sum_{m=1}^{N}\om^{mc}E_{1+m,a+m}\,,~c=(0,\ldots,N-1)\,,~~\om^N=1\,.
$$
In particular,
$$
\gt^c_1=\f1{\sqrt{N}}
\left(
 \begin{array}{ccccc}
  0 & \om^c & 0 & \ldots & 0 \\
 0 & 0 & \om^{2c} & \ldots & 0 \\
 \vdots & \ddots & \ddots & \ddots & \vdots \\
 0 &  &  &  & \om^{(N-1)c} \\
  1 & 0 & \ldots & \ldots & 0 \\
\end{array}
 \right)\,.
$$

There is only one orbit in the Cartan subalgebra $\gH$ under the action of $\la_{N-1}$
$$
e_1\to e_2\to e_3,\to\ldots,e_N,\,.
$$
For the $A_{N-1}$ roots it is convenient to pass to the
canonical over-complete  basis
$(e_1,e_2,\ldots,e_N)$. Then the basis (5.33.I) on $\gH$ is
$$
\gh^c=\f1{\sqrt{N}}\sum_{m=0}^{N-1}\om^{mc}e_{m+1}=\f1{\sqrt{N}}\di(1,\om^c,\ldots,\om^{(N-1)c})
\,, ~(c=1,\ldots,N-1)\,.
$$
Essentially, $(\gt^c_a,\gh^c)$ form a basis of the sin-algebra \cite{FFZ}.

If $N$ is a primitive number then the center $\clZ(\SLN)=\mu_N$ has not
nontrivial subgroups. Therefore, one can put in (3.9.I) $\xi\in P$.
This case leads to the sin-basis. Another options is $\xi\in Q$.
In this case $\la=Id$ and we come to the Chevalley basis (see Remark 5.1.I).


\subsubsection*{Generalized Sin-Basis in $\SLN$.}

Let $N=pl$ and $\xi=\varpi_{N-p}$ generates $\La_{N-p}=\La^p$.
 All orbits of $\mu_l$ have the length $l$. In the space of bases $\clE=\{E_{jk}\,,$ $j\neq k$
 there are $p(N-1)$ orbits passing through the matrices
\beq{sa}
E_{s,s+a}\,,~~(s=1,\ldots ,p\,,\,a=1,\ldots,N\,,\,\,(mod\,N))\,.
\eq
The off-diagonal basis is
\beq{atk}
\gt^c_{s,k}=\f1{\sqrt{l}}\sum_{m=0}^{l-1}\om^{mpc}E_{s+mp,s+k+mp}\,,~~\om^N=1\,,
\eq
$$
c=0,\ldots,l-1\,,~s=1,\ldots ,p\,,\,k=1,\ldots,N-1\,,(mod\,N))\,.
$$
 The pairing in this basis (5.8.I) assumes the form
\beq{stp}
(\gt^{a}_{s_1,k},\gt^{b}_{s_2,j})=
\de_{k,-j}\de^{(a,-b\,,mod\,l)}\de_{(s_1+k,s_2+rp)}\om^{rpb}\,,~~(s_1+k-s_2=rp)\,.
\eq

Let  $(e_1,\ldots,e_N)$ be the canonical basis in $\gH$. There are $p$ orbits
 of length $l$
$$
\begin{array}{l}
\ti{e}_1=e_1+e_{p+1}+\ldots+e_{(l-1)p+1}   \\
  \ti{e}_2=e_2+e_{p+2}+\ldots+e_{(l-1)p+2}  \\
  \ldots \\
\ti{e}_{p}=e_{p}+e_{2p}+\ldots+e_{lp} \,.
\end{array}
$$
 Let
$\ti\gH_0\subset\gH$ be a Cartan subalgebra $$
\ti\gH=\{\ti\bfu=\sum_{j=1}^pu_j\ti{e}_j\,|\,\sum_{j=1}^pu_j=0\}\,.
$$
with the basis of simple coroots
\beq{slco}
\ti\Pi^\vee=\{\ti\al^\vee_k=\ti e_k-\ti e_{k+1}\}\,.
\eq
 The  simple roots
$$
\ti\Pi=\{\ti\al_k=\f1{l}(\ti e_k-\ti e_{k+1})\}
$$
along with $\ti\Pi^\vee$ generate $A_{p-1}$ type Cartan matrix
 $a_{jk}=\lan\ti\al^\vee_j,\ti\al_k\ran$. The simple coroots and the subspaces
  \beq{cbp}
\ti E_{i,a}=\sum_{m=0}^{l-1}E_{i+mp,i+a+mp}\,,~~(1\leq
i\leq p\,,~  a=\pm 1,\pm 2,\ldots\pm(p-1))
 \eq
  form the Chevalley basis in the invariant subalgebra $\ti\gg_0={\bf\rm sl}(p,\mC)$
(see Proposition 5.1).

It follows from (\ref{atk}) that the basis (5.29.I) in the space $V$ (5.30.I)
takes the form
\beq{abv}
V=\{\gt^0_{s,a}=\f1{\sqrt{l}}\sum_{m=0}^{l-1}E_{s+mp,s+a+mp}\,,
~(1\leq s\leq p\,,~p\leq a\leq N)\}\,.
\eq
Together with $\ti\gg_0$,  $V$ forms the invariant subalgebra
$$
\gg_0=\ti\gg_0=\underbrace{{\bf\rm sl}(p,\mC) \oplus\ldots\oplus{\bf\rm sl}(p,\mC)}_{l}
\oplus(\underbrace{\mC\oplus\ldots\oplus\mC)}_{l-1}
$$
 Its structure is obtained  from $\Pi^{ext}$ by dropping $l$ roots $(\al_p,$
 $\ldots,\al_{(l-1)p},\al_0)$. This procedure defines an automorphism of $\gg_{A_{N-1}}$ of
 order $l=N/p$.

The Killing form in $\ti\gg_0$  is
\beq{siap}
(\ti e_k,\ti e_j)=l\de_{kj}\,,~~(\ti E_{i,a},\ti E_{j,b})=l\de_{a,-b}\de_{i+a,j\,,~(mod\,p)}
 \eq
 and commutation relations
$$
[\ti e_k,\ti E_{i,a}]=(\de_{i,k}-\de_{i-k+a,0\,mod\,p})\ti E_{i,a}\,,
$$
$$
       [\ti{E}_{i,a},\ti{E}_{j,b}]=
\de_{i+a,j\,mod\,p}\ti E_{i,a+b}-\de_{i,j+b\,mod\,p}\ti E_{j,a+b}\,,~(b\neq p-a)\,,
$$
$$
[\ti E_{i,a}, \ti{E}_{i+a,p-a}]=\ti e_i-\ti e_j\,.
$$

\bigskip

The basis in $\gH$ is generated by $\ti e_j$ and by
\beq{ahk}
\gh_{j}^c=\f1{\sqrt{l}}\sum_{m=0}^{l-1}\om^{mpc}e_{j+mp}\,,~~
c=1,\ldots,l-1\,, ~~j=1,\ldots,p
\eq
with the pairing
\beq{pahk}
(\gh_{j}^{c_1},\gh_{k}^{c_2})=\de_{j,k}\de^{c_1,-c_2}\,.
\eq

The subspace $\gg_c$ in (5.1.I) is formed by the basis
$\{\gh_{j}^c ~(\ref{ahk})\,,~~\gt^c_{s,a}~(\ref{atk})\,,~c\neq 0\}$.

The general commutation relations in this basis
\beq{scr}
[\gt^{c_1}_{i,a},\gt^{c_2}_{j,b}]=\f1{\sqrt{l}}
\left(\om^{(i-j+a)c_2}\de_{j,i+a\,(mod\,p)}\gt^{c_1+c_2}_{i,a+b}-
\om^{(j-i+b)c_1}\de_{i,j+b\,(mod\,p)}\gt^{c_1+c_2}_{j,a+b}\right)\,,
\eq
\beq{scr1}
[\ti e_{i},\gt^{c}_{i,a}]=
\de_{j,i}\gt^{c}_{i,a}-
\de_{i,j+b\,(mod\,p)}\gt^{c}_{j,b}\,,
\eq
\beq{scr2}
[\gh^{c_1}_{i},\gt^{c_2}_{j,b}]=\f1{\sqrt{l}}
\left(\om^{(i-j)c_2}\de_{j,i}\gt^{c_1+c_2}_{i,b}-
\om^{(j-i+b)c_1}\de_{i,j+b\,(mod\,p)}\gt^{c_1+c_2}_{j,b}\right)\,,
\eq


\subsection*{The Lax operators and the Hamiltonians.}

\subsubsection*{Trivial bundles}

For trivial bundles $\xi\in Q$, $\La=Id$ and $\gH\subset Ker\La$.
One can consider trivial
$\SLN$ or $\PSLN$ bundles. They differ by their moduli spaces
(3.21.I) or (3.22.I). At the case at hand they assume the following form.
Let
$$
C^+=\{\ti\bfu\in\mC^N\,|\,u_1\geq u_2\geq\ldots\geq u_N\}
$$
be a positive Weyl chamber (A.8.I). Then (3.21.I) and (3.22.I)
take the form
\beq{csca}
C^{SL} =\{\bfu\in C^+\,|\,
u_j\sim u_j+\tau n_j+m_j\,,~n_j,m_j\in\mZ\,,~\sum_j n_j=\sum_j m_j=0\}\,.
\eq
\beq{cada}
C^{PSL} =\{\bfu\in C^{SL}\,|\,n_j,m_j\in\f1{N}\mZ\,,~
n_j-n_k\in\mZ\,,~m_j-m_k\in\mZ~\}\,.
\eq

The Lax operator  is the SL$(N)$, (or PSL$(N)$) and the Hamiltonian  CM system are well-known
\cite{GH,LX,Wo}
$$
L^{CM}_{SL(N)}=\sum_{j=1}^N(v_j+S_j)e_j+\sum_{j\neq k}S_{jk}\phi(u_k-u_j,z)E_{jk}\,,
$$
$$
H^{CM}_{SL(N)}=\oh\sum_{j=1}^Nv_j^2-\sum_{j\neq k}S_{jk}S_{kj}E_2(u_j-u_k)\,.
$$


\subsubsection*{Nontrivial bundles}

Define a moduli space of bundles with characteristic class $\om^p$ (\ref{co0}).
Evidently,\\
 $\ti\gH=Ker(\La_{N-p}-1)|_{\ti\gH_0}$.
Let $Q_l\subset Q$ be an invariant coroot lattice in $\ti\gH_0\subset{\rm\bf sl}(p,\mC)$
$$
Q^\vee_l=\{\ga=\sum_{j=1}^pm_j\ti e_j\,,~~m_j\in\mZ\,,~\sum_{j=1}^pm_j=0\}\,,
$$
generated by the simple coroots (\ref{slco}). It is an invariant sublattice of
$Q^\vee$ with respect to the $\La_{N-p}$ action.

The fundamental coweights $(\lan\ti\varpi^\vee_j,\ti\al_k\ran=\de_{jk})$
$$
\begin{array}{l}
   \ti\varpi^\vee_1=\frac{p-1}p\ti e_1-\f1{p}\ti e_2-\ldots \f1{p}\ti e_{p}\,,\\
  \ti\varpi^\vee_2=\frac{p-2}p\ti e_1+\frac{p-2}{p}\ti e_2-\ldots -\frac{2}{p}\ti e_{p}\,, \\
 \ldots\ldots\\
  \ti\varpi^\vee_p=\frac{1}p\ti e_1+\f1{p}\ti e_2+\ldots \frac{1-p}{p}\ti e_{p}\,
\end{array}
$$
form a basis in the coweight lattice
$$
P^\vee_l=\{\ga=\sum _{j=1}^pn_j\ti e_j\,,~~n_j\in\f1{p}\mZ\,,~~
\sum _{j=1}^pn_j=0\,,~~n_j-n_k\in\mZ\}\,.
$$
 It is an invariant sublattice of $P^\vee$.

Let $\ti W$ is a permutation group of $\ti{e}_1,\ldots,\ti{e}_{p}$.
It is a Weyl group of $\ti R(\ti\Pi)$. Define the semidirect products (3.28.I), (3.31.I)
$$
\ti W_{BS}=\ti W\ltimes(\tau Q^\vee_l+Q^\vee_l)\,,~~~
\ti W^{ad}_{BS}=\ti W\ltimes(\tau P^\vee_l+P^\vee_l)\,.
$$
 The moduli space is defined as in (3.29.I), (3.31.I)
\beq{mscl}
 C^{(l) \,sc}=\ti\gH/ \ti W_{BS}\,,~~~
 C^{(l) \,ad}=\ti\gH/\ti W^{ad}_{BS}\,.
\eq
Thus, for
$$
C^{(l) \,sc}\,:\, u_j\sim u_j+\tau m_j+n_j\,,~n_j,m_j\in\mZ\,,~
\sum_{j=1}^p n_j=\sum_{j=1}^p m_j=0\,,
$$
and for
$$
C^{(l) \,ad}\,:\, u_j\sim u_j+\tau m_j+n_j\,,~~~~n_j,m_j\in\f1{p}\mZ\,,
$$
$$
\sum_{j=1}^p n_j=\sum_{j=1}^p m_j=0\,,~~~n_j-n_k\in\mZ\,,~~~m_j-m_k\in\mZ\,.
$$

Define $\ti\bfu\in\gH$ such that
\beq{u}
u_{i}=u_s~{\rm if}~i=s+mp\,.
\eq
It means that $\ti\bfu\in\ti\gH_0$, and  $\ti\bfu=\sum_{i=1}^pu_i\ti e_i$.
The Lax operator $L=\ti L_0+L'_0+\sum_{a=1}^{l-1}L_a$
(6.16.I), (6.17.I), (6.18.I)
 takes the form, (see  (\ref{rhoa}))
$$
  L_a(z)=\sum_{j=1}^pS^{a}_{j}\phi(\frac{a}l,z)\gh^a_{j}+
\sum_{i,k=1}^NS^{a}_{i,k}\bfe(zk/N)
\phi(u_{i+k}-u_i+\tau k/N+\frac{a}l,z)\gt^a_{i,k}\,,
$$
$$
\ti L_0(z)=\sum_{i=1}^p(v_{i}+ S_{i}E_1(z))\ti e_{i}+
\sum_{i,k}\ti S_{i,k}\bfe(zk/N)
\phi(u_{i+k}-u_i+\tau k/N,z)\ti E_{i,k}\,,
$$
\beq{sLM}
  L'_0(z)=
\sum_{i,k}S^{'}_{i,k}\bfe(zk/N)
\phi(u_{i+k}-u_i+\tau k/N,z)\gt^0_{i,k}\,.
\eq

To come to an integrable system  impose the moment constraints $\ti S^{\ti\gH}_{i}=0$
$(i=1,\dots,p)$ and the gauge fixing constraints. The calculation of the
quadratic Hamiltonian is based on the pairing relations (\ref{stp}), (\ref{pahk}), (\ref{siap}).
 Using (5.1.I) and (\ref{stp})
we come to the Hamiltonian
$$
H=\ti H_0+H_0'+\sum_{k=1}^{[l/2]}H_k\,.
$$
Here $\ti H_0$ is the sl$(p)$ Calogero-Moser Hamiltonian
$$
\ti H_0=l\Bigl(\frac{1}2\sum_{i=1}^pv_i^2-\sum_{i=1}^p\sum_{k=1}^{p-1}\ti S_{i,k}\ti S_{i+k,-k}E_2(u_i-u_{i+k})\Bigr)\,.
$$
For $H_a$ and $H_0'$ we have
$$
H_0'=-\oh\sum_{i=1}^p\sum_{k=p+1}^NS^0_{i,k}S^{0}_{i+k+rp,-k}
E_2(u_{i+k}-u_i+\tau k/N)\,,
$$
$$
H_a=-\frac{1}2\sum_{s=1}^p\Bigl(
S^a_{s}S^{-a}_{s}E_2(a/l)+\sum_{k=1}^N\om^{-aip}
S^{a}_{s,k}S^{-a}_{s+k+rp,-k}
E_2(u_{s+r}-u_s+a/l+\tau k/N)\Bigr)\,.
$$
These Hamiltonians and the corresponding Lax operators were obtained in \cite{LZ}.
The basis, used there is not the GS basis. The corresponding Hamiltonians describe $p$ interacting EA tops with inertia tensors depending on $\ti\bfu$. This interpretation is specific for
$\SLN$ $(N=pl)$. In particular,
 if $\xi=\varpi_{N-1}$, $\ti\gH_0=\emptyset$ and we deal with the sin-basis. The corresponding bundle has no moduli ($p=1$, $l=N$). The invariant Hamiltonian $\ti H_0=0$ and $H_0'$, $H_a$ describe the so-called the Elliptic top
\cite{KLO,RSTS}.


\section{SO$(2n+1)$\,, Spin$(2n+1)\,$, $\,B_n$ root system.}
\setcounter{equation}{0}

\subsection*{Roots and weights.}

For the Lie algebra  $B_n$ the universal covering group $\bG$ is Spin$(2n+1)$ and
$G^{ad}=$SO$(2n+1)$.
The simple roots of $B_n$ are
\beq{bsr1}
\Pi_{B_n}=\{\al_j=e_j-e_{j+1}\,,~j=1,\dots,n-1\,,~\al_n=e_n\}\,.,
\eq
and the minimal root is
\beq{mbr1}
\al_0=-e_1-e_2=-(\al_1+2\al_2+\ldots+2\al_n)\,.
\eq

\unitlength 1mm 
\linethickness{0.4pt}
\ifx\plotpoint\undefined\newsavebox{\plotpoint}\fi 
\begin{picture}(64.96,44.893)(0,0)
\put(26.087,9.881){\texttt{Fig.2:} \large B$_{n}$ and action of $\la_{1}$}
\put(22,36.568){\circle{.42}}
\put(22,36.717){\circle{2.102}}
\put(22,24.825){\circle*{2.081}}
\put(28.987,30.919){\circle{1.808}}
\put(37.014,31.068){\circle{1.808}}
\put(55.149,30.919){\circle{2.264}}
\put(63.92,30.919){\circle{2.081}}
\put(56.339,31.663){\line(1,0){6.838}}
\put(56.636,30.473){\line(1,0){6.243}}
\multiput(60.055,32.406)(.03344645,-.03344645){40}{\line(0,-1){.03344645}}
\multiput(61.244,31.068)(-.03303353,-.03716272){36}{\line(0,-1){.03716272}}
\put(22.892,35.974){\line(6,-5){5.351}}
\multiput(30.027,30.919)(1.218937,.02973){5}{\line(1,0){1.218937}}
\put(37.906,30.919){\line(1,0){3.716}}
\multiput(50.839,31.068)(.624334,-.02973){5}{\line(1,0){.624334}}
\multiput(28.095,30.325)(-.037976505,-.033636333){137}{\line(-1,0){.037976505}}
\put(17.838,40.136){$\alpha_1$}
\put(17.095,21.852){$\alpha_0$}
\put(30.027,34.636){$\alpha_2$}
\put(54.704,35.676){$\alpha_{l-1}$}
\put(64.069,35.974){$\alpha_l$}
\thicklines
\put(21.554,28.244){\vector(0,-1){.07}}\put(21.554,33.298){\vector(0,1){.07}}\put(21.554,33.298){\line(0,-1){5.0541}}
\end{picture}

The root system $R_{B_n}$ contains $2n$ short roots $\pm e_j\,$ $(j=1,\dots n)$ and $2n(n-1)$
long roots $(e_j\pm e_k)$ $(j\neq k)$. The positive roots are
\beq{prb1}
R^+=\{(e_j\pm e_k)\,,~(j>k)\,,~~e_j\,,~~(j,k=1,\ldots,n)\}\,.
\eq
The Weyl group of $R_{B_n}$ is the semi-direct product of the permutation group $S_n$
acting on $e_j$ and the signs changing   $e_j\to -e_j$
\beq{web}
W_{B_n}=S_n\ltimes(\sum\mu_2)\,.
\eq

The coroot basis  $H_\al$ in $\gH_{B_n}$  is formed by simple roots
of the dual system
\beq{src}
\Pi^\vee_{B_n}=\Pi_{C_n}=\{\al_j=e_j-e_{j+1}\,,~j=1,\ldots,n-1\,,~~\al_n=2e_n\}\,.
\eq
$\Pi^\vee_{B_n}$ generates the coroot lattice
\beq{bcrl3}
Q_{B_n}^\vee=Q_{C_n}=\{\ga=\sum_{j=1}^nm_je_j\,|\,\sum_{j=1}^nm_j\,{\rm is~even}\}\,.
\eq
The half-sum of positive coroots is $\rho_{B_n}^\vee=ne_2+(n-1)e_2+\ldots+2e_{n-1}+e_n$,
and since  $h=2n$
\beq{iv1}
\ka_{B_n}=\rho/h=\oh e_1+(\oh-\f1{2n})e_2+\ldots+\f1{n}e_{n-1}+\f1{2n}e_n)\,.
\eq

The dual to $\Pi_{B_n}$ the system of the fundamental coweights takes the
form
$$
\{\varpi_j^\vee=e_1+e_2+\ldots+e_j\,,~j=1,\ldots,n\}\,.
$$
It generates the coweight lattice
\beq{bcrl1}
P^\vee_{B_n}=P_{C_n}=\{\ga=\sum_{j=1}^nm_je_j\,|\,m_j\in\mZ\}\,.
\eq
The factor group $P_{B_n}^\vee/Q_{B_n}^\vee\sim\mu_2$ is isomorphic to the center of $Spin(2n+1)$.
The center is generated by the coweight $\varpi^\vee_1\,$ $(\zeta=\bfe\,(\varpi^\vee_1))$.
It follows from  (A.17.I) and (\ref{mbr1}) that
 the fundamental alcove  has the vertices
$$
 C_{alc}=(0,\varpi^\vee_1,\oh\varpi^\vee_2,\ldots,\oh\varpi^\vee_n)\,.
 $$
 We use this expression to find $\la_1$ corresponding  to $\xi=\varpi^\vee_1$:
 (3.10.I)
 \beq{lab}
 \la_1:\,(e_1\to -e_1\,,e_j\to e_j\,,~(j=2,\ldots,n))\,.
 \eq
It acts on $\Pi^{ext}_{B_n}$ as
$$
\la_1:\,(\al_1\to\al_0\,,~\al_j\to\al_j\,,~(j=2,\ldots,n))\,.
 $$
 It is clear that $\ti\gH=\gH_{B_{n-1}}$ and the moduli vector is  $\ti\bfu=(u_2,\ldots,u_n)$.


\subsection*{Bases.}

\subsubsection*{The Chevalley basis}

 The Chevalley basis in $\gg_{B_n}$ is generated by the simple coroots
 $\Pi^\vee_{C_n}=\Pi_{B_n}$
(\ref{bsr1}) in $\gH_{B_n}$ and by the root subspaces.
To describe the Chevalley basis for the classical groups we use their fundamental representations.
 For $\gg_{B_n}$ it is
 a fundamental representation $\pi_1$
 corresponding to the weight $\varpi_1=e_1$. It has dimension $2n+1$. $\gg_{B_n}$
   becomes the Lie algebra
 of matrices of order  $2n+1$ satisfying the constraints
$$
 Zq+qZ^T=0\,,
 $$
where $q$ in the bilinear  form
$$
q=\left(
    \begin{array}{ccc}
      0 & 0& J \\
      0&1&0\\
      J &0&   0 \\
    \end{array}
  \right)\,,
 $$
and $J$ is an $n$-th order anti-diagonal matrix
 \beq{15}
J=\left(
    \begin{array}{cccc}
        &  &    & 1 \\
       &     & 1 &  \\
       &  &  &  \\
      1&  &  &  \\
    \end{array}
  \right)
\eq
 Then $Z$ takes the form
\beq{31b}
Z=\left(
  \begin{array}{ccc}
    A &\al& B \\
   \ti\be^T&0&-\ti\al^T\\
    C &-\be& -\ti A \\
  \end{array}
\right)\,,~~
B=\ti{B}\,,~C=\ti{C}\,,~~\ti{X}=JX^TJ\,,
\eq
$$
\al=(\al_1,\ldots,\al_n)\,,~~~\be=(\be_1,\ldots,\be_n)\,,~~~~\ti\al=\al J\,.
$$

The basis in $\gH_{B_n}$ is generated  by the canonical basis $(e_1,\ldots,e_n)$. Then the canonical basis in $\pi_1(\gH)$ is
 $$
 \di(e_1,\ldots,e_n,0,-e_n,\ldots,-e_1)\,.
 $$
 The  root subspaces in $\pi_1$ are
 \beq{chbb}
 (e_j-e_k)\,,~j\neq k\to\gG^-_{jk}=(E_{j,k}\in A-E_{2n+2-k,2n+2-j}\in A\,, \ti A)\,,
 \eq
 $$
j<k\sim({\rm positive ~roots})\,,~~j>k\sim({\rm negative ~roots})\,,
$$
$$
( e_j+ e_k)\,,\to\gG^+_{jk}=(E_{j,k+n+1}-E_{n+1-k,2n+2-j})\in B\,,C \,,
$$
$$
e_j\,,\to\gG^+_{j}=\sqrt{2}( E_{j,n+1}-E_{n+1.2n+2-j})\in\al \,,\,({\rm positive ~roots}) \,,
$$
$$
-e_j\,,\to \gG^-_{j}=\sqrt{2}(E_{n+1,j}-E_{2n+2-j,n+1})\in\be\,,\, ({\rm negative ~roots}) \,.
$$
The levels of positive roots  are
\beq{rlb1}
f_{e_j-e_k}=k-j\,,~f_{e_j}=n-j+1\,,~f_{e_j+e_k}=2n-k-j+22n\,.
\eq

The Killing form normalized as in (A.24.I), (A.25.I) takes the form
\beq{kfb1}
(Z_1,Z_2)=\f1{2}\tr\,Z_1Z_2\,.
\eq
Then
\beq{kfb}
(\gG_{jk}^\pm,\gG_{il}^\pm)=\de_{ki}\de_{jl}\,,~~~(\gG_j^+,\gG_k^-)=2\de_{jk}\,.
\eq


\subsubsection*{The GS-basis}

The Weyl transformation $\La_1$ in this basis is represented by the matrix
$$
\La_1=
\left(
  \begin{array}{ccc}
    0 & 0 & 1 \\
    0 & -Id_{2n-1} & 0 \\
    1 & 0 & 0 \\
  \end{array}
\right)\,.
$$
Taking into account its action on $Z$ (\ref{31b}) we find that $\Pi_1=\Pi_{B_{n-1}}$ and
$\ti\Pi^\vee=\Pi^\vee_{B_{n-1}}$.  Then
$$
\gg_{B_{n}}=\gg_0+\gg_1\,,~~~\gg_0=\ti\gg_0+V\,,~~\ti\gg_0=\gg_{B_{n-1}}\,,
$$

$$
\dim \,\gg_{B_{n-1}}=(n-1)(2n-1) \,,~~\dim \,V=2(n-1)+1\,,~~\dim \,\gg_1=2(n-1)+2\,,
$$
where $V$ is the vector representation of $\ti\gg_0=\bfso(2n-1)$.
The invariant subalgebra $\gg_0$ is isomorphic to $\bfso(2n)$.
Its Dynkin diagram is obtained from $\Pi^{ext}_{B_n}$ by dropping out the root $\al_n$  \cite{Ka}.

The space $V+\gg_1$ is represented by matrices of the form
$$
\left(
    \begin{array}{ccccc}
      a & \vec x & \al & \vec y & 0 \\
      (\vec z) J & 0 & 0 & 0 & -(\vec y)J  \\
      \be & 0 & 0 & 0 & -\al \\
      (\vec w) J & 0 & 0 & 0 & -(\vec x) J \\
      0 & -\vec w & -\be & -\vec z & -a \\
    \end{array}
  \right)\,.
$$
Since $\ti\gg_0$ is generated by a trivial orbit $(l=1)$,
the GS-basis is formed by the Chevalley basis in $\gg_{B_{n-1}}$  (\ref{chbb})
and the GS generators in $V+\gg_1$.
$$
V=\left\{
\begin{array}{ll}
\gt^0_{1,k}=\f1{\sqrt{2}}(E_{1,k}+E_{2n+1,k}-\ldots)\, &\gt^0_{1,n+1+k}=\f1{\sqrt{2}}(E_{1,n+1+k}+E_{2n+1,n+1+k}-\ldots)\,, \\
\gt^0_{j+n+1,1}=\f1{\sqrt{2}}(E_{j+n+1,1}+E_{j+n+1,2n+1}-\ldots)\,, &
\gt^0_{j,1}=\f1{\sqrt{2}}(E_{j,1}+E_{j,2n+1}-\ldots)\,, \\
\gt^0_{n+1,1}=(E_{n+1,1}+E_{n+1,2n+1}-\ldots)\,, &
\gt^0_{1,n+1}=(E_{1,n+1}+E_{2n+1,n+1}-\ldots)\,,
    \end{array}
    \right.
$$
$$
\gg_1=\left\{
\begin{array}{ll}
\gt^1_{1,k}=\f1{\sqrt{2}}(E_{1,k}-E_{2n+1,k}-\ldots)\, &\gt^1_{1,n+1+k}=\f1{\sqrt{2}}(E_{1,n+1+k}-E_{2n+1,n+1+k}-\ldots)\,, \\
\gt^1_{j+n+1,1}=\f1{\sqrt{2}}(E_{j+n+1,1}-E_{j+n+1,2n+1}-\ldots)\,, &
\gt^1_{j,1}=\f1{\sqrt{2}}(E_{j,1}-E_{j,2n+1}-\ldots)\,, \\
\gt^1_{n+1,1}=\f1{\sqrt{2}}(E_{n+1,1}-E_{n+1,2n+1}-\ldots)\,, &
\gt^1_{1,n+1}=\f1{\sqrt{2}}(E_{1,n+1}-E_{2n+1,n+1}-\ldots)\,,\\
\gh^1_1=\sqrt{2}\di(e_1,0\ldots,0-e_1) & \,.
    \end{array}
    \right.
$$

From (5.9.I) (5.35.I)  and (\ref{kfb1}) we find the dual basis
\beq{dbb}
\gT^a_{jk}=\gt^a_{kj}\,,~~j~{\rm or}~k\neq n+1\,,~~\gH^1_1=\gh^1_1\,,~~
\gT^a_{n+1,1}=\oh\gt^a_{1,n+1}\,,~~\gT^a_{1,n+1}=\oh\gt^a_{n+1,1}\,.
\eq
and
\beq{bpa}
(\gt^a_{1j},\gt^b_{i1})=\left\{
\begin{array}{cc}
  \de^{(ab)}\de_{(ji)} & i,j\neq n+1\,, \\
  2\de^{(ab)}\de_{(ji)}  &  i=n+1\,,
\end{array}
\right.~~~
(\gh^1_1,\gh^1_1)=4\,.
\eq


\subsection*{The Lax operators and the Hamiltonians}

\subsubsection*{Trivial bundles.}

For trivial bundles the moduli space is described by the vector
$\bfu=(u_1,\ldots,u_n)$.
For a trivial $\bar G=Spin(2n+1)$-bundles it means that
\beq{csc1}
\bfu\in\gH_{B_n}/W_{B_n}\ltimes(\tau Q^\vee\oplus Q^\vee)\,,
\eq
where $W_{B_n}$ is defined in (\ref{web}).
For trivial $SO(2n+1)$-bundles we have
\beq{csc2}
\bfu\in\gH_{B_n}/W_{B_n}\ltimes(\tau P^\vee\oplus P^\vee)\,.
\eq
The dual variables $\bfv=(v_1,\ldots,v_n)\,$ $v_j\in\mC$
are the same in the both cases.
The standard CM Lax operator in the Chevalley basis  is
\beq{tlb}
L^{CM}_{B_n}(z)=\sum_{j=1}^n(v_j+S_{0,j}E_1(z))\gE_j+\sum_{j\neq k}S_{jk}\phi(u_j-u_k,z)\gG^-_{jk}
\eq
$$
+\sum_{j\neq k}S_{j,-k}\phi(u_j+u_k,z)\gG^+_{j,k}+
\sum_{j}S^{\pm}_j\phi(u_j,z)\gG^{\pm} _j\,,
$$
where $\gE_j=\di(0, \ldots,1,0,\dots,0,-1,0,\ldots,0) $.
The quadratic Hamiltonian after the reduction with respect to the Cartan subgroup takes the form (see (\ref{kfb}))
\beq{bth}
H^{CM}_{B_n}=\oh\sum_{j=1}^nv^2_j-\oh\sum_{j\neq k}((S_{jk}S_{kj}E_2(u_j-u_k)+
S_{j,-k}S_{k,-j}E_2(u_j+u_k))-\sum_{j}S^+_jS^-_jE_2(u_j,z)\,.
\eq
Thus, we have two types of the standard CM systems with the same Hamiltonians and
different configuration spaces described by (\ref{csc1}) and (\ref{csc2}).

\subsubsection*{Nontrivial bundles}

The GS-basis is described above.  Now $\ti\bfu=(u_2,u_3,\ldots,u_n)$
belongs to $\gH_{B_{n-1}}$. In fact, $\ti\bfu$ belongs to one of fundamental domains
in $\gH_{B_{n-1}}$ (\ref{csc1}), (\ref{csc2})
 under the action of the coweight and the coroot lattices.
 Taking into account (\ref{rlb1}) and $h=2n$ we find from the general
 prescription
$$
L_1(z)=S^1_1\phi(\oh,z)\gh^1_1+S^1_{n+1,1}\bfe\,(\frac{z}2)\phi(\frac{1+\tau}2,z)\gt_{1,n+1}^1+
S^1_{1,n+1}\bfe\,(-\frac{z}2)\phi(\frac{1+\tau}2,z)\gt_{n+1,1}^1+
$$
$$
\sum_{k=2}^n\Bigl(S^1_{k,1}\bfe\,((\frac{k-1}{2n})z)\phi(\frac{(k-1)\tau}{2n}-u_k+\oh,z)\gt_{1,k}^1
+S^1_{1,k}\bfe\,((\frac{1-k}{2n})z)\phi(\frac{(1-k)\tau}{2n}+u_k+\oh,z)\gt_{k,1}^1
$$
$$
+S^1_{n+1+k,1}\bfe\,(\frac{2n+1-k}{2n}z)\phi(\frac{(2n+1-k)\tau}{2n}-u_k+\oh,z)\gt_{1,n+1+k}^1
$$
$$
+S^1_{1,n+1+k}\bfe\,(-\frac{2n+1-k}{2n}z)\phi(u_k-\frac{(2n+1-k)\tau}{2n}+\oh,z)\gt_{k,1}^1
\Bigr)\,,
$$
$$
L'_0(z)=S'_{n+1,1}\bfe\,(\frac{z}2)\phi(\frac{\tau}2,z)\gt_{1,n+1}^0+
S'_{1,n+1}\bfe\,(-\frac{z}2)\phi(-\frac{\tau}2,z)\gt_{n+1,1}^0+
$$
$$
\sum_{k=2}^n\Bigl(S'_{k,1}\bfe\,((\frac{k-1}{2n})z)\phi(\frac{(k-1)\tau}{2n}-u_k,z)\gt_{1,k}^0
+S'_{1,k}\bfe\,((\frac{1-k}{2n})z)\phi(\frac{(1-k)\tau}{2n}+u_k,z)\gt_{k,1}^0
$$
$$
+S'_{n+1+k,1}\bfe\,(\frac{2n+1-k}{2n}z)\phi(\frac{(2n+1-k)\tau}{2n}-u_k,z)\gt_{1,n+1+k}^0
$$
$$
+S'_{1,n+1+k}\bfe\,(-\frac{2n+1-k}{2n}z)\phi(u_k-\frac{(2n+1-k)\tau}{2n},z)\gt_{k,1}^0
\Bigr)\,.
$$
The Lax operator $\ti L_0(z)$ coincides with (\ref{tlb}) after the corresponding replacement
of indices. Taking into account the pairing (\ref{bpa})  after the reduction we come to the Hamiltonian
$$
H=H^{CM}_{B_{n-1}}+H'+H_1\,,
$$
$$
-H'=S'_{n+1,1}S'_{1,n+1}E_2(\frac{\tau}2)+\oh\sum_{k=2}^n\Bigl(S'_{k,1}S'_{1,k}
E_2(u_k-\frac{(k-1)\tau}{2n})+S'_{n+1+k,1}S'_{1,n+1+k}E_2(\frac{(2n+1-k)\tau}{2n}-u_k)
\Bigr)\,,
$$
$$
-H_1=S^1_{n+1,1}S^1_{1,n+1}E_2(\frac{1+\tau}2)+\oh (S^1_1)^2E_2(\oh)+
$$
$$
\oh\sum_{k=2}^n\Bigl(S^1_{k,1}S^1_{1,k}
E_2(u_k-\frac{(k-1)\tau}{2n}-\oh)+S'_{n+1+k,1}S'_{1,n+1+k}E_2(\frac{(2n+1-k)\tau}{2n}-u_k-\oh)
\Bigr)\,.
$$
Again, we have two types of  systems with the same Hamiltonians and
different configuration spaces described by (\ref{csc1}) and (\ref{csc2}), where
$u_1$ is omitted.


\section{Sp$(n)$ - $C_n$ root system.}
\setcounter{equation}{0}

\subsection*{Roots, weights and bases.}

The algebra Lie $\gg_{C_n}$ has rank $n$ and $\dim\,(\gg_{C_n})=2n^2+n$.
The system of simple roots $\Pi_{C_n}$ is defined in (\ref{src}).
The minimal root is
\beq{mrcn}
-\al_0=2e_1=2\sum_{j=1}^{n-1}\al_j+\al_n\,.
\eq
There are $2n$ long roots $2e_{\pm j}$ and $2n(n-1)$ short roots
$\pm e_j\pm e_k$, $j\neq k$.
The Weyl group $W_{C_n}$  of $R(C_n)$ coincides with $W_{B_n}$ (\ref{web}).

The levels of positive roots (A.3.I) are
\beq{rlb}
f_{e_j-e_k}=k-j\,,~f_{2e_j}=2n-2j+1\,,~f_{e_j+e_k}=2n-k-j+1\,.
\eq

The simple coroots $\Pi^\vee_{C_n}=\Pi_{B_n}$
(\ref{bsr1})   generates the coroot lattice
\beq{bcrl}
Q_{C_n}^\vee=Q_{B_n}=\{\ga=\sum_{j=1}^nm_je_j\,|\,m_j\in\mZ\}\,.
\eq
 Since$
\rho_{C_n}^\vee=(n-\oh)e_1+(n-\frac{3}2)e_2+\ldots+\frac{3}2e_{n-1}+\oh e_n$,
and  $h=2n$
\beq{icv}
\ka_{C_n}=\rho/h=(\oh-\frac{1}{4n})e_1+(\oh-\frac{3}{4n})e_2+\ldots+
\frac{3}{4n}e_{n-1}+\frac{1}{4n}e_n\,.
\eq

The dual to $\Pi_{C_n}$ the system of the fundamental coweights takes the
form
$$
\{\varpi_j^\vee=e_1+e_2+\ldots+e_j\,,~j=1,\ldots,n-1\,,~
\varpi_n^\vee=\oh\sum_{j=1}^ne_j\}\,.
$$
It generates the coweight lattice
\beq{bcwl}
P_{C_n}^\vee=\{\ga=\sum_{j=1}^n\mZ e_j+\mZ\bigl(\oh\sum_{j=1}^n e_j\bigr)\}\,.
\eq
The factor-group $P^\vee_{C_n}/Q^\vee_{C_n}\sim\mu_2$ is isomorphic to the center of $Sp(n)$.
The center is generated by the coweight $\varpi^\vee_n\,$ $(\zeta=\bfe\,(\varpi^\vee_n))$.
It follows from  (A.17.I) and (\ref{mrcn}) that
 the fundamental alcove  has the vertices
$$
 C_{alc}=(0,\oh\varpi^\vee_1,\ldots,\oh\varpi^\vee_{n-1},\varpi^\vee_{n})\,.
 $$
 We use this expression to find $\la_n$ corresponding  to $\xi=\varpi^\vee_n$
 (3.10.I). $\la_n$ acts on the vertices of $ C_{alc}$ and, in this way, on $\varpi^\vee_j$, as
$$
 \la_n:\,(\varpi^\vee_1\leftrightarrow\varpi^\vee_{n-1}\,,\,\varpi^\vee_2\leftrightarrow\varpi^\vee_{n-2}\,,
 \ldots,\varpi^\vee_n\leftrightarrow  0 )\,.
$$
Then its action on $\Pi^{ext}_{C_n}$ assumes the form
$$
\la_n:\,(\al_0\leftrightarrow\al_{n}\,,\al_2\leftrightarrow\al_{n-2}\,,\ldots\al_n\leftrightarrow~\al_{0},)\,.
 $$
 The action of $\la_n$ on the canonical basis $(e_1,e_2,\ldots, e_n)$ takes
 the form
 \beq{ccb}
  \la_n:\,(e_1\leftrightarrow -e_n\,,e_2\leftrightarrow -e_{n-1}\,,\ldots)\,.
 \eq
If $\bfu=\sum_ju_je_j$, then
$ \la_n:\,(u_1\leftrightarrow -u_n\,,u_2\leftrightarrow -u_{n-1}\,,\ldots)$.
We define the invariant vector $\ti\bfu$ in the invariant basis
$\ti e_1=e_1-e_n\,,\,\ti e_2=e_2-e_{n-1}\,,\ldots,\ti e_l=e_l-e_{l+1}$,
$l=\Bigl[\frac{n}2\Bigr]$
 \beq{modc}
\ti\bfu  =\sum_{j=1}^lu_j\ti e_j\,.
  \eq

\begin{flushright}
\unitlength 1mm 
\linethickness{0.4pt}
\ifx\plotpoint\undefined\newsavebox{\plotpoint}\fi 
\begin{picture}(62.142,100.099)(0,0)
\put(-74.426,46.563){\circle*{3.64}}
\put(-72.426,47.813){\line(1,0){9.5}}
\multiput(-72.426,45.563)(1.125,.03125){8}{\line(1,0){1.125}}
\put(-63.426,45.813){\line(-1,0){.5}}
\put(-61.676,46.563){\circle{3.162}}
\put(-59.926,46.563){\line(1,0){10.5}}
\put(-47.176,46.563){\circle{3.041}}
\put(-6.176,47.313){\circle{3.202}}
\multiput(-4.426,48.563)(-.03125,-.03125){8}{\line(0,-1){.03125}}
\multiput(-4.676,48.313)(.03125,-.125){8}{\line(0,-1){.125}}
\put(-4.426,47.313){\line(-1,0){.25}}
\multiput(-4.676,47.313)(1.625,.03125){8}{\line(1,0){1.625}}
\put(10.574,47.313){\circle{3.536}}
\multiput(12.824,48.063)(1,.03125){8}{\line(1,0){1}}
\put(20.824,48.313){\line(1,0){3.25}}
\multiput(12.824,46.813)(1.34375,.03125){8}{\line(1,0){1.34375}}
\put(25.324,47.563){\circle{3.041}}
\multiput(-69.176,48.313)(.07222222,-.03333333){45}{\line(1,0){.07222222}}
\multiput(-65.926,46.813)(-.05416667,-.03333333){60}{\line(-1,0){.05416667}}
\multiput(15.574,47.563)(.07083333,.03333333){60}{\line(1,0){.07083333}}
\multiput(16.324,47.313)(.09210526,-.03289474){38}{\line(1,0){.09210526}}
\qbezier(-70.426,56.063)(-29.551,101.938)(17.824,54.313)
\qbezier(-54.926,56.313)(-30.426,77.688)(1.074,57.563)
\qbezier(-38.426,55.563)(-28.176,64.688)(-14.926,55.313)
\put(-74.426,34.313){$\alpha_0$}
\put(-61.426,35.313){$\alpha_1$}
\put(10.074,37.563){$\alpha_{n-1}$}
\put(24.824,39.063){$\alpha_n$}
\multiput(-68.176,60.063)(-.03365385,-.05769231){52}{\line(0,-1){.05769231}}
\multiput(-69.926,56.063)(.04104478,.03358209){67}{\line(1,0){.04104478}}
\multiput(-52.926,59.313)(-.03333333,-.05){60}{\line(0,-1){.05}}
\multiput(-54.926,56.313)(.07894737,.03289474){38}{\line(1,0){.07894737}}
\multiput(-37.176,57.563)(-.03289474,-.04605263){38}{\line(0,-1){.04605263}}
\multiput(-38.176,55.813)(.1086957,.0326087){23}{\line(1,0){.1086957}}
\multiput(-18.676,56.813)(.07894737,-.03289474){38}{\line(1,0){.07894737}}
\multiput(-17.176,58.563)(.03333333,-.05555556){45}{\line(0,-1){.05555556}}
\multiput(-3.176,59.063)(.08552632,-.03289474){38}{\line(1,0){.08552632}}
\multiput(.324,58.063)(-.03353659,.03658537){82}{\line(0,1){.03658537}}
\multiput(14.574,56.563)(.04477612,-.03358209){67}{\line(1,0){.04477612}}
\multiput(15.574,58.063)(.03333333,-.06666667){45}{\line(0,-1){.06666667}}
\multiput(17.074,55.063)(-.0333333,.0333333){15}{\line(0,1){.0333333}}
\put(-39.747,46.993){\line(1,0){1.8958}}
\put(-35.955,46.993){\line(1,0){1.8958}}
\put(-32.163,46.993){\line(1,0){1.8958}}
\put(-28.372,46.993){\line(1,0){1.8958}}
\put(-24.58,46.993){\line(1,0){1.8958}}
\put(-20.788,46.993){\line(1,0){1.8958}}
\put(-51.097,28.414){\makebox(0,0)[cc]{Fig.3 C$_n$ and $\lambda_n$ action}}
\end{picture}
\end{flushright}


\subsection*{Bases.}

\subsubsection*{The Chevalley basis}
 The Chevalley basis in $\gg_{C_n}$ is generated by the simple coroots
 $\Pi^\vee_{C_n}=\Pi_{B_n}$
(\ref{bsr1}) in $\gH$ and by the root subspaces.
It is convenient
to define the Chevalley basis using a fundamental representation $\pi_1$
 corresponding to the
weight $\varpi_1=e_1$. It has dimension $2n$. We define it as the Lie algebra
 of matrices $\{Z\}$ satisfying the constraints
$$
 Zq+qZ^T=0\,,~~Z\in\pi_1\,,
 $$
where $q$ in the bilinear  form
$$
q=\left(
    \begin{array}{cc}
      0 &  J \\
      -J &  0 \\
    \end{array}
  \right)\,,
 $$
and $J$ is an $n$-th order anti-diagonal matrix (\ref{15}).
 Then $Z$ takes the form
\beq{31a}
Z=\left(
  \begin{array}{cc}
    A & B \\
    C & -\ti A \\
  \end{array}
\right)\,,~~
B=\ti{B}\,,~C=\ti{C}\,,~~\ti{X}=JX^TJ\,.
\eq
The basis in $\gH_{C_n}$ is generated  by the canonical basis $(e_1,\ldots,e_n)$. Then the canonical basis in $\pi_1(\gH)$ is
 $$
 \di(e_1,\ldots,e_n,-e_n,\ldots,-e_1)\,.
 $$
 The Killing form is similar to the $B_n$ case (\ref{kfb1})
 $(Z_1,Z_2)=\f1{2}\tr\,Z_1Z_2$.

 The  root subspaces in $\pi_1$ are
 \beq{chbc}
\begin{array}{cl}
 (e_j-e_k)&\to\gG_{jk}^-=(E_{j,k}\in A-E_{2n+1-k,2n+1-j}\in \ti A),,\\
( e_j+ e_k) &\to\gG_{jk}^+=(E_{j,k+n}\in B+E_{n+1-k,2n+1-j}\in C)\,,
\end{array}
\eq
$$
j<k~ \sim~({\rm positive ~roots})\,,~~~j>k~\sim~({\rm negative ~roots})
$$
$$
2e_j\to \gG^+_j=E_{j,j+n}\in B\,,~({\rm positive ~roots}) \,, ~~
-2e_j\to \gG_j^-= E_{j+n,j}\in C ~({\rm negative ~roots})\,.
$$
The Killing form in this basis is
\beq{cbkfc}
(\gG_{jk}^\pm,\gG_{il}^\pm)=\de_{ki}\de_{jl}\,,~~~
(\gG^+_k,\gG_j^-)=\oh\de_{jk}\,.
\eq


\subsubsection*{The GS-basis}
 The transformation $\la_n$ (\ref{ccb}) in $\pi_1$ takes the form
 \beq{lncn}
 \La_n=
 \left(
   \begin{array}{cc}
     0 & i\,Id_n \\
     i\,Id_n & 0 \\
   \end{array}
 \right)\,,~~\la_n(Z)=
 \left(
  \begin{array}{cc}
   -\ti A  & C \\
    B &  A \\
  \end{array}
\right)\,
 \eq
 Since $\la_n^2=Id$
 \beq{cgr2}
 \gg_{C_n}=\gg_0+\gg_1\,,
 \eq
 where
 $$
 \gg_0=\left\{
 \left(
   \begin{array}{cc}
     X & Y \\
     Y & X \\
   \end{array}
 \right)\,\begin{array}{lc}
            | & \ti X=-X \\
            | & \ti Y=Y
          \end{array}
 \right\}\,,
~~~
 \gg_1=\left\{
 \left(
   \begin{array}{cc}
     X & Y \\
     -Y & -X \\
   \end{array}
 \right)\,\begin{array}{lc}
            | & \ti X=X \\
            | & \ti Y=Y
          \end{array}
 \right\}\,,
 $$
\beq{cgr3}
\gg_0=\ti\gg_0+V\,,~~\ti\gg_0
=\left\{
 \left(
   \begin{array}{cc}
     X & 0 \\
     0 & X \\
   \end{array}
 \right)\,\right\}\,,~~V=\left\{
 \left(
   \begin{array}{cc}
     0 & Y \\
     Y & 0 \\
   \end{array}
 \right) \right\}\,,
\eq
$$
\dim\,\gg_{C_n}=n(2n+1)\,,~~\dim\,\ti\gg_0=\oh n(n-1)\,,~~\dim\,V=\oh n(n+1)\,,~~
\dim\,\gg_1=n(n+1)\,.
$$

A type of $\ti\gg_0$ depends on a parity of $n$.
Note, that $\Pi_1=A_{n-1}$ (\ref{dpi1}) is
generated by roots $(\al_1,\ldots,\al_{n-1})$.
 Let $\la_n|_{\gH_{A_{n-1}}}=\ti\la$.
We prove general Lemma about automorphisms of $\gg_{A_{n-1}}$ that will be applied to $C_n$ and $D_n$ algebras.
Let $(e_1,\ldots,e_{n})$ be a canonical basis in $\gH_{A_{n-1}}$ and
$E_{j,k},$ $\,(1\leq j<k\leq n)$ is the root basis $\gg_{A_{n-1}}$
=sl$(n,\mC)$.
It follows from  (\ref{ccb}) and (\ref{lncn}) that
the action of $\ti\la$ on the Chevalley basis of $\gH_{A_{n-1}}$
takes the form
\beq{laca}
\ti\la\,:\,
\left\{
\begin{array}{l}
  (e_1,e_2\ldots,e_{n})\,\to\,(-e_{n},-e_{n-1},\ldots,-e_1)\, \\
  \,E_{jk}\,\to\,-E_{n-k+1,n-j+1}\,.
\end{array}
.\right.
\eq
\begin{lem}
 Under the action (\ref{laca}) $\gg_{A_{n-1}}$
is decomposed as
$$
\gg_{A_{n-1}}=\ti\gg_0+\ti\gg_1\,,
$$
where\\
$\bullet$
$$
\ti\gg_0=\left\{
\begin{array}{cc}
  \gg_{D_{\frac{n}2}} & n-{\rm even}\,, \\
  \gg_{B_{\frac{n-1}2}} & n-{\rm odd}\,.
\end{array}
\right.
$$
with the defined below Chevalley bases (\ref{bsoe}), (\ref{bsoo}).\\
$\bullet$ $\ti\gg_1$ is a space of traceless symmetric matrices of order $n$
$(\dim\,(\ti\gg_1)=\oh n(n+1)$\,,
and $\ti\gg_0$ acts on $\ti\gg_1$  by commutators.
\end{lem}
\emph{Proof}\\
Let $X\in$sl$(n,\mC)$ be a traceless matrix of order $n$ and
$$
\ti X=JX^TJ\,,~~ J_{ik}=\de_{i,n-1-k}\,.
$$
It follows from  (\ref{laca}) that
 $\ti\la(X)=-\ti X$ and the invariant subalgebra
$$
\ti\gg_0=\gg_{A_{n-1}}+\ti\la(\gg_{A_{n-1}})
$$
is the algebra of anti-invariant matrices
with respect to the symmetric form $J_{jk}$
$$
XJ+JX^T=0\,.
$$
Thus,  $\ti\gg_0$ is the Lie algebra of orthogonal matrices SO$(n,\mC)$
and we come to the first statement.

Similarly, for $B\in\ti\gg_1$
\beq{B}
YJ-JY^T=0\,.
\eq
It means that $\ti\gg_1$ is the space of complex symmetric matrices with
respect to the secondary diagonal.
In these terms the commutation relations in $A_{n-1}=$sl$(n,\mC)$ assume the form
$$
[\ti\gg_0,\ti\gg_0]\subset\ti\gg_0\,,~~
[\ti\gg_0,\ti\gg_1]\subset\ti\gg_1\,,~~
[\ti\gg_1,\ti\gg_1]\subset\ti\gg_0\,.
$$

We construct the Chevalley basis in the invariant subalgebras.
The $\ti\la$ on the simple roots of $A_{n-1}$ is
$$
\al_1\leftrightarrow\al_{n-1}\,,\al_2\leftrightarrow\al_{n-2}\,,\ldots,
\left\{
\begin{array}{lc}
   \al_{\frac{n}2-1}\leftrightarrow\al_{\frac{n}2+1}&  n \,{\rm even}\,, \\
  \al_{\frac{n}2}\leftrightarrow\al_{\frac{n}2+1}&  n \,{\rm even}\,, \\
  \al_{\frac{n-1}2-1}\leftrightarrow\al_{\frac{n-1}2+2}&  n \,{\rm odd}\,, \\
  \al_{\frac{n-1}2}\leftrightarrow\al_{\frac{n-1}2+1}&  n \,{\rm odd}\,, \\
\end{array}
\right.
$$
For $n=2l$ $\,\ti\gg_0=\gg_{D_l}={\bf so}(2l,\mC)$ the Chevalley basis  is formed by the
 invariant coroots $\{\tial_j^\vee,$ $(j=1,\dots,l)\}$, constructed from $A_{2l-1}$ roots
$\al_1,\ldots,\al_{2l-1}$, and the root spaces basis
\beq{kc}
\ti\Pi^\vee_{D_l}=\{\tial_1^\vee=\al_1+\al_{2l-1}\,,\ldots,\tial^\vee_{l-1}=\al_{l-1}+\al_{l+1}\,,\,
\tial_l^\vee=\al_{l-1}+2\al_l+\al_{l+1}\}=
\eq
$$
\{\tial_1^\vee=e_1-e_2+e_{n-1}-e_n\,,\ldots\tial^\vee_{l-1}=e_{\frac{n}2-1}-
e_{\frac{n}2}+e_{\frac{n}2+1}-e_{\frac{n}2+2}\,,\tial^\vee_{l}=
e_{\frac{n}2-1}+e_{\frac{n}2}-e_{\frac{n}2+1}-e_{\frac{n}2+2}\}\,.
$$
The simple roots dual to coroots are
$\tial_j=\oh \tial_j^\vee$. The matrix $a_{jk}=\lan\tial_j,\tial_k^\vee\ran$
is the Cartan matrix $D_l$. The Chevalley generators corresponding to the simple roots are
\beq{bsoe}
\ti E_{\tial_j}=E_{j,j+1}-E_{n-j,n-j+1}\,,~ j<l\,,
~~~\ti E_{\tial_l}=E_{l-1,l+1}-E_{l,l+2}\,.
\eq

For $n=2l+1$ the Chevalley basis of the $\ti\gg_0=\gg_{B_l}={\bf so}(2l+1,\mC)$-algebra
(see Section 9) takes the form
\beq{cibb}
\ti\Pi^\vee_{B_l}=\{\tial_1^\vee=\al_1+\al_{2l}\,,\ldots,
\tial_{l-1}^\vee=\al_{l-1}+\al_{l+2}\,,
\tial_l^\vee=2(\al_l+\al_{l+1})\}=
\eq
$$
\{\tial_1^\vee=e_1-e_2+e_{n-1}-e_n\,,\ldots\tial^\vee_{l-1}=e_{\frac{n-1}2-1}-
e_{\frac{n-1}2}+e_{\frac{n-1}2+1}-e_{\frac{n-1}2+2}\,,\tial^\vee_{l}=
2(e_{\frac{n-1}2-1}-e_{\frac{n-1}2+2})\}\,.
$$
\beq{bsoo}
\ti E_{\tial_j}=E_{j,j+1}-E_{n-j,n-j+1}\,,~ j<l\,,
~~~\ti E_{\tial_l}=E_{l,l+1}-E_{l+1,l+2}\,,~ j=l\,.
\eq
The dual root system
$$
\ti\Pi_{B_l}=\{\tial_1=\oh\tial_1^\vee\,,\ldots,\tial_{l-1}=\oh\tial_{n-1}^\vee\,,
\tial_l=\f1{4}\tial_l^\vee\}\,.
$$
leads to the $B_l$ Cartan matrix $a_{jk}=\lan\ti\al_j,\ti\al_k^\vee\ran$.
$\Box$

\begin{rem}
In this Lemma we have found the coroot basis in the invariant subalgebra $\ti \gH_0$
for the special cases $C_l$, $D_l$ . The expressions (\ref{kc} ), (\ref{cibb} )
for $\ti\al^\vee_l$ replace the general formula (5.26.I).
\end{rem}
\bigskip

\emph{\textbf{Basis in $V$ and $\gg_1$}}

We have constructed a basis in $\ti \gg_0$. Consider other component in (\ref{cgr2}), (\ref{cgr3}).
The basis in $V=\{B\}$, where $B$ satisfies (\ref{B}), has the form
\beq{BV}
\gt^0_{j,k+n}=\f1{\sqrt{2}}(E_{j,k+n}+E_{n+j,k}+E_{n-k+1,2n-j+1}+E_{2n-k+1,n-j+1})\,,~(j\neq k)\,,
\eq
$$
\gt^0_{j,n+j}=\f1{\sqrt{2}}(E_{j,n+j}+E_{n+j,j}) \,.
$$

It is easy to find that the invariant subalgebra $\gg_0=\ti\gg_0+V$ is isomorphic to
${\bf gl}(n)$. The form of  $\gg_0$ is read off from
 from the extended Dynkin diagram for $\Pi^{ext}_{C_n}$
(Fig.3)  by dropping out the roots $\al_0$ and $\al_n$  (see \cite{Ka}).

The GS basis in $\gg_1$ is represented as
$$
\gh^1_j=\f1{\sqrt{2}}(e_j+e_{n-j+1}-
e_{2n-j+1}-e_{n+j})\,,~~(j=1,\dots\Bigl[\frac{n}2\Bigr])\,,
$$
\beq{Bg1}
\gt^1_{j,n+k}=\f1{\sqrt{2}}(E_{j,n+k}+E_{n-k+1,2n-j+1}-E_{n+j,k}-E_{2n-k+1,n-j+1})\,,~(j\neq k)\,,
\eq
$$
\gt^1_{j,k}=\f1{\sqrt{2}}(E_{j,k}+E_{n-k+1,n-j+1}-E_{n+j,n+k}-E_{2n-k+1,2n-j+1})\,,~~(j\neq k)\,,
$$
$$
\gt^1_{j,n+j}=\f1{\sqrt{2}}(E_{j,n+j}-E_{n+j,j})\,.
$$

In terms of the GS basis the Killing form is
\beq{kfgsc}
(\gt^a_{j,n+j},\gt^b_{k,n+k})=\frac{(-1)^a}{2}\de_{jk}\de^{a,b}\,,~~(\gt^a_{s,r+n},\gt^b_{j,k+n})=
(-1)^a\de_{sk}\de_{rj}\de^{a,b}\,,
\eq
$$
(\gt^1_{s,r},\gt^1_{j,k})=\de_{sk}\de_{rj}\,,~~(\gh^1_j,\gh^1_k)=\de_{jk}\,.
$$

\bigskip


\subsection*{The Lax operators and the Hamiltonians.}

\subsubsection*{Trivial bundles.}

For trivial bundles the moduli space is described by the vector
$\bfu=(u_1,\ldots,u_n)$.
If $E$ is a $\bar G=Sp(n)$-bundle
\beq{cscc1}
\bfu\in\gH_{C_n}/W_{C_n}\ltimes(\tau Q_{C_n}^\vee\oplus Q_{C_n}^\vee)\,.
\eq
For trivial $Sp(n)/\mu_2$-bundles
\beq{cscc2}
\bfu\in\gH/W_{C_n}\ltimes(\tau P_{C_n}^\vee\oplus P_{C_n}^\vee)\,.
\eq
The dual variables $\bfv=(v_1,\ldots,v_n)\,$ $v_j\in\mC$
are the same in the both cases.
The standard CM Lax operator in the Chevalley basis  is
\beq{tlc}
L(z)=\sum_{j=1}^n(v_j+S_{0,j}E_1(z))e_j+\sum_{j\neq k}S_{j,k}\phi(u_j-u_k,z)\gG^-_{j,k}
\eq
$$
+\sum_{j<k}S_{j,k+n}\phi(u_j+u_k,z)\gG^+_{j,k}
+\sum_{j>k}S_{j+n,k}\phi(-u_j-u_k,z)\gG^+_{j,k}
$$
$$
\sum_{j}(S_{j,j+n}\phi(2u_j,z)\gG^-_{j}+S_{j+n,j}\phi(-2u_j,z)\gG^+_{j})\,.
$$
The quadratic Hamiltonian after the reduction with respect to the Cartan subgroup takes the form (see (\ref{cbkfc}))
$$
H^{CM}_{C_n}=\oh\sum_{j=1}^nv^2_j-\oh\sum_{j\neq k}((S_{j,k}S_{k,j}E_2(u_j-u_k)+
S_{j,k+n}S_{j+n,k}E_2(u_j+u_k))-\f1{4}\sum_{j}S_{j,j+n}S_{j+n,j}E_2(2u_j,z)\,.
$$
Thus, we have two types of the standard CM systems with the same Hamiltonians and
different configuration spaces described by (\ref{cscc1}) and (\ref{cscc2}).


\subsubsection*{Nontrivial bundles}

\emph{Moduli space}

Consider a bundle with a characteristic class defined by $\zeta=\bfe\,(\varpi^\vee_n)$.
The moduli space is defined by vectors  $\ti\bfu\in\ti\gH_0$ (\ref{modc}).

Let $n=2l+1$ be an odd number. As it was explained above,
 $\ti\gH_0$ is a Cartan subalgebra of $\gg_{B_l}$.
 The coroot lattice in $\ti\gH_0$ (\ref{bcrl})
 \beq{crsb}
\ti Q^\vee=\{\ga=\sum_j^lm_j\ti e_j\,,~\sum_j^lm_j~{\rm is~even}\}
\eq
is invariant sublattice in $Q^\vee_{C_n}$ (\ref{bcrl}).
The coweight lattice in $\ti\gH_0$ (\ref{bcrl1})
$$
\ti P^\vee=\{\ga=\sum_{j=1}^lm_j\tie_j\,|\,m_j\in\mZ\}\,.
$$
Let $\ti W=W_{B_l}$ be the Weyl group. A closer of the positive
Weyl chamber for $\ti\bfu=(u_1,\ldots,u_l)$ is \beq{wcc} u_1\geq
u_2\geq\ldots\geq u_l\geq 0\,. \eq There are two types of the
moduli spaces defined as the quotient of $\ti\gH_0$
$$
\ti\gH_0/(\ti W\ltimes(\tau\ti Q^\vee+\ti Q^\vee)\,,
~~~
\ti\gH_0/(\ti W\ltimes(\tau\ti P^\vee+\ti P^\vee)\,.
$$

For $n=2l$ the invariant subalgebra is $\gg_{D_l}$. The structure of lattices
for this algebra will be described in Section 5.
The invariant coroot sublattice is the same as above (\ref{crsb}).
The form of the coweight sublattice depends on a parity of $l$.

  Let $l$ be odd. In this case
  $$
 \ti P^\vee=\ti Q^\vee+\mZ(\oh\sum_j^l\ti e_j)\,.
 $$
 There is an intermediate sublattice
 \beq{isl}
\ti P_2^\vee=\{\ga=\sum_{j=1}^lm_j\tie_j\,|\,m_j\in\mZ\}\,.
\eq
Therefore, there are three types of the moduli spaces:
$$
\ti\gH_0/(\ti W\ltimes(\tau\ti Q^\vee+\ti Q^\vee)\,,
~~~
\ti\gH_0/(\ti W\ltimes(\tau\ti P^\vee+\ti P^\vee)\,,
~~~
\ti\gH_0/(\ti W\ltimes(\tau\ti P_2^\vee+\ti P_2^\vee)\,.
$$

For $l$ even there are two type of coweight lattices
 $$
 \ti P^{L\vee}=\ti Q^\vee+\mZ(\oh\sum_j^l\ti e_j)\,,
 ~~~
 \ti P^{R\vee}=\ti Q^\vee+\mZ\oh(\sum_j^{l-1}\ti e_j-\tie_l)
 $$
 and the intermediate sublattice $\ti P_2^\vee$ (\ref{isl}).
 In this case there are four types of the moduli spaces.

\bigskip
\emph{Lax operator}

Using the general prescription for Lax operators
(6.15.I), (6.17.I), (6.18.I), the GS basis
(\ref{BV}), (\ref{Bg1}),  and (\ref{rlb}) we define $L$ in our case.
The invariant operator $\ti L_0(z)$ is the Lax operator of the
SO$(n)$ CM system (\ref{tlb}) and (\ref{trso}).
$$
L_1(z)=\sum_{j=1}^n\Bigl(S_j\phi(\oh,z)\gh^1_j+
\oh S_{j,j+n}\bfe\,(\frac{2n+1-2j}{2n}z)\phi(\frac{(2n+1-2j)\tau}{2n}+\oh-2u_j,z)
\gt^1_{j,j+n}\Bigr)
$$
\beq{l1sp}
+\sum_{j\neq k}^n\Bigl(S_{j,n+k}\bfe\,(\frac{2n+1-j-k}{2n}z)
\phi(\frac{(2n+1-j-k)\tau}{2n}-u_j-u_k+\oh,z)\gt^1_{j,n+k}\Bigr.
\eq
$$
\Bigl.
+S_{j,k}\bfe\,(\frac{k-j}{n}z)
\phi(\frac{(k-j)\tau}{n}-u_j+u_k+\oh,z)\gt^1_{k,j}\Bigr)\,,
$$
$$
L_0'(z)=\sum_{j\neq k}^nS_{j,k+n}^{'}\bfe\,(\frac{2n+1-j-k}{2n}z)
\phi(\frac{(2n+1-j-k)\tau}{2n}-u_j-u_k,z)\gt^0_{j,k+n}\Bigr.
$$
\beq{l0sp}
+\oh\sum_{j=1}^nS_{j,n+j}^{'}\bfe\,(\frac{2n+1-2j}{2n}z)
\phi(\frac{(2n+1-2j)\tau}{2n}-2u_j,z)\gt^0_{j,n+j}\,.
\eq
In these expressions the identification $u_j=-u_{n+1-j}$ as a result
of $\la_n$ action is assumed.

\bigskip
\emph{Hamiltonian}\\
Due to the gradation the Hamiltonian contain three terms
$$
H=\ti H_0+H'_0+H_1\,,
$$
where $\ti H_0$ is the CM Hamiltonian related to SO$(n,\mC)$ (\ref{bth}), (\ref{dh}).
To define $H'_0$ and $H_1$ one should take into account the Killing
form (\ref{kfgsc}).
Then from (\ref{l1sp}) and (\ref{l0sp}) we find
$$
H_1=
\oh\sum_{j=1}^n\Bigl(-S_{j}^2E_2(\oh)+ S^2_{j,j+n}E_2(\frac{(2n+1-2j)\tau}{2n}-2u_j+\oh)
\Bigr)
$$
$$
-\oh\sum_{j<k}^n\Bigl(-S_{j,k+n}S_{k,j+n}E_2(\frac{(2n+1-j-k)\tau}{2n}-u_j-u_k-\oh)+
S_{j,k}S_{k,j}E_2(\frac{(k-j)\tau}{n}-u_j+u_k-\oh)\Bigr)\,,
$$
$$
H'_0=-\oh\sum_{j<k}^n\Bigl(S'_{j,k+n}S'_{k,j+n}E_2(\frac{(2n+1-j-k)\tau}{2n}-u_j-u_k)
+\oh (S'_{j,j+n})^2E_2(\frac{(2n+1-2j)\tau}{2n}-2u_j)\Bigr)\,.
$$
As above $u_j=-u_{n-j+1}$.


\section{SO$(2n)$\,, Spin$(2n)\,$.
General construction}
\setcounter{equation}{0}

\subsection*{Roots and weights.}

The  Lie algebra  $D_n$ has dimension $2n^2-n$ and rank $n$.
The universal covering group $\bG$ is Spin$(2n)$ and
$G^{ad}=$SO$(2n)/\mu_2$.
 In terms of the canonical basis on $\gH_{D_n}$
$(e_1,e_2,\ldots,e_n)$ simple roots of $D_n$  are
$\Pi_{D_n}=\{\al_j=e_j-e_{j+1}\,,~j=1,\dots,n-1\,,~\al_n=e_1+e_n\}$,
and the minimal root is
\beq{mbr}
\al_0=-e_1-e_2=-(\al_1+2\al_2+\ldots+2\al_{n-2}+\al_{n-1}+\al_{n})\,.
\eq

The roots of system $R_{D_n}$ have the same length and $\sharp R=2n(n-1)$.
 The positive roots are \\
 $R^+=\{(e_j\pm e_k)\,,~j,k=1,\ldots,n\,,~(j>k)\}$.
The levels of positive roots  are
\beq{rld}
f_{e_j-e_k}=k-j\,,~~f_{e_j+e_k}=2n-k-j\,.
\eq

The Weyl group $W_{D_n}$ of $R_{D_n}$ is the semidirect product of permutations
$S_n$ and the  sign changes
\beq{wed}
 S_n\ltimes{\,\rm  (the~sign~changes\,}e_j\to-e_j)\,,~(\prod_{j=1}^n(\pm 1)_j=1)\,.
 \eq

The root and coroot systems coincide and
$R_{D_n}$ generates the root lattice
\beq{drl}
Q_{D_n}=\{\ga=\sum_{j=1}^nm_je_j\,|\,\sum_{j=1}^nm_j\,{\rm is~even}\}\,.
\eq
The half-sum of positive $D_n$ coroots is $\rho_{D_n}=(n-1)e_1+(n-2)e_2+\ldots+2e_{n-2}+e_{n-1}$,
and since the Coxeter number $h=2n-2$
\beq{iv}
\ka=\rho/h=\oh e_1+(\oh-\f1{2(n-1)})e_2+\ldots+(\oh-\frac{j-1}{2n-2})e_{j}+\ldots\f1{2n-2}e_{n-1}\,.
\eq

The system of the fundamental weights takes the form
$$
\varpi_j=e_1+e_2+\ldots+e_j\,,~j=1,\ldots,n-2\,,
$$
$$
\varpi_{n-1}=\oh(e_1+\ldots+e_{n-1}-e_n)\,,~~
\varpi_{n}=\oh(e_1+\ldots+e_{n-1}+e_n)\,.
$$

The weights $\varpi_1$, $\varpi_{n-1}$, and $\varpi_{n}$ are the highest weights
of the vector $\underline{2n}$, right spinors $(\underline{2^{n-1})}^R$ and left spinors
$(\underline{2^{n-1})}^L$ representations.

We find from
(A.17.I) and (\ref{mbr}) that
 the fundamental alcove  has the vertices
\beq{alcd}
 C_{alc}=(0,\varpi_1,\oh\varpi_2,\ldots,\oh\varpi_{n-2},
 \varpi_{n-1}, \varpi_n)\,.
 \eq


\subsection*{The Chevalley  basis}

The Chevalley  basis is convenient
to define using a fundamental representation $\pi_1$
 corresponding to the
weight $\varpi_1=e_1$. It has dimension $2n$.
If the symmetric form is represented by the anti-diagonal matrix as for $B_n$  then $Z\in${\bf so}$(2n)$ takes the form
\beq{31}
Z=
\left(
  \begin{array}{cc}
    A & B \\
    C & -\ti A \\
  \end{array}
\right)\,,~~
B=-\ti{B}\,,~C=-\ti{C}\,,~~\ti{X}=JX^TJ\,.
\eq

The basis in $\gH_{D_n}$ is generated by the simple roots
 $\Pi_{D_n}$  in $\gH$ (or by the canonical basis $(e_1,\ldots,e_n)$).
 In  $\pi_1$ the canonical basis in $\gH_{D_n}$ is
 $\di(e_1,\ldots,e_n,-e_n,\ldots,-e_1)$.

 The  root subspaces in $\pi_1$ are
 \beq{chbd}
\begin{array}{ccc}
 (e_j-e_k)\,,~j<k&\to\gG_{jk}^-=(E_{j,k}(\in A)-E_{2n+1-k,2n+1-j}(\in \ti A),,&({\rm positive ~roots})\\
 (e_j-e_k)\,,~j>k&\to\gG_{jk}^-=(E_{j,k}(\in A)-E_{2n+1-k,2n+1-j}(\in \ti A)),,&({\rm negative ~roots})\,,
\end{array}
\eq
$$
\begin{array}{ccc}
( e_j+ e_k)\,,~j<k &\to\gG_{jk}^+=(E_{j,k+n}-E_{n+1-k,2n+1-j})\in B &({\rm positive ~roots}) \,, \\
( e_j+ e_k)\,,~j>k &\to\gG_{jk}^+= (E_{j+n,k}-E_{2n+1-k,n+1-j})\in C& ({\rm negative ~roots})\,.
\end{array}
$$

The Killing form  (A.25.I) takes the form
\beq{kfcd}
(\gG_{jk}^\pm,\gG^\pm_{il})=\de_{jl}\de_{ik}\,.
\eq


\section{SO$(2n)$\,, Spin$(2n)\,$, $n=2l+1$.}

\setcounter{equation}{0}

\subsection*{Lattices and characteristic classes.}
It follows from (\ref{drl}) that
$$
\varpi_j\in Q(D_n)~{\rm for}~ 1<j<n-1\,,~~
 \varpi_1\sim 2\varpi_n\,,~\varpi_{n-1}\sim 3\varpi_n\,,~
 mod\,Q(D_n)\,.
$$
It means that
\beq{pdno}
P(D_n)=Q(D_n)+\mZ\varpi_n\sim Q(D_n)+\mZ\varpi_{n-1}\,.
\eq
The weight lattice $P(D_n)$ contains a sublattice
$P_2(D_n)$ of index two
\beq{P2D}
P_2(D_n)=Q(D_n)+\mZ \varpi_1=\{\sum_{j=1}^nm_je_j\,|\,m_j\in\mZ\}\,.
\eq
This lattice is self-dual $P_2(D_n)=^LP_2(D_n)$ and isomorphic to the group of cocharcters $P_2(D_n)=t(SO(2n))$, (A.43.I), where $SO(2n)=Spin(2n)/\mu_2$. On the other hand, the weight lattice $P(D_n)$  is
dual to the  root lattice $^LP(D_n)=Q(D_n)$.

The center of $Spin(2n)$ for odd $n$ is $\clZ(\bar G)=P_{D_n}/Q_{D_n}\sim\mu_4$. The group element $\zeta=\bfe\,(\xi)$ for
 $\xi=\varpi_n$ generates $\mu_4$.  Putting $\xi=\varpi_n$
 and  $C_{alc}$ (\ref{alcd}) in (3.10.I)
we find $\la_n$
 $$
 \la_n\,:\,~
 \left\{
 \begin{array}{l}
  \varpi_n\to 0\to \varpi_{n-1}\to\varpi_1\to\varpi_n\,,   \\
   \varpi_j\to\varpi_{n-j}\,,~1<j<n-1\,.
 \end{array}
\right\}\,,
~~~
 \left\{
 \begin{array}{l}
\al_0\to \al_n\to \al_{1}\to\al_{n-1}\to\al_0\,,\\
\al_j\to\al_{n-j}\,,~~1<j<n-1\,.
 \end{array}
\right\}\,.
$$
  In $\pi_1$ the transformation takes the form
$$
\la_n=
  \left(
    \begin{array}{cccc}
      0 & 1 & 0 & 0 \\
      0 & 0 & 0 & Id_{n-1} \\
      Id_{n-1}& 0 & 0 & 0 \\
      0 & 0 & 1 & 0 \\
    \end{array}
  \right)
$$
It acts on the basis elements in the space of $\underline{2n}$ representation as
\beq{indo}
\left(
\begin{array}{cccccccccccc}
  1\,,&2\,,&\ldots,&n-1\,,& n\,,&n+1\,,&n+2\,,&\ldots, &2n-1\,,& 2n \\
  n\,,&n+2 \,,&\ldots,&2n-1&2n\,,&1\,,&2\,,&\ldots, &n-1\,,& n+1
\end{array}
\right)\,.
\eq

 In the canonical basis $(e_1,e_2,\ldots,e_n)$ $\la_n$ is represented by the matrix
\beq{cbdl}
\la_n\to \La_{jk}=
\left\{
 \begin{array}{c}
\de_{j,n-k+1}\,,~j<n\\
-\de_{j,n-k+1}\,,~j=n\,.
 \end{array}
\right.
\eq
This transformation is an element of the Weyl group $W(SO(2n))$.

\vspace{0.5cm}

\unitlength 1mm 
\linethickness{0.4pt}
\ifx\plotpoint\undefined\newsavebox{\plotpoint}\fi 
\begin{picture}(59.806,33.967)(0,0)
\put(23.636,23.784){\circle{1.808}}
\put(30.771,19.473){\circle{2.081}}
\put(37.906,19.77){\circle{2.081}}
\put(51.88,19.77){\circle{1.808}}
\put(58.866,23.635){\circle{1.88}}
\put(58.718,15.757){\circle{1.808}}
\put(23.636,15.905){\circle*{1.88}}
\put(24.528,23.338){\circle*{.297}}
\put(32.109,19.621){\circle*{.297}}
\put(32.109,19.621){\line(1,0){4.757}}
\put(39.096,19.77){\line(1,0){5.054}}
\put(48.461,19.621){\line(1,0){.149}}
\put(48.461,19.77){\line(1,0){2.378}}
\multiput(57.826,23.189)(-.05679775,-.03340449){89}{\line(-1,0){.05679775}}
\multiput(52.623,18.878)(.06937333,-.03369333){75}{\line(1,0){.06937333}}
\multiput(29.879,18.73)(-.06937333,-.03369333){75}{\line(-1,0){.06937333}}
\multiput(24.379,23.338)(.055175258,-.033721649){97}{\line(1,0){.055175258}}
\put(20.812,26.905){\makebox(0,0)[cc]{$\alpha_1$}}
\put(30.474,15.608){\makebox(0,0)[cc]{$\alpha_2$}}
\put(49.65,16.203){\makebox(0,0)[cc]{$\alpha_{n-2}$}}
\put(56.934,27.203){\makebox(0,0)[cc]{$\alpha_{n-1}$}}
\put(55.15,12.784){\makebox(0,0)[cc]{$\alpha_n$}}
\put(19.771,11.743){\makebox(0,0)[cc]{$\alpha_0$}}
\put(50.541,11.595){\vector(4,1){.07}}\qbezier(28.541,12.041)(39.541,8.399)(50.541,11.595)
\put(27.203,24.527){\vector(-1,0){.07}}\qbezier(56.933,19.027)(49.501,26.386)(27.203,24.527)
\put(53.663,26.311){\vector(2,-1){.07}}\qbezier(26.163,27.054)(40.359,33.967)(53.663,26.311)
\put(25.865,18.284){\vector(-3,-2){.07}}\qbezier(54.406,23.784)(35.082,24.156)(25.865,18.284)
\put(38.649,5.054){\makebox(0,0)[cc]{Fig.4 $D_n$, $n$-odd, $\lambda_{n}$}}
\end{picture}

\vspace{0.5cm}

In a similar way the fundamental weight $\varpi_{n-1}$ generates the Weyl
transformation
$$
\la_{n-1}\,:\,
 \left\{
 \begin{array}{l}
  \varpi_n\to\varpi_1\to\varpi_{n-1}\to 0\to\varpi_n\,,   \\
   \varpi_j\to\varpi_{n-j}\,,~1<j< n-1\,.
 \end{array}
\right.
$$
Thus, $\la_{n-1}=\la_n^{-1}$ and we will not consider this case.

\bigskip
Consider a subgroup of order two of $\clZ(\bG)$ generated by $\xi=\varpi_1$.
Acting as above we find
$$
\la_1\,:\,
 \left\{
 \begin{array}{l}
  \varpi_1\leftrightarrow 0\,,   \\
  \varpi_{n-1}\leftrightarrow\varpi_n\,,\\
   \varpi_j\leftrightarrow\varpi_{j}\,,~1<j< n-1\,.
 \end{array}
\right.
$$
In terms of roots the action assumes the form
\beq{ldr1}
\la_1\,:\,~
 \left\{
 \begin{array}{l}
  \varpi_1\leftrightarrow 0\,,   \\
  \varpi_{n-1}\leftrightarrow\varpi_n\,,\\
   \varpi_j\leftrightarrow\varpi_{j}\,,~1<j< n-1\,.
 \end{array}
\right\}\,,~~~
 \left\{
 \begin{array}{l}
\al_0\leftrightarrow \al_{1}\,,\\
\al_n\leftrightarrow \al_{n-1}\,,\\
\al_j\leftrightarrow\al_{j}\,,~~1<j\leq n\,.
 \end{array}
\right\}.
\eq
Explicitly, in $\pi_1$
$$
\la_1=
\left(
  \begin{array}{cccccc}
    0 &  &  &  &  & 1 \\
     & Id_{n-2} &  &  & 0 &  \\
     &  & 0 & 1 &  &  \\
     &  & 1 & 0 &  &  \\
     & 0 &  &  & Id_{n-2} &  \\
    1 &  &  &  &  & 0 \\
  \end{array}
\right)\,,
$$
 or
 \beq{indo1}
\left(
\begin{array}{cccccccccccccc}
  1\,,&2\,,&\ldots,&n-1\,,& n\,,&n+1\,,&n+2\,,&\ldots, &2n-1\,,& 2n \\
  2n\,,&2 \,,&\ldots,&n-1\,,& n+1\,,&n\,,&n+2\,,&\ldots, &2n-1\,,& 1
\end{array}
\right)\,.
\eq
 In the canonical basis $(e_1,e_2,\ldots,e_n)$ $\la_n$ is represented by the matrix
\beq{cbdl1}
\la_1\to \La_{jk}=
\left\{
 \begin{array}{c}
-\de_{j,k}\,,~j=1,n\\
\de_{j,k}\,,~j\neq 1,n\,.
 \end{array}
\right.
\eq


\subsection*{The GS-basis.}

For $n$ odd
\beq{gdo}
\gg_{D_n}=\gg_0+\gg_1+\gg_2+\gg_3\,,~~(\la_n(\gg_k)=i^k\gg_k)\,,
\eq
where
$$
\gg_0=\ti\gg_0+V\,.
$$
 In $\Pi^{ext}$ there are one orbit of length 4 and $\frac{n-3}2$ orbits
 of  length 2. The former orbit passes through $\al_0$ and the latter orbits contain $\Pi_1=A_{n-3}$. Since $n-3$ is even
it follows from Lemma 10.1 that $\ti\gg_0=\gg_{B_{\frac{n-3}2}}$.
We will demonstrate it explicitly.

The subspaces in (\ref{gdo}) can be read off from the $\la_n$-action.
First we find that
$$
\ti\gg_0=\left\{Z=
  \left(
    \begin{array}{ccccc}
      0 &  &  &  & 0 \\
       & X &  & 0 &  \\
       &  & 0 &  &  \\
       &  0&  & X &  \\
      0 &  &  &  & 0 \\
    \end{array}
  \right)\,,~~~X= A^{(n-2)}-\ti A^{(n-2)}
  \right\}\,,
         $$
 where $A^{(n-2)}$ is a matrix of order $n-2$,
$\ti A^{(n-2)}=J_{n-2}A^{(n-2)}J_{n-2}$.
 $$
 \dim\,(\ti\gg_0)=\frac{(n-2)(n-3)}2\,.
 $$
It is easy to see that $\ti\gg_0$ has a type $B_{\frac{n-3}2}$.
Namely, $\{X= A^{(n-2)}-\ti A^{(n-2)}\}={\bf so}(n-2)$. The invariant
Cartan subalgebra $\ti\gH\subset\ti\gg_0$ has a basis
\beq{hdno}
 \ti e_2=e_2-e_{n-1}\,,\,
\ti e_3=e_3-e_{n-2}\,,
  \ldots,
\ti e_{l}=
e_{l}-e_{l+2}+e_{n+l}\,,
~~~(l=\frac{n-3}2)\,.
\eq
An arbitrary element from $\ti\gH_0$ has the form
\beq{diu}
\ti\bfu=\di\,(0,u_2,\dots,u_l,0,-u_l,\dots,-u_2,0,0,u_2,\dots,u_l,0,-u_l,\dots,-u_2,0)\,.
\eq

The space $V$ is generated by its GS basis.
 In following formulas $j=2,\ldots,n-1$
\beq{gt0d}
\gt^0_{1j}=\f1{2}(E_{1,j}+E_{n,n+j}+E_{2n,j}+E_{n+1,n+j})-\ldots\,,~~(j=2,\ldots,n-1)
\eq
$$
\gt^0_{1,n}=\f1{2}(E_{1,n}+E_{n,2n}+E_{2n,n+1}+E_{n+1,1})-\dots\,,
$$
$$
\gt^0_{j,1}=\f1{2}(E_{j,1}+E_{n+j,n}+E_{j,2n}+E_{n+j,n+1})
-\ldots\,,~~(j=2,\ldots,n-1)\,,
$$
$$
\gt^0_{j,n+k}=\f1{\sq2}(E_{j,n+k}+E_{n+j,k})-E_{n-k+1,2n-j+1}
-E_{2n-k+1,n-j+1}\,,~~(j,k=2,\ldots,n-1)\,,
$$
where $\ldots$ means the antisymmetric part of $Z$ (\ref{31}).
The latter generators  form the adjoint representation of $\ti\gg_0={\bf so}(n-2)$.
We have
$$
\dim\,(V)=\oh(n-2)(n-3)+2\cdot (n-2)+1\,.
$$
Let $\underline{(n-2)}$, $\underline{1}$
 be a vector and a scalar representations of ${\bf so(n-2)}$.
Then $\gg_0$ is decomposed as
$$
\gg_0=(\ti\gg_0={\bf so(n-2)})+2\times\underline{(n-2)}+\underline{1}+
\underline{\oh(n-2)(n-3)}\,,~~\dim\,(\gg_0)=n^2-3n+3\,.
$$
It is isomorphic to ${\bf so}(n-1)+{\bf so}(n-1)+\underline{1}$. This algebra is obtained from the
extended Dynkin diagram by dropping two middle roots. This procedure generates the
automorphism of order four \cite{Ka}.

The Killing form in $V$ is defined by (5.8.I). We write  down its nonzero components
\beq{kfvdo}
(\gt^0_{j_1,n+k_1},\gt^0_{j_2,n+k_2})=\de_{j_1,k_2}\de_{k_1,j_2}\,,~~
(\gt^0_{1,j},\gt^0_{j,1})=1\,,~~(\gt^0_{1,n},\gt^0_{1,n})=-1\,.
\eq
Formally, the last pairing vanishes. However, one can consider instead
 nontrivial pairing $(\gt^0_{1,n},\gt^0_{n,1})$. The basic element $\gt^0_{n,1}$
is defined by the equivalent orbit passing through $E_{n,1}$.
 The sign minus in the last pairing arises due to the antisymmetry of matrix $Z$.

Consider the GS basis in $\gg_1$
\beq{gt1d}
\gh^1_{1}=\f1{\sq2}(e_{1}+ie_{n})-\ldots\,,
\eq
$$
\gt^1_{1,j}=\f1{2}(E_{1,j}+iE_{n,n+j}-E_{2n,j}-iE_{n+1,n+j})
\ldots\,,~~(j=2,\ldots,n-1)\,,
$$
$$
\gt^1_{1,n}=\f1{2}(E_{1,n}+iE_{n,2n}-E_{2n,n+11}-iE_{n+1,1})-\dots\,,
$$
$$
\gt^1_{j,1}=\f1{2}(E_{j,1}+iE_{n+j,n}-E_{j,2n}-iE_{n+j,n+1})
-\ldots\,,~~(j=2,\ldots,n-1)\,,
$$
$$
\dim\,\gg_1=2(n-2)+2=2n-2\,.
$$

The GS basis in $\gg_2$ is
$$
\gh^1_{1}=\f1{\sq2}(e_{1}-e_{n})-\ldots\,,
$$
\beq{gt2d}
\gt^2_{1j}=\f1{2}(E_{1,j}-E_{n,n+j}+E_{2n,j}-E_{n+1,n+j})
\ldots\,.
\eq
$$
\gt^2_{1n}=\f1{2}(E_{1,n}-E_{n,2n}+E_{2n,n+1}-E_{n+1,1})-\dots\,,
$$
$$
\gt^2_{j,1}=\f1{2}(E_{j,1}-E_{n+j,n}+E_{j,2n}-E_{n+j,n+1})
-\ldots\,,~~(j=2,\ldots,n-1)\,,
$$
$$
\gt^2_{j,n+k}=\f1{\sq2}(E_{j,n+k}-E_{n+j,k})-\ldots\,,~~(j,k=2,\ldots,n-1)\,,
$$
$$
\gt^2_{j,k}=\f1{\sq2}(E_{j,k}+E_{n+1-k,n+1-j})-\ldots\,,~~(j,k=2,\ldots,n-1)\,.
$$
$$
\dim\,\gg_2=2(n-2)+2+\frac{(n-2)(n-3)}2+\frac{(n-2)(n-1)}2=n^2-2n+1\,.
$$

Finely, the GS basis in $\gg_3$ is
\beq{gt3d}
\gh^3_{1}=\f1{\sq2}(e_{1}-ie_{n})-\ldots\,,
\eq
$$
\gt^3_{1j}=\f1{2}(E_{1,j}-iE_{n,n+j}-E_{2n,j}+iE_{n+1,n+j})
\ldots\,,~~(j=2,\ldots,n-1)\,,
$$
$$
\gt^3_{1n}=\f1{2}(E_{1n}-iE_{n,2n}-E_{2n,n+1}+iE_{n+1,1})-\dots\,,
$$
$$
\gt^3_{j,1}=\f1{2}(E_{j,1}-iE_{n+j,n}-E_{j,2n}+iE_{n+j,n+1})
-\ldots\,,~~(j=2,\ldots,n-1)\,,
$$
$$
\dim\,\gg_3=2(n-2)+2=2n-2\,.
$$

It follows from (5.8.I) that there is a nontrivial pairing $(\gg_2,\gg_2)$ and
$(\gg_1,\gg_3)$. For the basis  we have
\beq{kfvdok1}
(\gh^1_{1},\gh^3_{1})=1\,,~~(\gt^1_{1,j},\gt^3_{j,1})=1
\,,~~(\gt^3_{1,j},\gt^1_{1,j})=1
\,,~~(\gt^1_{1,n},\gt^3_{1,n})=-1\,,
\eq
$$
(\gh^2_{1},\gh^2_{1})=1\,,~~(\gt^2_{1j},\gt^2_{j,1})=1
\,,~~(\gt^2_{1,n},\gt^2_{1,n})=-1\,,
$$
\beq{kfvdok2}
(\gt^2_{j_1,k_1},\gt^2_{j_2,k_2})=\de_{j_1,k_2}\de_{j_2,k_1}\,,~~
(\gt^2_{j_1,n+k_1},\gt^2_{j_2,n+k_2})=-\de_{j_1,k_2}\de_{j_2,k_1}\,.
\eq

\bigskip

Consider the GS-basis corresponding to $\la_1$ (\ref{indo1}).
In this case
$$
\gg_{D_n}=\gg_0+\gg_1\,,~~(\la_1(\gg_k)=(-1)^k\gg_k)\,,
$$
$$
\gg_0=\ti\gg_0+V\,,~~\ti\gg_0=\gg_{B_{n-2}}={\bf so}(2n-3)\,.
$$
Then  $Z\in\ti\gg_0$ if
$$
Z=
\left(
           \begin{array}{cccccc}
             0 & 0 & 0 & 0 & 0 & 0 \\
          0 & A^{(n-2)} & a_{jn} &a_{jn}  & B^{(n-2)} & 0 \\
    0 & a_{nj} & 0 & 0 & -a^T_{jn} & 0 \\
    0 & a_{nj} & 0 & 0 & -a^T_{jn} & 0 \\
     & C^{(n-2)} & -a^T_{nj} & -a^T_{nj} & -\ti A^{(n-2)} & 0 \\
    0 &  0 & 0 & 0 & 0 & 0 \\
           \end{array}
         \right)\in {\bf so}(2n-3)\,.
$$
Here $A^{(n-2)}$, $B^{(n-2)}$ and $C^{(n-2)}$ are matrices of order $n-2$,
$\,\ti A^{(n-2)}=J(A^{(n-2)})^TJ$ and\\
 $\ti B^{(n-2)}=-B^{(n-2)}$,  $\,\ti C^{(n-2)}=-C^{(n-2)}$.

The Cartan subalgebra $\ti\gH_0\sim\gH_{B_{n-2}}$ has the form
\beq{csbn}
\ti\gH_0=\{\ti\bfu=\di(0,u_2,\ldots,u_{n-1},0,0,-u_{n-1},\ldots,-u_2,0\}\,.
\eq

For $Z\in V$ we have
$$
Z=
\left(
  \begin{array}{cccccc}
    0      & a_{1j} & a_{1n}    & b_{11}    & -a^T_{j1}  & 0 \\
    a_{j1} & 0      & 0         & 0         & 0       & a_{j1} \\
    -a_{1n} & 0      & 0         & 0         & 0       & -b_{11} \\
    -b_{11} & 0      & 0         &0          & 0       & -a_{1n} \\
    -a^T_{1j} & 0      & 0         & 0         & 0       & -a^T_{1j} \\
    0 &  a_{1j}     & b_{11}   & a_{1n}   & -a^T_{j1}      & 0 \\
  \end{array}
\right)\,,
$$
$$
V=\underline{2(n-2)+1}+\underline{1}\,.
$$
Here $\underline{2(n-2)+1}$ is a vector representation of $B_{n-2}=${\bf so$(2n-3)$}.
The GS-basis in $V$ takes the form
\beq{gt1d2}
\gt^0_{1,n}=\f1{\sq2}(E_{1,n}+E_{2n,n+1})-\ldots\,,~~
\gt^0_{1,j}=\f1{\sq2}(E_{1,j}+E_{2n,j})-\ldots
\,,~(j=2,\ldots,n-1)\,,
\eq
$$
\gt^0_{n,1}=\f1{\sq2}(E_{n,1}+E_{n+1,2n})-\ldots\,,~~
\gt^0_{j,1}=\f1{\sq2}(E_{j,1}+E_{j,2n})-\ldots\,,~(j=2,\ldots,n-1)\,.
$$
As above, $\ldots$ means the antisymmetric partners of the generators.
Thus
$$
\dim\,\gg_0=\dim\,\gg_{B_{n-2}}+\dim\,V=2n^2-5n+4\,.
$$
The invariant subalgebra $\gg_0$ is isomorphic to ${\bf so(2n-2)}+\underline{1}$.
It is obtained from the extended Dynkin diagram for ${\bf so(2n)}$ by dropping two
roots roots $\al_1$ and $\al_0$. This procedure provides the second order automorphism
\cite{Ka}
that we consider.

Non-vanishing components of the Killing form on $V$ are
\beq{kfd1}
(\gt^0_{1,n},\gt^0_{n,1})=-1\,,~~
(\gt^0_{1,j},\gt^0_{j,1})=1\,.
\eq

Consider the GS-basis in $\gg_1$
\beq{gt1d1}
\gh^1_{1}=e_{1}\,,~~\gh^1_{n}=e_{n}\,,
\eq
$$
\gt^1_{1,j}=\f1{\sq2}(E_{1,j}-E_{2n,j})
\ldots\,,
~~
\gt^1_{1,n}=\f1{\sq2}(E_{1,n}-E_{2n,n+1})-\ldots\,,
$$
$$
\gt^1_{n,1}=\f1{\sq2}(E_{n,1}-E_{n+1,2n})-\ldots\,,
~~
\gt^1_{j,1}=\f1{\sq2}(E_{j,1}-E_{j,2n})
-\ldots\,,
$$

The Killing form on this basis assumes the form
\beq{kfd2}
(\gh^1_{1},\gh^1_{1})=1\,,~~
(\gh^1_{n},\gh^1_{n})=1\,,~~
(\gt^1_{1,n},\gt^1_{n,1})=1\,,~
~(\gt^1_{1,j},\gt^1_{j,1})=1\,.
\eq


\subsection*{The Lax operators and the Hamiltonians.}

\subsubsection*{Trivial bundles.}

The moduli space for the trivial $\bG=Spin(2n)$- bundles is the
quotient $\gH_{D_n}/(W_{D_n}\ltimes(\tau Q(D_n)+Q(D_n)))$, where $W_{D_n}$ is defined in (\ref{wed}).
It implies that
\beq{tdbg}
u_j\sim u_j+m_j+n_j\tau\,,~~n_j\,,m_j\in\mZ\,,~\sum_{j=1}^nm_j\,{\rm~and ~}~
\sum_{j=1}^nn_j\,{\rm~ is~even}
\eq
(see (\ref{drl})).

For trivial $G^{ad}=SO(2n)/\mu_2$-bundles the moduli space is
$\gH_{D_n}/(W\ltimes(\tau P(D_n)+P(D_n)))$, where $P(D_n)$ is the $D_n$-weight lattice
(\ref{pdno}). In this case  instead of the shifts (\ref{tdbg}) the moduli
space is defined up to the shifts
\beq{tdbg1}
u_j\sim u_j+m_j+n_j\tau\,,~~n_j\,,m_j\in\mZ {~\rm or}~\in\oh+\mZ ,.
\eq

The intermediate situation arises for trivial $SO(2n)$-bundles.
In this case $\bfu$ is defined up to the shifts
\beq{tdbg2}
u_j\sim u_j+\tau m_j+n_j\,,~~m_j,n_j\in\mZ\,.
\eq

For all  $Spin(2n)$, $SO(2n)$ $SO(2n)/\mu_2$  trivial bundles we have the same Lax operator
\beq{trso}
L(z)=\sum_{j=1}^n(v_j+S_{0,j}E_1(z))e_{j}+\sum_{j\neq k}^n S_{k,j}\phi(u_j-u_k,z)\gG^-_{j,k}
+\sum_{j\neq k}^nS_{k+n,j}\phi(u_j+u_k,z)\gG^+_{j,k+n}\,,
\eq
and
 the same CM Hamiltonian
\beq{dh}
H=\oh\sum_{j=1}^nv^2_j-\sum_{j\neq k}^n S_{j,k} S_{k,j}E_2(u_j-u_k)
-\sum_{j\neq k}^nS_{j,k+n}S_{j+n,k}E_2(u_j+u_k)\,,
\eq
but different configuration spaces (\ref{tdbg}), (\ref{tdbg1}),
(\ref{tdbg2}).

\subsubsection*{Nontrivial bundles.}

\textbf{The moduli space.}\\
Consider  a  bundle with a nontrivial characteristic class generated by
the weight $\varpi_n$. The moduli space is a quotient of $\ti\gH_0=\{
\ti\bfu=\sum_{j=1}^lu_j\ti e_j\},~$ $l=\frac{n-3}2$, (see (\ref{diu})),
where $\ti\gH_0$ is a Cartan subalgebra of the invariant algebra $\ti\gg_0=\gg_{B_l}
={\bf so}(n-3)$.

 Following (\ref{bcrl3}) and (\ref{bcrl1}) we
construct invariant coroot and  coweight sublattices
$$
\ti Q^\vee(B_l)=\{\sum_{j=1}^lm_j\tie_j\,,~m_j\in\mZ\,,~\sum_{j=1}^lm_j-{\rm even}\}\,,
~~~
\ti P^\vee(B_l)=\{\sum_{j=1}^lm_j\tie_j\,,~m_j\in\mZ\}\,.
$$
where $\{\tie_j\}$ is the invariant basis (\ref{hdno}).
Therefore, there are two types of the moduli spaces. They are defined up to the shifts by a vector $\ti\bfu$
$$
\ti\bfu\sim\ti\bfu+\tau\ga_1+\ga_2\,,~~\ga_j\in \ti Q^\vee(B_l)\,,~{\rm or}~\ti P^\vee(B_l)\,.
$$
The nontrivial $Spin_{2n}$, SO$(2n)$, SO$(2n)/\mu_2$ - bundles
with the characteristic class $\varpi_n$ have the same moduli space as trivial
 $Spin_{n-2}$, SO$(n-2)$-bundles. Two fundamental domains describe these moduli spaces.

Let us define the moduli space of  bundles with nontrivial characteristic class generated by the weight $\varpi_1$.
In this case $\Pi_1=\Pi_{D_{n-1}}$ and $\ti\Pi=\Pi_{B_{n-2}}$ $(\ti\gg_0={\bf so}(2n-3))$.
Then $\ti\bfu\in\gH_{B_{n-2}}$
(\ref{csbn}).
Two lattices $\ti Q^\vee(B_{n-2})$, $\ti P^\vee(B_{n-2})$ in $\gH_{B_{n-2}}$
(\ref{bcrl3}), (\ref{bcrl1}) defines two fundamental domains
$$
\ti\bfu\sim\ti\bfu+\tau\ga_1+\ga_2\,,~~\ga_j\in \ti Q^\vee(B_{n-2})\,,
~{\rm or}~\ti P^\vee(B_{n-2})\,.
$$
They are the moduli spaces of trivial  $Spin_{2n-3}$, SO$(2n-3)$-bundles.

\textbf{The Lax operators and Hamiltonians.}\\
In the GS-basis corresponding to
 the characteristic class  $\varpi_n$ the Lax operator is decomposed as
 $$
 L=\ti L_0+L'_0+L_1+L_2+L_3\,,
 $$
 where $\ti L_0$ is a Lax operator corresponding to the trivial SO$(n-2)$ bundle
 (\ref{tlb}). The other components take the form
 $$
 L'_0=
 S'_{1,n}\bfe\,(z/2)\phi(\oh,z)\gt^0_{1,n}+
 $$
 $$
 \sum_{j=2}^{n-1}\Bigl(
 S'_{j,1}\bfe\,(\frac{j-1}{2n-2}z)\phi(\frac{j-1}{2n-2}\tau-u_j,z)\gt^0_{1,j}
  +S'_{1,j}
\bfe\,(\frac{1-j}{2n-2}z)\phi(\frac{1-j}{2n-2}\tau+u_j,z)\gt^0_{j,1}
\Bigr.
$$
$$
\Bigl.        +\sum_{k=1}^{n-2}S'_{k,n+j}
\bfe\,(\frac{2n-k-j}{2n-2}z)\phi(\frac{2n-k-j}{2n-2}\tau-u_j-u_k,z)\gt^0_{j,k+n}
 \Bigr)\,.
 $$
 Here $u_j$ are elements of the diagonal matrix (\ref{diu}) $u_i=-u_{n-i}$, $(i=2,\ldots,l)$.
$$
L_{k=1,3}=S^{4-k}_{1}\phi(\frac{k}{4},z)\gh^k_{1}+S^{4-k}_{1,n}\bfe(zk/4)\phi(k/4,z)\gt^k_{1,n}
$$
$$
+\sum_{j=2}^{n-1}\Bigl(S^{4-k}_{j,1}\bfe\,(\frac{j-1}{2n-2}z)\phi(\frac{j-1}{2n-2}\tau-u_j
+\frac{k}{4},z)\gt^k_{1,j} +S^{4-k}_{1,j}
\bfe\,(\frac{1-j}{2n-2}z)\phi(\frac{1-j}{2n-2}\tau+u_j-\frac{k}{4},z)\gt^k_{j,1}
\Bigr)\,,
$$
 $$
L_2=S^2_{1,n}\bfe(z/2)\phi(1/2,z)\gt^2_{1,n}+S^2_{1}\bfe(z/2)\phi(1/2,z)\gh^2_{1}+
$$
$$
\sum_{j=2}^{n-1}\Bigl(S^2_{j,1}\bfe\,(\frac{j-1}{2n-2}z)\phi(\frac{(j-1)\tau}{2n-2}-u_j
+\frac{1}{2},z)\gt^2_{1,j} +S^2_{1,j}
\bfe\,(\frac{1-j}{2n-2}z)\phi(\frac{1-j}{2n-2}\tau+u_j-\frac{1}{2},z)\gt^2_{j,1}+
\Bigr.
$$
$$
\Bigl.   \sum_{m=2,m\neq j}^{n-1}S^2_{m,j+n}
\bfe\,(\frac{2n-m-j}{2n-2}z)\phi(\frac{2n-m-j}{2n-2}\tau-u_j-u_m-\frac{1}{2},z)\gt^2_{j,m+n}
 $$
 $$
 +\sum_{m=2,m\neq j}^{n-1}S^2_{m,j}
\bfe\,(\frac{m-j}{2n-2}z)\phi(\frac{(m-j)}{2n-2}\tau-u_j+u_m-\frac{1}{2},z)\gt^2_{j,m}\Bigr)\,.
$$

 After the symplectic reduction with respect to the Cartan subgroup $\ti{\cal H}$ we come to Hamiltonians of integrable systems
 $$
 H=\ti H_0+H_0'+H_2\,,
 $$
 where $\ti H_0$ is the Hamiltonian of $B_l$ CM system (\ref{bth}),
 $$
H_0'=\oh(L'_0(z),L'_0(z))|_{\rm const.~part}\,,~~
 H_2=(L_1(z),L_3(z))+\oh(L_2(z),L_2(z))|_{\rm const.~part}\,.
 $$
 From (\ref{kfvdo}) and (\ref{kfvdok2}) we find
 $$
-H_0'=\oh\Bigl(-(S'_{1,n})^2E_2(\oh)+\sum_{j=2}^{n-1}
 S'_{1,j}S'_{j,1}E_2(\frac{j-1}{2n-2}\tau-u_j)
 +\sum_{k\neq j}^{n-1}S'_{j,n+k}S'_{k,n+j}E_2(\frac{2n-k-j}{2n-2}\tau-u_j-u_k)
 \Bigr)\,,
 $$
 $$
  H_2=(S^1_1S^3_1-S^1_{1,n}S^3_{1,n})E_2(1/4)-\oh(S^2_{1,n})^2E_2(\oh)
  +\oh\sum_{j\neq m}^{n-1}S^2_{j,m}S^2_{m,j}
  E_2(\frac{m-j}{2n-2}\tau-u_j+u_m-\oh)
  $$
  $$
  +\sum_{j=2}^{n-1}(S^1_{j,1}S^3_{j,1}+S^3_{j,1}S^1_{j,1})
  E_2(\frac{1-j}{2n-2}\tau+u_j-\f1{4})+\oh\sum_{j\neq m}^{n-1}(S^2_{j,m+n}S^2_{m+n,j}
  E_2(\frac{2n-m-j}{2n-2}\tau-u_j-u_m-\oh)\,.
 $$

\bigskip

Consider systems corresponding to the characteristic class $\varpi_1$.
In this case using the GS-basis (\ref{gt1d2}), (\ref{gt1d1}) we define
$L=\ti L_0+L_0'+L_1\,,$
where $\ti L_0$ is the Lax operator of $B_{n-2}$  (\ref{tlb}),
$$
L'_0(z)=S'_{n,1}\bfe\,(\frac{1}{2}z)\phi(\tau/2,z)\gt^0_{1,n}+
S'_{1,n}\bfe\,(-\frac{1}{2}z)\phi(\tau/2,z)\gt^0_{n,1}+
 $$
 $$
 \sum_{j=2}^{n-1}\Bigl(
 S'_{1,j}\bfe\,(\frac{j-1}{2n-2}z)\phi(\frac{(j-1)\tau}{2n-2}+u_j,z)\gt^0_{1j}
  +S'_{j,1}
\bfe\,(\frac{1-j}{2n-2}z)\phi(\frac{(1-j)\tau}{2n-2}-u_j,z)\gt^0_{j,1}
\Bigr)\,,
$$
$$
L_1(z)=(S^1_{1}\gh^1_{1}+
S^1_{n}\gh^1_{n})\phi(\frac{1}{2},z)+(S^1_{1,n}\gt^1_{n,1}+S^1_{n,1}\gt^1_{1,n})
\bfe\,(\oh z)\phi(\frac{1}{2}(1+\tau),z)
$$
$$
+\sum_{j=2}^{n-1}\Bigl(S^k_{1,j}\bfe\,(\frac{j-1}{2n-2})\phi(\frac{(j-1)\tau}{2n-2}-u_j
+\oh,z)\gt^1_{1j} +S^1_{j,1}
\bfe\,(\frac{1-j}{2n-2})\phi(\frac{(1-j)\tau}{2n-2}+u_j+\oh,z)\gt^1_{j,1}
\Bigr)\,.
$$
Then we come to the Hamiltonians $\ti H_0=H_{B\frac{n-3}2}$ (\ref{bth}) and
$$
H'_0=-\oh S'_{1,n}S'_{n,1}\wp(\tau/2)+\sum_{j=2}^{n-1}S'_{1,j}S'_{j,1}
E_2(\frac{(j-1)\tau}{2n-2}-u_j)\,,
$$
$$
-2H_1=((S^1_1)^2+(S^1_n)^2)E_2(\oh)+
\sum_{j=2}^{n-1}S^1_{1,j}S^1_{j,1}E_2(\frac{(j-1)\tau}{2n-2}+\oh-u_j)
+S^1_{1,n}S^1_{n,1}E_2(\frac{1}{2}(1+\tau)) \,.
$$


\section{SO$(2n)$\,, Spin$(2n)\,$, $n=2l$.}

\setcounter{equation}{0}

\subsection*{Lattices and characteristic classes}

For $n=2l$
\beq{lfen}
P(D_n)=Q(D_n)+\mZ\varpi_a+\mZ\varpi_{b}\,,~~a,b=(1,n-1,n)\,,~a\neq b\,.
\eq
The weight lattice $P(D_n)$ contains apart from $Q(D_n)$ three sublattices
of index $2$ generated by these weights
\beq{wlen}
P^V(D_n)=Q(D_n)+\mZ\varpi_1\,,
\eq
$$
P^R(D_n)=Q(D_n)+\mZ\varpi_{n-1}\,,~~
P^L(D_n)=Q(D_n)+\mZ\varpi_{n}\,.
$$
The sublattice $P^V(D_n)$ is self-dual.
If $n=8l$ then $^L(P^R(D_n))=P^L(D_n)$. For $n=8l+4\,$ $\,^L(P^R(D_n))=P^R(D_n)$, and
$^L(P^L(D_n))=P^L(D_n)$. In other words, we have the following hierarchy of the lattices
$$
\begin{array}{ccccc}
   &   &P  &  \\
   & \swarrow & \downarrow & \searrow &  \\
  P^{R} &  & P^V &  & P^{L} \\
   & \searrow & \downarrow & \swarrow &  \\
   &  &Q  &  &
\end{array}
$$

The center of $Spin(2n)$ for even $n$ is $\clZ(\bar G)\sim\mu_2\times\mu_2$.
The group element $\zeta=\bfe\,(\xi)$ for
 $\xi=\varpi_n$ generates one of the subgroups $\mu_2$.  Putting $\xi=\varpi_n$
 and  $C_{alc}$ (\ref{alcd}) in (3.10.I)
we find $\la_n$
 $$
 \la_n\,:\,
 \left\{
 \begin{array}{l}
  0\leftrightarrow \varpi_{n}\,,~~\varpi_1\leftrightarrow\varpi_{n-1}\,,   \\
   \varpi_j\leftrightarrow\varpi_{n-j}\,,~1<j<n-1\,.
 \end{array}
\right.
$$
In terms of roots the action assumes the form
$\al_0\leftrightarrow\al_n \,,~\al_1\leftrightarrow \al_{n-1}\,,\ldots$
$\al_j\leftrightarrow\al_{n-j}\,,$\\ $1<j<n-1$.
 In the representation $\pi_1$ (\ref{31}) the $\la_n$ action
takes the form
$$
\la_n=\left(
       \begin{array}{cc}
         0 & Id_n \\
         Id_n & 0 \\
       \end{array}
     \right)\,.
$$
Its action on the indices of the basis $(e_1,e_2,\ldots,e_{2n})$ is
\beq{soel}
\la_n\,:\,\left(
\begin{array}{cccccc}
  1 & j & n & n+1 & n+j & 2n \\
  n+1 & n+j & 2n & 1 & j & n
\end{array}
\right)\,.
\eq
 In the canonical basis $(e_1,e_2,\ldots,e_n)$ $\la_n$ in $\gH_{D_{2l}}$ is represented by the matrix
$$
\la_n\to E_{jk}=-\de_{j,n-k+1}\,.
$$

\vspace{0.5cm}
\unitlength 1mm 
\linethickness{0.4pt}
\ifx\plotpoint\undefined\newsavebox{\plotpoint}\fi 
\begin{picture}(57.427,33.149)(0,0)
\put(21.257,23.636){\circle{1.808}}
\put(28.392,19.325){\circle{2.081}}
\put(35.527,19.622){\circle{2.081}}
\put(49.501,19.622){\circle{1.808}}
\put(56.487,23.487){\circle{1.88}}
\put(56.339,15.609){\circle{1.808}}
\put(21.257,15.757){\circle*{1.88}}
\put(22.149,23.19){\circle*{.297}}
\put(29.73,19.473){\circle*{.297}}
\put(29.73,19.473){\line(1,0){4.757}}
\put(36.717,19.622){\line(1,0){5.054}}
\put(46.082,19.473){\line(1,0){.149}}
\put(46.082,19.622){\line(1,0){2.378}}
\multiput(55.447,23.041)(-.05679775,-.03340449){89}{\line(-1,0){.05679775}}
\multiput(50.244,18.73)(.06937333,-.03369333){75}{\line(1,0){.06937333}}
\multiput(27.5,18.582)(-.06937333,-.03369333){75}{\line(-1,0){.06937333}}
\multiput(22,23.19)(.055175258,-.033721649){97}{\line(1,0){.055175258}}
\put(18.433,26.757){\makebox(0,0)[cc]{$\alpha_1$}}
\put(28.095,15.46){\makebox(0,0)[cc]{$\alpha_2$}}
\put(47.271,16.055){\makebox(0,0)[cc]{$\alpha_{n-2}$}}
\put(54.555,27.055){\makebox(0,0)[cc]{$\alpha_{n-1}$}}
\put(52.771,12.636){\makebox(0,0)[cc]{$\alpha_n$}}
\put(17.392,11.595){\makebox(0,0)[cc]{$\alpha_0$}}
\put(49.352,27.054){\vector(2,-1){.07}}\qbezier(23.933,26.757)(35.602,33.149)(49.352,27.054)
\put(46.528,23.041){\vector(3,-1){.07}}\qbezier(30.027,23.041)(38.426,26.311)(46.528,23.041)
\put(46.974,13.676){\vector(4,1){.07}}\qbezier(26.163,13.527)(37.163,10.777)(46.974,13.676)
\put(35.974,5.351){\makebox(0,0)[cc]{Fig.5 $D_n$, $n$-even, $\lambda_n$}}
\end{picture}
\vspace{0.5cm}

Let us take  $\xi=\varpi_{n-1}$. Then the corresponding Weyl transformation
 $\la_{n-1}$ acts on $C_{alc}$ as
 $$
 \la_{n-1}\,:\,
 \left\{
 \begin{array}{l}
  0\leftrightarrow\varpi_{n-1}\,,~~\varpi_1\leftrightarrow\varpi_{n}\,,   \\
   \varpi_j\leftrightarrow\varpi_{n-j}\,,~1<j<n-1\,,
 \end{array}
\right.
$$
or $\al_{n-1}\leftrightarrow \al_0\,,$  $ \al_{n}\leftrightarrow\al_1\,,$
 $\al_j\leftrightarrow\al_{n-j}\,,$ $1<j<n-1$\,.

\vspace{0.5cm}
\unitlength 1mm 
\linethickness{0.4pt}
\ifx\plotpoint\undefined\newsavebox{\plotpoint}\fi 
\begin{picture}(59.806,33.967)(0,0)
\put(23.636,23.784){\circle{1.808}}
\put(30.771,19.473){\circle{2.081}}
\put(37.906,19.77){\circle{2.081}}
\put(51.88,19.77){\circle{1.808}}
\put(58.866,23.635){\circle{1.88}}
\put(58.718,15.757){\circle{1.808}}
\put(23.636,15.905){\circle*{1.88}}
\put(24.528,23.338){\circle*{.297}}
\put(32.109,19.621){\circle*{.297}}
\put(32.109,19.621){\line(1,0){4.757}}
\put(39.096,19.77){\line(1,0){5.054}}
\put(48.461,19.621){\line(1,0){.149}}
\put(48.461,19.77){\line(1,0){2.378}}
\multiput(57.826,23.189)(-.05679775,-.03340449){89}{\line(-1,0){.05679775}}
\multiput(52.623,18.878)(.06937333,-.03369333){75}{\line(1,0){.06937333}}
\multiput(29.879,18.73)(-.06937333,-.03369333){75}{\line(-1,0){.06937333}}
\multiput(24.379,23.338)(.055175258,-.033721649){97}{\line(1,0){.055175258}}
\put(20.812,26.905){\makebox(0,0)[cc]{$\alpha_1$}}
\put(30.474,15.608){\makebox(0,0)[cc]{$\alpha_2$}}
\put(49.65,16.203){\makebox(0,0)[cc]{$\alpha_{n-2}$}}
\put(56.934,27.203){\makebox(0,0)[cc]{$\alpha_{n-1}$}}
\put(55.15,12.784){\makebox(0,0)[cc]{$\alpha_n$}}
\put(19.771,11.743){\makebox(0,0)[cc]{$\alpha_0$}}
\put(27.203,24.527){\vector(-1,0){.07}}\qbezier(56.933,19.027)(49.501,26.386)(27.203,24.527)
\put(25.865,18.284){\vector(-3,-2){.07}}\qbezier(54.406,23.784)(35.082,24.156)(25.865,18.284)
\put(38.649,5.054){\makebox(0,0)[cc]{Fig.6 $D_n$, $n$-even, $\lambda_{n-1}$}}
\end{picture}

 In the representation $\pi_1$ (\ref{31}) the $\la_{n-1}$ action
takes the form
$$
\la_{n-1}=
  \left(
    \begin{array}{cccccc}
      0 &0  & 1 & 0 & 0 & 0 \\
      0 & 0 & 0 & 0  & Id_{n-2}&0 \\
      1 &0  &  & 0 & 0 & 0 \\
      0 &0  &  & 0 & 0 & 1 \\
      0& Id_{n-2} & 0 & 0& 0 & 0 \\
      0 & 0 & 0 & 1 & 0 & 0\\
    \end{array}
  \right)\,.
$$
It is conjugated to $\la_{n}$ by the matrix
$$
 \left(\begin{array}{cccc}
  Id_{n-1} & 0 & 0 & 0 \\
  0 & 0 & 1 & 0 \\
  0 & 1 & 0 & 0  \\
  0 & 0 & 0 &  Id_{n-1}
\end{array}
 \right)\,.
$$
We will not consider the corresponding GS basis.

The case $\xi=\varpi^\vee_1$ was already considered (\ref{indo1}). In this case $\ti\gg_0={\bf so}(2n-3)$.



\subsection*{The GS-basis.}

The $\la_n$ action on $Z$ in $\pi_1$ takes the form
$$
\la_n(Z)=\la_n
\left(
  \begin{array}{cc}
    A & B \\
    C & -\ti A \\
  \end{array}
\right)=\left(
  \begin{array}{cc}
   -\ti A & C \\
    B &  A \\
  \end{array}
\right)\,,
$$
Since $\la_n^2=Id$
 \beq{cgr}
 \gg_{D_n}=\gg_0+\gg_1\,,
 \eq
 where
 $$
 \gg_0=\left\{
 \left(
   \begin{array}{cc}
     X & Y \\
     Y & X \\
   \end{array}
 \right)\,\begin{array}{lc}
            | & \ti X=-X \\
            | & \ti Y=-Y
          \end{array}
 \right\}\,,
~~~
 \gg_1=\left\{
 \left(
   \begin{array}{cc}
     X & Y \\
     -Y & -X \\
   \end{array}
 \right)\,\begin{array}{lc}
            | & \ti X=X \\
            | & \ti Y=-Y
          \end{array}
 \right\}\,,
 $$
where
\beq{cgr1}
\gg_0=\ti\gg_0+V\,,~~\ti\gg_0
=\left\{
 \left(
   \begin{array}{cc}
     X & 0 \\
     0 & X \\
   \end{array}
 \right)\,\right\}\,,~~V=\left\{
 \left(
   \begin{array}{cc}
     0 & Y \\
     Y & 0 \\
   \end{array}
 \right) \right\}\,.
\eq
According to Lemma 9.1 $\ti\gg_0=\gg_{D_{\frac{n}2}}={\bf so}(n)$.
 The Cartan subalgebra $\ti\gH_0$ in $\gg_0$ has the basis
 \beq{bne}
 \begin{array}{l}
   \tie_1=\di(1,0,\ldots,0,-1,1,0,\ldots,0,-1)\,, \\
   \tie_2=\di(0,1,\ldots,-1,0,0,1,\ldots,-1,0)\,,  \\
   \ldots\ldots,\\
   \tie_l=\di(0,\ldots,1,-1,\dots,0,0,\ldots,1,-1,\dots,0)\,.
 \end{array}
 \eq
It defines the moduli space of the SO$(2n,\mC)$-bundles with characteristic class corresponding to $\varpi_n$
 \beq{csinv}
 \ti\gH_0=\{\ti\bfu=\sum_{j=1}^lu_j\tie_j\}\,.
\eq

From (\ref{soel}) the GS-basis in $V$ takes the form
\beq{Vn}
\gt^0_{j,n+k}=\f1{\sq2}(E_{j,n+k}+E_{n+j,k}-E_{n-k+1,2n-j+1}
-E_{2n-k+1,n-j+1})\,,~~(j,k=1,\ldots,n\,,~j\neq k)
\eq
with the Killing form
\beq{pvn}
(\gt^0_{j_1,n+k_1},\gt^0_{j_2,n+k_2})=\de_{j_1,k_2}\de_{k_1,j_2}\,.
\eq

The GS-basis in $\gg_1$ is
$$
\gh_{j}^1=\f1{\sq2}(e_{j}-e_{n+j}+e_{n-j+1}-e_{2n-j+1})\,,~~(j=1,\ldots,l)\,,
$$
\beq{gs1e}
\gt^1_{j,k}=\f1{\sq2}(E_{j,k}-E_{n+j,n+k}+E_{n-k+1,n-j+1}-E_{2n-k+1,2n-j+1})\,,~~(j,k=1,\ldots,l)\,,
\eq
$$
\gt^1_{j,n+k}=\f1{\sq2}(E_{j,n+k}-E_{n+j,k}-E_{n-k+1,2n-j+1}
+E_{2n-k+1,n-j+1})\,,~~(j,k=1,\ldots,l\,,~j\neq k)\,,
$$
with the Killing form
\beq{pg1}
(\gh_{j}^1,\gh_{k}^1)=\de_{j.k}\,,~~
(\gt^1_{j_1,k_1},\gt^1_{j_2,k_2})=\de_{j_1,k_2}\de_{j_2,k_1}\,,
\eq
$$
(\gt^1_{j_1,n+k_1},\gt^1_{j_2,n+k_2})=-\de_{j_1,k_2}\de_{j_2,k_1}\,.
$$

\bigskip

The GS-basis and the Lax operator related to the characteristic class
defined by  $\varpi_{n-1}$ are conjugated to the GS-basis and the Lax operator related to the characteristic class defined by  $\varpi_{n}$. Therefore the
corresponding Hamiltonians coincide.


\subsection*{Lax operators and Hamiltonians.}

Consider  a  bundle with nontrivial characteristic class generated by
the weight $\varpi_n$. The moduli space is a quotient of $\ti\gH_0=\{
\ti\bfu=\sum_{j=1}^lu_j\ti e_j\},~$ $l=\frac{n}2$,
where $\ti\gH_0$ is a Cartan subalgebra of the invariant algebra
$\ti\gg_0=\gg_{D_l}$ (\ref{csinv})

 For $l$ odd there are three types of invariant sublattices
 $\ti Q(D_l)$ (\ref{drl}), $\ti P(D_l)$ (\ref{pdno}) and  $\ti P_2(D_l)$ (\ref{P2D}).
 They define three types of moduli spaces for these bundles as the fundamental domains
\beq{lod}
\ti\bfu\sim\ti\bfu+\ga_1\tau+\ga_2\,, ~~\ga_j\in\ti Q(D_l)\,,~{\rm or}~
 \ti P(D_l)\,,~{\rm or}~\ti P_2(D_l)\,.
 \eq

For $l$ even the  invariant sublattices are  $\ti Q(D_l)$, $\ti P^L(D_l)$,
 $\ti P^R(D_l)$ and $\ti P^V(D_l)$ (\ref{wlen}).
 Therefore, in this case there are four types of the moduli spaces.
\beq{lev}
\ti\bfu\sim\ti\bfu+\ga_1\tau+\ga_2\,, ~~\ga_j\in\ti Q(D_l)\,,~{\rm or}~
 \ti P^L(D_l)\,,~{\rm or}~\ti P^R(D_l)~{\rm or}~\ti P^V(D_l)\,.
 \eq

Consider the Lax operator for bundles with the characteristic class defined by
$\varpi_n$ in the GS-basis (\ref{Vn}), (\ref{gs1e}) $L=\ti L_0+L_0'+L_1$.
Here $\ti L_0$ is the Lax operator of $D_{\frac{n}2}$,
$$
L'_0(z)=
 \sum_{j\neq k}^{n}
S'_{k,n+j}
\bfe\,(\frac{2n-j-k}{2n-2}z)\phi(\frac{2n-j-k}{2n-2}\tau-u_j-u_k,z)\gt^0_{j,n+k}\,,
$$
$$
L_1(z)=\sum_{j=1}^{n}S^1_{j}\phi(\frac{1}{2},z)\gh^1_{j}+
\sum_{j\neq k}^{n}
S^1_{j,k}
\bfe\,(\frac{k-j}{2n-2}z)\phi(\frac{k-j}{2n-2}\tau-u_j+u_k+\oh,z)\gt^1_{k,j}
$$
$$
-\sum_{j\neq k}^{n}S^1_{j,n+k}\bfe\,(\frac{2n-j-k}{2n-2}z)
\phi(\frac{2n-j-k}{2n-2}\tau-u_j-u_k-\oh,z)\gt^1_{k,n+j}\,.
$$
From (\ref{pvn}) and (\ref{pg1}) after the diagonal reduction we come to the Hamiltonians
$$
H'_0=-\oh\sum_{j\neq k}^{n}S^1_{j,n+k}S^1_{k,n+j}
E_2(\frac{2n-j-k}{2n-2}\tau-u_j-u_k)\,,
$$
$$
H_1=-\oh\sum_{j=1}^n(S^1_{j})^2E_2(\frac{1}{2})-\oh\sum_{j\neq k}^{n}S^1_{j,k}S^1_{k,j}E_2(\frac{k-j}{2n-2}\tau-u_j+u_k-\oh)
$$
$$
+\oh\sum_{j\neq k}^{n}S^1_{j,n+k}S^1_{n+k,j}E_2(\frac{2n-j-k}{2n-2}\tau-u_j-u_k-\oh)\,.
$$
Note that in all expressions $u_j=-u_{n+1-j}$.

Summarizing, the Hamiltonian $H^{CM}_{D_l}+H_0'+H_1$ describes the integrable systems corresponding to the bundles with characteristic classes defined by $\varpi_n$,
or $\varpi_{n-1}$ with the moduli spaces (\ref{lod}) for $l$ odd, or (\ref{lev}) for $l$ even.


\section{E$_6$\,.}
\setcounter{equation}{0}

\subsection*{Roots and weights}
The Cartan subalgebra of the Lie algebra ${\bf e_6}$  is the space
\beq{cse6}
\gH({\bf e_6})=\{\bfu\in\mC^7\,|\, u_5+u_6+u_7=0\}\,.
\eq
The root system $R({\bf e_6})$ is related to the root system  $R(\bas8)$
 of   $\bas8$
$$
R(\bas8)=\sum_{j\neq k}^4\pm e_j\pm e_k\,,~~\sharp\,(R(\bas8))=24\,.
$$
Let
$$
P^L=\{\varpi^L_a\}\,,~P^R=\{\varpi_a^R\}\,,~P^V=\{\varpi_a^V\}\,,~~(a=1,\ldots,8)
$$
 be the weights
of the left (L) and right (R) spinor representations $\underline{8}^L$,
$\underline{8}^R$ and the vector representation (V) $\underline{8}^V$
of $\bas8$.
They are equal to the following combinations of the basic vectors
\beq{we}
\begin{array}{l}
   \varpi^L_a\to\oh\sum_{k=1}^4\pm e_k \,,~~{\rm even~number~of~negative~terms}\,,\\
   \varpi^R_a\to\oh\sum_{k=1}^4\pm e_k \,,~~{\rm odd~number~of~negative~terms}\,,  \\
\varpi^V_a\to\pm e_k \,.
        \end{array}
\eq
In these terms
\beq{are6r}
R({\bf e_6})=R(\bas8)\cup (P^L\pm\f1{\sqrt{2}}(e_5-e_7))
\cup (P^R\pm\f1{\sqrt{2}}(e_5-e_6))\cup (P^V\pm\f1{\sqrt{2}}(e_6-e_7))=
\eq
$$
R(\bas8)\cup \{
\al_{a,\pm}^L=(\varpi_a^L\pm\f1{\sqrt{2}}(e_5-e_7))\,,
\al_{a,\pm}^R =(\varpi_a^R\pm\f1{\sqrt{2}}(e_5-e_6)) \,,
\al_{a,\pm}^V=(\varpi_a^V\pm\f1{\sqrt{2}}(e_6-e_7))\}\,.
$$
The systems of ${\bf e_6}$ roots  is self-dual and the corresponding Dynkin diagram is simply-laced.

Since $\sharp(P^L)=\sharp(P^R)=\sharp(P^V)=16$ the number of roots is equal
$\sharp(R({\bf e_6}))=24+16+16+16=72$.
We have from here
$\dim\,{\bf e_6}={\rm rank}\,{\bf e_6}+\sharp(R({\bf e_6}))=78$.

It follows from here that ${\bf so(8)}$ is subalgebra of ${\bf e_6}$.
It can be found that
\beq{e6d}
{\bf e_6}={\bf so(8)}\oplus \clJ=\underline{28}+\underline{50}\,,
\eq
where $\clJ$ is a representation space of ${\bf so(8)}$. It is  decomposed
on irreducible components as
\beq{clJ}
\clJ=2\times\underline{1}+2\times\underline{8}^L+2\times\underline{8}^R+
2\times\underline{8}^R\,.
\eq
Here two scalar representations complete the Cartan subalgebra
$\gH(\bas8)$ to  the Cartan subalgebra $\gH({\bf e_6})$ (\ref{cse6}).

The basis  of simple roots can be chosen as
\beq{sr}
\Pi=\left\{
\begin{array}{l}
  \al_1=\oh(e_4-e_3-e_2-e_1)+\f1{\sqrt{2}}(e_5-e_6)\,, \\
  \al_2=e_3-e_4 \,,\\
  \al_3=e_2-e_3\,, \\
  \al_4=e_1-e_2\,, \\
  \al_5=-e_1+ \f1{\sqrt{2}}(e_6-e_7)\,,\\
  \al_6=e_4+e_3\,.
\end{array}
\right.
\eq
The subsystem of simple roots
\beq{p1o8}
\Pi_1=\{\al_2,\al_3,\al_4,\al_6\}
\eq
is a system of simple roots of subalgebra ${\bf so(8)}$.

The Weyl group $W_{\bf e_6}$ of $R({\bf e_6})$ is generated by $W_{\bf so(8)}$
(\ref{wed}) and by the reflections $s_{\al_1}$, $s_{\al_5}$
\beq{wee6}
W_{\bf e_6}=\{W_{\bf so(8)}\,,~(e_1\leftrightarrow-e_1\,,e_6\leftrightarrow e_7)\,,
~(e_j\leftrightarrow-e_j\,, \,j=1,\ldots,6)\}\,.
\eq
The defined below $\la_1$ (\ref{lae6}) is an element from $W_{\bf e_6}$

The simple roots (\ref{sr}) define the fundamental Weyl chamber
$$
C=\{\bfu\in\gH\,|\,u_1>u_2>u_3>u_4>0\,,~u_5-u_6>\sqrt{2}\sum_{j=1}^4u_j\,,
~u_6-u_7>\sqrt{2}u_1\}\,.
$$
The minimal root is
\beq{mre6}
\al_0=-\oh(e_4+e_3+e_2+e_1)-\f1{\sqrt{2}}(e_5-e_7)
\eq
$$
=-(\al_1+2\al_2+3\al_3+2\al_4+\al_5+2\al_6)\,.
$$
Then
\beq{wce6}
C_{alc}=\{\bfu\in\gH~|~u_1>u_2>u_3>u_4>0\,,
\eq
$$
 u_5-u_6>\sqrt{2}\sum_{j=1}^4u_j\,,
~u_6-u_7>\sqrt{2}u_1\,,~
\oh\sum_{j=1}^4u_j+\sqrt{2}(u_5-u_7)<1\}\,.
$$

The subset of positive roots corresponding to $\Pi_{\bf e_6}$ assumes the form
\beq{pre6}
R^+({\bf e_6})=R^+(\bas8)\cup (P^L+\f1{\sqrt{2}}(e_5-e_7))
\cup (P^R+\f1{\sqrt{2}}(e_5-e_6))\cup (P^V+\f1{\sqrt{2}}(e_6-e_7))\,.
\eq

This data allows one to construct the Chevalley basis in ${\bf e_6}$.

It follows from (\ref{sr}) that the root lattice of ${\bf e_6}$ is
\beq{rle6}
Q({\bf e_6})=Q(\bas8)+\mZ(-e_1+\f1{\sqrt2}(e_6-e_7))+
\mZ(\oh(e_4-e_3-e_2-e_1)+\f1{\sqrt2}(e_5-e_6))\,.
\eq

The fundamental weights dual to $\Pi_{\bf e_6}$ (\ref{sr}) are
\beq{fwe6}
\left\{
\begin{array}{l}
  \varpi_1 =\frac{\sqrt 2}{3}(2e_5-e_6-e_7)=
  \f1{3}(4\al_1+5\al_2+6\al_3+4\al_4+2\al_5+3\al_6),,\\
   \varpi_2=\oh(e_1+e_2+e_3-e_4)+\f1{3\sqrt 2}(5e_5-e_6-4e_7)=
   \f1{3}(5\al_1+10\al_2+12\al_3+8\al_4+4\al_5+6\al_6)\,,  \\
     \varpi_3=e_1+e_2+\sqrt 2(e_5-e_7)=
     2\al_1+4\al_2+6\al_3+4\al_4+2\al_5+3\al_6 \,,\\
 \varpi_4=e_1+\f1{3\sqrt 2}(4e_5+e_6-5e_7)=
 \f1{3}(4\al_1+8\al_2+12\al_3+10\al_4+5\al_5+6\al_6)\,,  \\
  \varpi_5=\frac{\sqrt 2}{3}(e_5+e_6-2e_7)=
  \f1{3}(2\al_1+4\al_2+6\al_3+5\al_4+4\al_5+3\al_6)\,,  \\
  \varpi_6=\oh(e_1+e_2+e_3+e_4)+\f1{\sqrt 2}(e_5-e_7)
  =\al_1+2\al_2+3\al_3+2\al_4+\al_5+2\al_6 \,.
\end{array}
\right.
\eq

The vector $\rho_{\bf e_6}$ takes the form
$\rho_{\bf e_6}=\rho_{\bf so(8)}+\frac{8}{\sqrt{2}}(e_5-e_7)=3e_1+2e_2+e_3+
\frac{8}{\sqrt{2}}(e_5-e_7)$ and since $h=12$
\beq{re6}
\ka_{\bf e_6}=\f1{12}\rho_{\bf so(8)}+\frac{2}{3\sqrt{2}}(e_5-e_7)=\f1{4}e_1+\f1{6}e_2+\f1{12}e_3+
\frac{2}{3\sqrt{2}}(e_5-e_7)\,.
\eq

The weight lattice $P({\bf e_6})$ is generated by the root lattice $Q({\bf e_6})$
(\ref{wle6}) and one of the fundamental weights $\varpi_k\,$ $\,(k=1,2,4,5)$
\beq{wle6}
P({\bf e_6})=Q({\bf e_6})+\mZ\varpi_1\,.
\eq
The factor group $P/Q$ being isomorphic to the center of the universal
covering group $E_6$  is
\beq{cent}
P({\bf e_6})/Q({\bf e_6})\sim\mu_3\,.
\eq

It follows from (\ref{mre6}) that the fundamental alcove (\ref{wce6}) has the vertices
$$
C_{alc}=(0,\varpi_1,\oh\varpi_2,\f1{3}\varpi_3,\oh\varpi_4,\varpi_5,\oh\varpi_6)\,.
$$
The transformation $\la_1\in\G_{C_{alc}}$ generated by $\varpi_1$ acts on
the extended Dynkin diagram as
\beq{lae6}
\la_1=\left\{
\begin{array}{l}
  \al_1\to \al_0 \,,\\
  \al_2\to \al_6\,, \\
  \al_3\to \al_3\,,\\
  \al_4\to \al_2\,, \\
  \al_5\to \al_1\,, \\
  \al_6\to \al_4\,, \\
  \al_0\to \al_5\,,
\end{array}
~~~~
\begin{array}{l}
  e_1\to \oh(e_1+e_2+e_3-e_4)\,,\\
  e_2\to \oh(e_1+e_2-e_3+e_4)\,,\\
  e_3\to \oh(e_1-e_2+e_3+e_4)\,,\\
  e_4\to \oh(e_1-e_2-e_3-e_4)\,,\\
  e_5\to e_6\,,\\
  e_6 \to e_7\,, \\
  e_7 \to e_5\,.
\end{array}
\right.
\eq
\unitlength 1mm 
\linethickness{0.4pt}
\ifx\plotpoint\undefined\newsavebox{\plotpoint}\fi 
\begin{picture}(60.684,68.413)(0,0)
\put(22.804,37.477){\circle{3.182}}
\put(33.057,37.3){\circle{2.55}}
\put(41.012,37.653){\circle{2.475}}
\put(50.205,37.3){\circle{2.828}}
\put(59.397,37.477){\circle{2.574}}
\put(41.012,27.931){\circle{2.5}}
\put(24.042,37.3){\line(1,0){7.248}}
\put(31.289,37.3){\line(1,0){.53}}
\multiput(34.471,37.477)(.883883,.029463){6}{\line(1,0){.883883}}
\multiput(42.25,37.653)(1.090123,-.029463){6}{\line(1,0){1.090123}}
\multiput(51.972,37.477)(.972272,.029463){6}{\line(1,0){.972272}}
\put(41.012,36.062){\line(0,-1){7.071}}
\put(40.659,17.854){\circle*{2.574}}
\put(40.659,26.693){\line(0,-1){7.248}}
\put(58.867,42.426){\vector(3,-4){.07}}\qbezier(22.981,42.78)(39.333,68.413)(58.867,42.426)
\put(49.321,43.841){\vector(3,-4){.07}}\qbezier(32.704,43.134)(38.891,56.922)(49.321,43.841)
\put(44.371,29.345){\vector(-2,-1){.07}}\qbezier(50.558,33.234)(48.702,31.289)(44.371,29.345)
\put(44.548,20.329){\vector(-3,-2){.07}}\qbezier(57.983,32.704)(55.154,26.517)(44.548,20.329)
\put(33.057,34.648){\vector(-1,4){.07}}\qbezier(36.239,29.345)(33.941,30.229)(33.057,34.648)
\put(24.042,32.35){\vector(-1,2){.07}}\qbezier(35.532,20.683)(28.549,23.335)(24.042,32.35)
\put(20.153,40.835){\makebox(0,0)[cc]{$\alpha_1$}}
\put(29.875,41.012){\makebox(0,0)[cc]{$\alpha_2$}}
\put(39.421,41.366){\makebox(0,0)[cc]{$\alpha_3$}}
\put(48.437,41.543){\makebox(0,0)[cc]{$\alpha_4$}}
\put(57.276,41.189){\makebox(0,0)[cc]{$\alpha_5$}}
\put(35.002,25.633){\makebox(0,0)[cc]{$\alpha_6$}}
\put(31.997,17.678){\makebox(0,0)[cc]{$\alpha_0$}}
\put(38.184,11.137){\makebox(0,0)[cc]{Fig.7 ${\bfe_6}$,$~\lambda_1$}}
\end{picture}

Note that being restricted on $\bas8$ $\,\la$ acts on the fundamental weights as
$$
\varpi^V \to\varpi^R\to\varpi^L\,.
$$


\subsection*{Chevalley basis}

We start with the Chevalley basis in subalgebra ${\bf so(8)}$. It is generated
 by the canonical basis $(e_1,e_2,e_3,e_4)$ in $\gH(\bas8)$ and
 the  root subspaces
\beq{rso8}
\begin{array}{ll}
\al_{jk}=(e_j-e_k)\to E_{j,k}\,, ~~j\neq k  \,,& \lan\ka_{\bf e_6},\al_{jk}\ran=\frac{j-k}{12} \,,\\
\al_{j,k+4}= (e_j+e_k)\to   E_{j,k+4}\,,~~j\neq k  \,, & \lan\ka_{\bf e_6},\al_{jk}\ran=\frac{j+k-8}{12} \,,  \\
\al_{j+4,k}=(-e_j-e_k)\to   E_{j+4,k}\,,~~j\neq k ,k+4\,,~~j\neq k  \,, & \lan\ka_{\bf e_6},\al_{jk}\ran=\frac{-j-k+8}{12} \,, \,.
\end{array}
\eq
The root subspaces in $R({\bf e_6})\setminus R(\bas8)$ are representations
$2\times\underline{8}^L+2\times\underline{8}^R+2\times\underline{8}^R$ of
$\bas8$ (\ref{clJ})
$$
\al_{a,\pm}^L=(\varpi_a^L\pm\f1{\sqrt{2}}(e_5-e_7))\to E_{a,\pm}^L \,,~~a=1,\ldots,8\,,~~
\lan\ka_{\bf e_6}, \al_{a,\pm}^L\ran=\pm\frac{2}3+\lan\ka_{\bf so(8)}, \varpi^L_{a}\ran\,,
$$
\beq{sroe6}
\al_{a,\pm}^R =(\varpi_a^R\pm\f1{\sqrt{2}}(e_5-e_6))\to E_{a,\pm}^R \,,~~a=1,\ldots,8\,,
~~
\lan\ka_{\bf e_6}, \al_{a,\pm}^R\ran=\pm\frac{1}3+\lan\ka_{\bf so(8)}, \varpi^R_{a}\ran\,,
\eq
$$
\al_{a,\pm}^V=(\varpi_a^V\pm\f1{\sqrt{2}}(e_6-e_7))\to E_{a,\pm}^V \,,~~a=1,\ldots,8
~~
\lan\ka_{\bf e_6}, \al_{a,\pm}^V\ran=\pm\frac{1}3+\lan\ka_{\bf so(8)}, \varpi^V_{a}\ran\,.
$$
It follows from (\ref{we}) that $-\varpi_a^{L,R,V}$ is again a spinor weight
of the same type. We denote by $E_{a,\pm}^A$ the Chevalley generator related
to $\varpi_a^A$.

The Killing form in the Chevalley basis assumes the form
$$
(e_j,e_k)=\de_{j,k}\,,~~~( E_{j,k},E_{l,m})=\de_{k,l}\de_{j,m}\,,~~(j,k=1,\ldots,4)\,,
$$
\beq{kfc6}
(E_{a,\pm}^A,E_{b,\mp}^B)=\de_{a,-b}\de^{A,B}\,,
~~A,B=L,R,V\,.
\eq


\subsection*{GS basis}

Consider the Weyl action $\la_1$  (\ref{lae6}) on the root spaces of ${\bf e_6}$.
It easy to see that $\la_1$ preserves ${\bf so(8)}$ and therefore $\clJ$
in (\ref{e6d}).


\subsubsection*{GS basis in ${\bf so(8)}$.}

Consider the action of $\la_1$ on ${\bf so(8)}$. Note first, that an orbit of $\la_1$
that does not contains the minimal root is $\Pi_1$ (\ref{p1o8}).
It is a basis in the Cartan subalgebra $\gH(\bas8)$. Since $\la_1^3=1$
\beq{g2c}
\gH(\bas8)=\ti\gH_0+\gH_1+\gH_2\,,~~~\la_1\,\gH_j=\om^j\gH_j\,,~~\om=\exp\,\frac{2\pi i}3\,.
\eq
Here $\ti\gH_0$ is a Cartan subalgebra in $\ti\gg_0$.
The basis $\ti\Pi^\vee$ of simple coroots is
\beq{srg2a}
\ti\Pi^\vee=\{\ti\al^\vee_3=\sum_{k=0}^2\la^k\al_3=3\al_3=
3(e_2-e_3)\,,~~
\ti\al^\vee_2=\sum_{k=0}^2\la^k\al_2=(e_1-e_2+2e_3)\}\,.
\eq
The dual system is a system of simple roots of ${\bf g_2}$
\beq{srg2}
\ti\Pi=\{\ti\al_3=\f1{3}(e_2-e_3)\,,~\ti\al_2=\f1{3}(e_1-e_2+2e_3)\}\,.
\eq
In this basis the positive ${\bf g_2}$ roots are
\beq{rg2}
\ti R^+({\bf g_2})=(\ti\Pi\,,~\f1{3}(2,1,1,0)\,,~\f1{3}(1,2,-1,0)\,,~\f1{3}(1,1,0,0)\,,~
\f1{3}(1,0,1,0))\,.
\eq
It is convenient to pass from the coroot basis $\ti\Pi^\vee$ in
$\gH(\bag2)$ to the canonical basis $(e_1,e_2,e_3)$
\beq{cbg2}
\ti\bfu=u_1e_1+u_2e_2+u_3e_3\,, ~~u_1=u_2+u_3\,,~~(u_4=0)\,.
\eq
According with (\ref{g2c})  the GS basis in $\gH(\bas8)$
is generated by (\ref{cbg2}) and by
\beq{gh0}
\left\{
\begin{array}{c}
  \gh^1_{\al_2}=\{\f1{\sqrt{3}}(\al_2+\om\al_4+\om^2\al_6)\}
=\{\f1{\sqrt{3}}(\om,-\om,1+\om^2,-1+\om^{2})\}\,, \\
  \gh^2_{\al_2}=\{\f1{\sqrt{3}}(\al_2+\om^2\al_4+\om\al_6)\}=
\{\f1{\sqrt{3}}(\om^2,-\om^2,1+\om,-1+\om)\}\,.
\end{array}
\right.
\eq
With respect to the $\la_1$ $\bas8$ is decomposed as
\beq{so8ex1}
{\bf so(8)}={\bf g}_2\oplus\underline{7}\oplus\underline{7}'\,,~~
\la_1({\bf g}_2)={\bf g}_2\,,~\la_1(\underline{7})=\om\cdot\underline{7}\,,
~\la_1(\underline{7}')=\om^2\cdot\underline{7}'\,,
\eq
where  $\underline{7}$,
$\,\underline{7}'$ are fundamental representations of ${\bf g}_2$.

According with the gradation we have the following gradation of the root
spaces
\vspace{2mm}
$$
\begin{tabular}{|l|l|l|}
  \hline
  ${\bf g_2}$ & $\underline{7}$ & $\underline{7}'$ \\
  \hline
  $E_{12}^0=E_{12}+E_{34}+E_{35}$ & $\gt_{12}^1=\f1{\sqrt{3}}(E_{12} +\om E_{34}+\om^2E_{35})  $&$\gt_{12}^2=\f1{\sqrt{3}}(E_{12} +\om^2 E_{34}+\om E_{35})  $\\
  $E_{13}^0=E_{13}+E_{24}+E_ {25} $&$ \gt_{13}^1=\f1{\sqrt{3}}(E_{13}+\om E_{24}+\om^2E_ {25})   $&$ \gt_{13}^2=\f1{\sqrt{3}}(E_{13}+\om^2 E_{24}+\om E_ {25}) $ \\
$E_{14}^0=E_{14}+E_{15}+E_ {26}  $&$ \gt_{14}^1=\f1{\sqrt{3}}(E_{14}+\om E_{15}+\om^2E_ {26} )   $&$ \gt_{14}^2=\f1{\sqrt{3}}(E_{14}+\om^2 E_{15}+\om E_ {26}) $\\
  $E_{23}^0=E_{23}\,,~E_{16}^0=E_{16}\,,$ & $\gt^1_{(2,3,4),1}$\,, $~\gh^1_{\al_2}$ & $\gt^2_{(2,3,4),1}$\,, $~\gh^2_{\al_2}$ \\
  $E^0_{17}=E_{17}$ &    &\\
  \hline
\end{tabular}
$$
\begin{center}
\textbf{Table1.}
GS basis in ${\bf so(8)}$.
\end{center}
\vspace{2mm}
The roots of ${\bf so(8)}$ that parameterized the GS basis are
$$
(e_1-e_m)\to \gt^a_{1,m}\,,~~(e_m-e_1)\to \gt^a_{m,1}\,.
$$
The left column of table 5  contains seven positive root subspaces of ${\bf g_2}$.
In particular, $E_{23}^0=\ti E_{\al_3}$ and $E_{12}^0=\ti E_{\al_2}$.

The Killing form on ${\bf g}_2$ is the canonical form on $\gH({\bf g}_2)$, and for the root subspaces
non-vanishing elements are
\beq{kfg2r}
(E_{12}^0,E_{21}^0)=(E_{13}^0,E_{31}^0)= (E_{14}^0,E_{41}^0)=3\,,
\eq
$$
(E_{23}^0,E_{32}^0)=(E_{16}^0,E_{61}^0)=(E^0_{17},E^0_{71})=1\,.
$$
For rest generators of the GS basis in  $\bas8$ we have
\beq{kfg8}
(\gh^1_{\al_2},\gh^2_{\al_2})=2\,,~~
(\gt^{k_1}_{jk},\gt^{k_2}_{lm})=\de^{(k_1+k_2,0)}\de_{j,m}\de_{k,l}\,.
\eq


\subsubsection*{GS basis in $\clJ$.}

The basis in the Cartan part of $\clJ$
$$
\gH(\clJ)=\{u_5e_5+u_6e_6+u_7e_7\,,~~u_5+u_6+u_7=0\}\,.
$$
is transformed as
$\la_1(e_5)=e_{7}\,, ~\la_1(e_6)=e_{5}\,, ~\la_1(e_7)=e_{6}$.

The basis $ E_{a,\pm}^A$  (\ref{sroe6}) is transformed as
$$
\la_1\,:~~E_{a,+}^L\to E_{a,-}^V\to E_{a,-}^R\,,~~~
E_{a,-}^L\to E_{a,+}^V\to E_{a,+}^R\,.
$$
There are 16 orbits of this type in  $\clJ$.
Thus, the GS basis in $\clJ$ takes the form
\beq{gsJ}
\gt^k_{a,+}=\f1{\sqrt{3}}(
E_{a,+}^L+\om^k E_{a,-}^V+\om^{2k} E_{a,-}^R)\,,
\eq
$$
\gt^k_{a,-}=\f1{\sqrt{3}}(E_{a,-}^L+\om^k E_{a,+}^V+\om^{2k} E_{a,+}^R)\,,
~~(k=0,1,2)\,,
$$
$$
\gh^k_5=\f1{\sqrt{3}}(e_5+\om^k e_6+\om^{2k} e_7)\,,~~(k=1,2)\,.
$$
Here $\gt^0_{a,5,7}$, $\gt^0_{a,7,5}$ generate the GS basis in $V$ (5.29.I).

The Killing form is
\beq{kfg9}
(\gt^{k_1}_{a,+},\gt^{k_2}_{b,-})=\de_{a,-b}\de^{(k_1+k_2,0)}\,,~~~
(\gh^{k_1}_5,\gh_5^{k_2})=\de^{(k_1+k_2,0)}\,.
\eq

\bigskip

In summary, under the $\la_1$ action ${\bf e_6}$ is decomposed as
\beq{mde}
{\bf e_6}=\gg_0+\gg_1+\gg_2\,,
\eq
where
$$
\gg_0={\bf g_2}+V\,, ~~V=\{\gt^0_{a,+}\,,~\gt^0_{a,-}\}\,,~~\dim\,V=16\,.
$$
The $\la_1$-invariant subalgebra $\gg_0$ is isomorphic to $\bas8+2\times\underline{1}$.
It is obtained from the ${\bf e_6}$  extended Dynkin diagram (Fig.7) by dropping out three  vertices $\al_0$, $\al_1$ and $\al_5$ .

The subspaces $\gg_1$ and $\gg_2$ have the GS basis of the form
\beq{GSe6}
\gg_1=\left\{\gt^1_{a,+}\,,~\gt^1_{a,-}\,,~\gh^1_{5}~
~\gh^1_{\al_2}\,,~\gt^1_{1,(2,3,4)}\,,~\gt^1_{(2,3,4),1}
\right\}\,,
\eq
$$
\gg_2=\{\gt^2_{a,+}\,,~\gt^2_{a,-}\,,~\gh^2_{5}\,,~
\gh^2_{\al_2}\,,~\gt^2_{1,(2,3,4)}\,,~\gt^2_{(2,3,4),1}\}\,.
$$
In correspondence with (\ref{mde}) we have
$$
\underline{78}=\underline{30}+\underline{24}+\underline{24}\,.
$$

The form of the dual basis (5.9.I), (5.15.I) follows from (\ref{kfg8}), (\ref{kfg9})
\beq{dbe6}
\gH^k_{\al_2}=\oh\gh_{\al_2}^{3-k}\,,
~~\gT^k_{jm}=\gt^{3-k}_{mj}\,,~~(k=1,2)\,,
\eq
$$
\gT^k_{a,+}=\gt^{3-k}_{-a,-}\,,~~\gT^k_{a,-}=\gt^{3-k}_{-a,+}\,,~~(k=1,2)\,,
$$
$$
\gH_5^k=\gh^{3-k}_5\,,~~(k=1,2)\,.
$$


\subsection*{Lax operators and Hamiltonians}

\subsubsection*{Trivial bundles}

The moduli space of trivial $\bG=E_6$ bundles is the quotient
$$
\gH_{\bf e_6}/(W_{\bf e_6}\ltimes(\tau Q({\bf e_6})+Q({\bf e_6}))\,,
$$
where $W_{\bf e_6}$ is defined in (\ref{wee6})
It means in particular that $\bfu\in\gH_{\bf e_6}$ is defined up to the shifts
\beq{e6sc}
\bfu\sim\bfu+\tau\ga_1+\ga_2\,,~~\ga_{1,2}\in Q(\bf e_6)\,.
\eq
where $Q({\bf e_6})$ is  (\ref{rle6}).

The moduli space of trivial $G^{ad}=E_6/\mu_3$ bundles is the quotient
$$
\gH_{\bf e_6}/(W_{\bf e_6}\ltimes(\tau P({\bf e_6})+P(\bf e_6)))\,.
$$
It means that
\beq{e6ad}
\bfu\sim\bfu+\tau\ga_1+\ga_2\,,~~\ga_{1,2}\in P(\bf e_6)\,,
\eq
and  $P(\bf e_6)$ is  (\ref{cent}).

For trivial bundles the Lax operator takes the form
\beq{lmtre}
L^{CM}_{e_6}(z)=L^{CM}_{so(8)}+\sum_{j=5}^7(v_j+S_{0,j}E_1(z))e_j
+
\eq
$$
\sum_{a=1}^8\Bigl(
S^L_{a,+}\phi((\bfu,\varpi_a^L)+\f1{\sqrt{2}}(u_5-u_7),z)E^L_{a,+}+
S^L_{a,-}\phi((\bfu,\varpi_a^L)+\f1{\sqrt{2}}(u_7-u_5),z)E_{a,-}^L+\Bigr.
$$
$$
S^R_{a,+}\phi((\bfu,\varpi_a^R)+\f1{\sqrt{2}}(u_5-u_6),z)E_{a,+}^R+
S^R_{a,-}\phi((\bfu,\varpi_a^R)+\f1{\sqrt{2}}(u_6-u_5),z)E_{a,-}^R+
$$
$$
\Bigl.
S^V_{a,+}\phi((\bfu,\varpi_a^V)+\f1{\sqrt{2}}(u_6-u_7),z)E_{a,+}^V+
S^V_{a,-}\phi((\bfu,\varpi_a^V)+\f1{\sqrt{2}}(u_7-u_6),z)E_{a,-}^V
\Bigr)\,.
$$

\bigskip

After symplectic reduction with respect to the action of the Cartan subgroup
we come to integrable $E_6$ hierarchy with two types of moduli space (\ref{e6sc}),
(\ref{e6ad}). The quadratic Hamiltonian is
\beq{he6}
H^{CM}_{e_6}=H^{CM}_{so(8)}+\oh\sum_{j=5}^7v_j^2+
\sum_{a=1}^8\Bigl(S^L_{a,+}S^L_{-a,-}E_2((\bfu,\varpi_a^L)+\f1{\sqrt{2}}(u_5-u_6))
\Bigr.
\eq
$$
\Bigl.
S^R_{a,+}S^R_{-a,-}E_2((\bfu,\varpi_a^R)+\f1{\sqrt{2}}(u_5-u_6))+
S^V_{a,+}S^V_{-a,-}E_2((\bfu,\varpi_a^V)+\f1{\sqrt{2}}(u_6-u_7))
\Bigr)\,.
$$


\subsubsection*{Nontrivial bundles}

\textbf{The moduli space.}

Now $\bfu=\ti\bfu\in\gH(\bag2)$
\beq{hg2}
\ti\bfu=u_1e_1+u_2e_2+u_3e_3\,,~~u_1=u_2+u_3\,.
\eq
 More exactly, $\ti\bfu$ belongs to fundamental
domains with respect to action of affine Weyl group corresponding to ${\bag2}$.
The Weyl group $W_{\bag2}$ is isomorphic to the dihedral group of order 12
with two generators
$$
(u_1,u_2,u_3)\to (u_1,u_3,u_2)\,,
$$
$$
\left(
  \begin{array}{c}
     u_1\\
    u_2 \\
    u_3 \\
  \end{array}
\right)
\to \f1{3}
\left(
  \begin{array}{rrr}
    2 & 1 & -2 \\
    1 & 2 & 2  \\
    -2 & 2 & -1\\
  \end{array}
\right)
\left(
  \begin{array}{c}
     u_1\\
    u_2 \\
    u_3 \\
  \end{array}
\right)\,.
$$

The invariant root sublattice $\ti Q^\vee=Q^\vee(\bag2)$ coincides with
$\ti P^\vee=P^\vee(\bag2)$. From (\ref{srg2a}) we find
$$
\ti Q^\vee(\bag2)=\{m_2e_1+(3m_3-m_2)e_2+(2m_2-3m_3)e_3\,|\,m_2,m_3\in\mZ\}\,.
$$
Then the moduli space of nontrivial $E_6$ bundle is defined as
\beq{mne6}
\gH(\bag2)/(W_{\bag2}\ltimes(\tau \ti Q^\vee(\bag2)+\ti Q^\vee(\bag2)))\,.
\eq
In other words
$$
u_1\sim u_1+\tau m_2+m_2'\,,~~u_2\sim u_2+\tau(3m_3-m_2)+3m'_3-m'_2 \,,
$$
$$
u_3\sim u_3+\tau(2m_2-3m_3)+2m_2'-3m_3'\,,~~m_{2,3}\in\mZ\,.
$$

\bigskip

\textbf{The Lax operators}

According with (\ref{mde})
$L(z)=\ti L_{0}(z)+L_0'(z)+L_1(z)+L_2(z)$.
Here $\ti L_0(z)=L^{CM}_{\bag2}(z)$. Let $\ti\bfv=(v_1,v_2,v_3)$
$\,(v_1=v_2+v_3)$ be the
Poisson dual vector to $\ti\bfu$ (\ref{hg2}). Following to (\ref{rg2}) and table 5 we find
$$
L^{CM}_{\bag2}(z)=\sum_{j=1}^3(v_j+S_{0,j}E_1(z))e_j+
S^0_{12}\phi(\f1{3}(u_1-u_2+2u_3),z)E_{12}^0
+
S^0_{21}\phi(\f1{3}(-u_1+u_2-2u_3),z)E_{21}^0
$$
$$
+
S^0_{23}\phi(\f1{3}(u_2-u_3),z)E_{23}^0+S^0_{32}\phi(\f1{3}(u_3-u_2),z)E_{32}^0+
S^0_{16}\phi(\f1{3}(u_1+u_3),z)E_{16}^0+S^0_{61}\phi(\f1{3}(-u_1-u_3),z)E_{61}^0
$$
$$
+S^0_{17}\phi(\f1{3}(u_1+u_2),z)E_{17}^0+S^0_{71}\phi(\f1{3}(-u_1-u_2),z)E_{71}^0+
S^0_{14}\phi(\f1{3}(2u_1+u_2+u_3),z)E_{14}^0
$$
$$
+S^0_{41}\phi(\f1{3}(-2u_1-u_2-u_3),z)E_{41}^0+
S^0_{13}\phi(\f1{3}(u_1+2u_2-u_3),z)E_{13}^0+
S^0_{31}\phi(\f1{3}(-u_1-2u_2+u_3),z)E_{31}^0\,.
$$
Define $\varphi^k_{\be}$ (6.14.I)
$$
\varphi^k_{\be}(\ti\bfu,z)=\bfe\,\Bigl(\lan\ka_{\bf e_6},\be\ran z\Bigr)
\phi(\lan\ti\bfu+\ka_{\bf e_6}\tau,\be\ran+k/3, z)\,,~~~(k=0,1,2)\,,
$$
where
 $\be=\al_{jk}$ and $\lan\ka_{\bf e_6},\be\ran$  (\ref{rso8}), or
$\be=\al_{(\pm a,\pm)}^L$  and $\lan\ka_{\bf e_6},\be\ran$  (\ref{sroe6}).

Then following (6.15.I) and  (6.17.I) we find
$$
L'_0(z)=\sum_{a=1}^8\Bigl(S^0_{a,-}\varphi^0_{\al_{(-a,-)}^L}(-\ti\bfu,z)
\gt^0_{a,+}+
S^0_{a,+}\varphi^0_{\al_{(a,+)}^L}(-\ti\bfu,z)\gt^0_{a,-}\Bigr)\,,
$$
and for $k=1,2$
$$
L_k(z)=S_5^{3-k}\phi(k/3, z)\gh_5^{k}+S_{\al_2}^{3-k}\phi(k/3, z)\gh_{\al_2}^{k}+
\sum_{m=2}^4(S^{3-k}_{m,1}\varphi^k_{\al_{1,m}}(-\ti\bfu,z)\gt^k_{1,m}+
S^{3-k}_{1,m}\varphi^k_{\al_{m,1}}(-\ti\bfu,z)\gt^k_{m,1})+
$$
$$
\sum_{a=1}^8(S^{3-k}_{-a,-}\varphi^k_{\al_{(a,+)}^L}(-\ti\bfu,z)\gt^{k}_{a,+}+
S^{3-k}_{-a,+}\varphi^k_{\al_{(a,-)}^L}(-\ti\bfu,z)\gt^{k}_{a,-})\,.
$$

After the symplectic reduction we come to the integrable system with quadratic Hamiltonian
$H_{\bf e_6}=H^{CM}_{\bag2}+H_0'+H_1$,
where $H_0'$ is defined by $\oh\tr(L_0^{'2})$, $H_1$ by $\tr(L_1L_2)$, and
$$
H^{CM}_{\bag2}=\oh\sum_{j=1}^3v_j^2-3S^0_{12}S^0_{21}E_2(\f1{3}(u_1-u_2+2u_3))-
S^0_{23}S^0_{32}E_2(\f1{3}(u_2-u_3))
$$
$$
-
S^0_{16}S^0_{61}E_2(\f1{3}(u_1+u_3))-
S^0_{17}S^0_{71}E_2(\f1{3}(u_1+u_2))-3S^0_{14}S^0_{41}E_2(\f1{3}(2u_1+u_2+u_3))
$$
$$
-
3S^0_{13}S^0_{31}E_2(\f1{3}(u_1+2u_2-u_3))\,,
$$
$$
H_{0}'=-
\sum_{a=1}^8S^L_{a,+}S^L_{-a,-}E_2(\varpi^L_a,\ti\bfu)\,,
$$
$$
-H_1=(S_5^1S^2_5+2S^1_{\al_2}S^2_{\al_2})E_2(1/3)+
\sum_{m=2}^4(S^1_{m,1}S^2_{1,m}+ S^1_{1,m}S^2_{m,1}) E_2(u_1-u_m)+
$$
$$
\sum_{a=1}^8(S^1_{a,-}S^2_{-a,+}+S^1_{a,+}S^2_{-a,-})E_2(\varpi_a^L,\ti\bfu)\,.
$$


\section{E$_7$\,.}
\setcounter{equation}{0}

\subsection*{Roots and weights}
The Cartan subalgebra can be identified with the space
$\gH_{\bf e_7}=\{\bfu\in\mC^7\}$.
The root system $R({\bf e_7})$ is related to the root system  $R({\bf e_6})$ (\ref{are6r})
as follows
\beq{are7}
R({\bf e_7})=R({\bf e_6})\cup \pm (\sqrt{2}e_{i+4}\,,~i=1,2,3)
\eq
$$
\cup
( P^R\pm\f1{\sqrt{2}}(e_5+e_6))\cup ( P^V\pm\f1{\sqrt{2}}(e_6+e_7))
\cup ( P^L\pm\f1{\sqrt{2}}(e_5+e_7))=
$$
$$
R({\bf e_6})\cup\{\al^+_{i+4}\,,\al^{V,+}_{a,\pm}\,,\al^{R,+}_{a,\pm}\,,
\al^{L,+}_{a,\pm}\}\,,~(a=1,\ldots,8\,,\,i=1,2,3)
$$
(compare with (\ref{are6r})).
Then $\sharp(R({\bf e_7}))=72+6+3\times 16=126$ and
$\dim\,{\bf e_7}={\rm rank}\,{\bf e_7}+\sharp(R({\bf e_7}))=133$.

The basis  of simple roots can be chosen as
\beq{sr7}
\Pi=\left\{
\begin{array}{l}
  \al_1=\oh(e_4-e_3-e_2-e_1)+\f1{\sqrt{2}}(e_5-e_6)\,, \\
  \al_2=e_3-e_4 \,,\\
  \al_3=e_2-e_3\,, \\
  \al_4=e_1-e_2\,, \\
  \al_5=-e_1+ \f1{\sqrt{2}}(e_6-e_7)\,,\\
  \al_6=e_4+e_3\,,\\
  \al_7=\sqrt{2}e_7\,.
\end{array}
\right.
\eq
The subsystem of simple roots
\beq{p17}
\Pi_1=\{\al_i\,,~~i=1,\ldots 6\}=\Pi({\bf e_6})
\eq
is a system of simple roots of subalgebra ${\bf e_6}$.
It follows from (\ref{are7}) that the positive roots, corresponding to $\Pi$ is
$$
R^+({\bf e_7})=R^+({\bf e_6})\cup  (\sqrt{2}e_{i+4}\,,~i=1,2,3)
$$
$$
\cup
( P^R+\f1{\sqrt{2}}(e_5+e_6))\cup ( P^V+\f1{\sqrt{2}}(e_6+e_7))
\cup ( P^L+\f1{\sqrt{2}}(e_5+e_7))
$$
Then the root lattice takes the form
\beq{rle7}
Q({\bf e_7})=Q({\bf e_6})+\mZ\sqrt{2}e_{7}+\mZ(-e_1+\f1{\sqrt{2}}(e_6+e_7))
+\mZ(\oh(e_1-e_2-e_3-e_4)+\f1{\sqrt{2}}(e_5+e_6))\,.
\eq
The simple roots (\ref{sr7}) define the fundamental Weyl chamber
$C=\{\bfu\in\gH\,|\,\lan\bfu,\al\ran>0\,,~\al\in \Pi({\bf e_7})\}$.

The minimal root is
\beq{mre7}
\al_0=-\sqrt{2}e_5=-(2\al_1+3\al_2+4\al_3+3\al_4+2\al_5+2\al_6+\al_7)\,.
\eq

The root system $R_{\bf e_7}$  is self-dual and the corresponding Dynkin diagram is simply-laced.

The half-sum of the positive roots
$\rho_{\bf e_7}=3e_1+2e_2+e_3+\f1{\sqrt{2}}(17e_5+9e_6+e_7)$
can be expressed in terms of roots of subalgebras
${\bf e_6}$ or $\bf f_4$ (see below)
$$
\rho_{\bf e_7}=\rho_{\bae}+\frac{9}{\sqrt{2}}(e_5+e_6+e_7)=
\rho_{\bf f_4}+\f1{4}(3e_1+e_2+e_3-e_4)+\frac{5}{2\sqrt{2}}(e_5-e_7)\,,
$$
where $\rho_{\bf f_4}=\f1{4}(9,7,3,1,\frac{11}{2\sqrt{2}},0,-\frac{11}{2\sqrt{2}})$.
The Coxeter number is equal to $h=18$.
Then
\beq{ka7}
\ka_{\bf e_7}=\f1{18}\ka_{f_4}+\f1{72}(3e_1+e_2+e_3-e_4)+
\frac{5}{36\sqrt{2}}(e_5-e_7)\,,~~~(\ka_{\bf f_4}=\f1{18}\rho_{\bf f_4})\,.
\eq


We define the fundamental weights dual to $\Pi$ (\ref{sr7})
\beq{fwe7}
\begin{array}{l}
  \varpi_1 =\sr2  e_5=
  2\al_1+3\al_2+4\al_3+3\al_4+2\al_5+2\al_6+\al_7\,,\\
   \varpi_2=\oh(e_1+e_2+e_3-e_4)+\f1{3\sqrt 2}(5e_5-e_6-4e_7)\\
   =\oh(4\al_1+8\al_2+12\al_3+9\al_4+6\al_5+7\al_6+3\al_7)\,,  \\
     \varpi_3=e_1+e_2+\sqrt 2(e_5-e_7)=
     3\al_1+6\al_2+8\al_3+6\al_4+4\al_5+4\al_6+2\al_7 \,,\\
 \varpi_4=e_1+\f1{3\sqrt 2}(4e_5+e_6-5e_7)=
 4\al_1+8\al_2+12\al_3+9\al_4+6\al_5+6\al_6+3\al_7\,,  \\
  \varpi_5=\sr2(e_5+e_6)=
  \f1{2}(6\al_1+12\al_2+18\al_3+15\al_4+10\al_5+9\al_6+5\al_7)\,,  \\
  \varpi_6=\oh(e_1+e_2+e_3+e_4)+\f1{\sqrt 2}(e_5-e_7)\\
  =2\al_1+4\al_2+6\al_3+5\al_4+4\al_5+3\al_6+2\al_7 \,,\\
  \varpi_7=\f1{\sr2}(e_5+e_6+e_7)=\
  \oh(2\al_1+4\al_2+6\al_3+5\al_4+4\al_5+3\al_6+3\al_7)\,.
\end{array}
\eq

It follows from (\ref{fwe7}) that
\beq{wle7}
P(\ble)=Q(\ble)+\mZ\varpi_7\,.
\eq
and the factor-group $P(\ble)/Q(\ble)$ is isomorphic $\mu_2$.

The minimal root (\ref{mre7}) defines the vertices of the fundamental alcove\\
$C_{alc}=(0,\oh\varpi_1,\f1{3}\varpi_2,\f1{4}\varpi_3,\f1{3}\varpi_4,
\oh\varpi_5,\oh\varpi_6,\varpi_7)$.
Then  $\varpi_7$ generates $\la_7$
\beq{lae7}
\la_7=\left\{
\begin{array}{l}
  \al_1\to \al_5 \,,\\
  \al_2\to \al_4\,, \\
  \al_3\to \al_3\,,\\
  \al_4\to \al_2\,, \\
  \al_5\to \al_1\,, \\
  \al_6\to \al_6\,, \\
  \al_7\to \al_0\,, \\
  \al_0\to \al_7\,,
\end{array}
~~~~
\begin{array}{l}
  e_1\to \oh(e_1+e_2+e_3-e_4)\,,\\
  e_2\to \oh(e_1+e_2-e_3+e_4)\,,\\
  e_3\to \oh(e_1-e_2+e_3+e_4)\,,\\
  e_4\to \oh(-e_1+e_2+e_3+e_4)\,,\\
  e_5\to -e_7\,,\\
  e_6 \to -e_6\,, \\
  e_7 \to -e_5\,.
\end{array}
\right.
\eq
\unitlength 1mm 
\linethickness{0.4pt}
\ifx\plotpoint\undefined\newsavebox{\plotpoint}\fi 
\begin{picture}(84.036,63.251)(0,0)
\put(40.731,38.204){\circle*{2.081}}
\put(47.866,38.204){\circle{1.88}}
\put(54.852,38.204){\circle{1.88}}
\put(61.988,38.352){\circle{1.904}}
\put(68.826,38.204){\circle{1.808}}
\put(76.109,38.054){\circle{1.904}}
\put(83.095,37.906){\circle{1.88}}
\put(61.987,31.068){\circle{2.081}}
\multiput(41.771,37.906)(1.040552,.02973){5}{\line(1,0){1.040552}}
\multiput(48.906,37.906)(1.010822,.02973){5}{\line(1,0){1.010822}}
\multiput(56.041,37.906)(.981092,.02973){5}{\line(1,0){.981092}}
\multiput(63.028,38.054)(.981092,-.02973){5}{\line(1,0){.981092}}
\put(70.014,37.757){\line(1,0){4.905}}
\put(77.298,37.757){\line(1,0){4.608}}
\put(61.987,37.163){\line(0,-1){4.608}}
\qbezier(40.73,43.852)(57.453,63.251)(81.609,44.892)
\qbezier(74.92,44.744)(59.534,55.075)(47.419,44.298)
\qbezier(54.852,44.298)(60.129,49.575)(67.487,43.852)
\thicklines
\qbezier(40.73,44)(42.142,46.676)(41.473,45.784)
\qbezier(41.473,45.784)(40.582,42.886)(40.879,43.852)
\qbezier(40.73,44.149)(42.96,45.041)(42.217,44.744)
\multiput(46.973,44.149)(.03344631,.04459508){40}{\line(0,1){.04459508}}
\multiput(47.271,44)(.0550557,.0330334){27}{\line(1,0){.0550557}}
\multiput(54.555,44)(.0335662,.0527469){31}{\line(0,1){.0527469}}
\multiput(54.852,44)(.0431565,.0335662){31}{\line(1,0){.0431565}}
\multiput(65.703,44.446)(.1167966,-.0318536){14}{\line(1,0){.1167966}}
\multiput(67.19,44.149)(-.0330334,.0440445){27}{\line(0,1){.0440445}}
\multiput(72.987,45.338)(.0991002,-.0330334){18}{\line(1,0){.0991002}}
\multiput(74.771,44.744)(-.0335662,.0431565){31}{\line(0,1){.0431565}}
\multiput(79.528,45.933)(.057542,-.0335662){31}{\line(1,0){.057542}}
\multiput(81.46,44.892)(-.03303339,.04542091){36}{\line(0,1){.04542091}}
\put(38.352,34.338){\makebox(0,0)[cc]{$\alpha_0$}}
\put(45.636,34.784){\makebox(0,0)[cc]{$\alpha_1$}}
\put(53.217,35.23){\makebox(0,0)[cc]{$\alpha_2$}}
\put(59.906,41.622){\makebox(0,0)[cc]{$\alpha_3$}}
\put(60.798,27.352){\makebox(0,0)[cc]{$\alpha_6$}}
\put(66.595,34.784){\makebox(0,0)[cc]{$\alpha_4$}}
\put(74.474,34.933){\makebox(0,0)[cc]{$\alpha_5$}}
\put(81.163,35.379){\makebox(0,0)[cc]{$\alpha_7$}}
\put(60.947,21.406){\makebox(0,0)[cc]{Fig.8 $E_7$, $\lambda_7$}}
\end{picture}


\subsection*{Chevalley basis}

The Lie algebra $\ble$ can be defined in terms of $\bae$ and its representations
\beq{e7e6}
\ble=\bae\oplus \clI\,,~~\clI=\underline{27}\oplus\underline{\overline{27}}\oplus\underline{1}\,.
\eq
Here $\underline{27}$ and $\underline{\overline{27}}$ are two fundamental representations
of $\bae$ corresponding to the weights $\varpi_1$ and $\varpi_5$ (\ref{fwe6}).
The scalar $\underline{1}$ corresponds to the generator $(e_5+e_6+e_7)$ in the Cartan subalgebra $\gH_{\ble}$.
 It allows us to use in $\gH(\ble)$ the unrestricted canonical basis $e_j$, $j=1,\ldots,7$.
It follows from  (\ref{are7}) that the rest generators correspond to new  54 root subspaces. Similarly to (\ref{sroe6}) they are
\beq{sroe7}
\begin{array}{ll}
\al_{(a,\pm)}^{(L,+)}=  (\varpi_a^L\pm\f1{\sqrt{2}}(e_5+e_7))\to E_{a,\pm}^{L,+} & a=1,\ldots,8\,, \\
\al_{(a,\pm)}^{(R,+)}=  (\varpi_a^R\pm\f1{\sqrt{2}}(e_5+e_6))\to E_{a,\pm}^{R,+} \,, & a=1,\ldots,8\,, \\
\al_{(a,\pm)}^{(V,+)}=   (\varpi_a^V\pm\f1{\sqrt{2}}(e_6+e_7))\to E_{a,\pm}^{R,+} \,, & a=1,\ldots,8\,, \\
\pm \al^{(+)}_j=  \pm\sqrt{2}e_j\to E_{j,\pm}\,, & j=5,6,7\,.
\end{array}
\eq
The positive root subspaces are $E_{a,+}^{A,+}$ and $E_{j,+}$.

The new generators are orthogonal to the $\bae$ generators and (see (A.25.I))
\beq{kfc7}
(E_{a,+}^{A,+},E_{b,-}^{B,+})=\de_{a,-b}\de^{A,B}\,,
~~A,B=L,R,V\,,
\eq
$$
(E_{j,\pm},E_{k,\mp})=\de_{jk}\,.
$$


\subsection*{GS basis}

Since $\la_7^2=1$,
\beq{gse7}
\ble=\gg_0\oplus\gg_1\,,~~\gg_0=\ti\gg_0+V\,.
\eq
It follows from (\ref{e7e6}) and Fig. 8 that $\gg_0\sim\bae+\underline{1}$
We will prove that $\ti\gg_0=\baf$.

The root subsystem $\Pi_1$ that does not contain an orbit of $\la$ passing
through $\al_0$ is $\Pi_{\bae}$ (\ref{p17}). It follows from (\ref{lae7}) that the $\la_7$-action
on $\Pi_{\bae}$ takes the form
$$
\ti\la_7=\left\{
\begin{array}{l}
  \al_1\to \al_5 \,,\\
  \al_2\to \al_4\,, \\
  \al_3\to \al_3\,,\\
  \al_4\to \al_2\,, \\
  \al_5\to \al_1\,, \\
  \al_6\to \al_6\,, \\
\end{array}
\right.
$$
The set of orbits in $\Pi_1^\vee=\Pi_1$ is
$$
\ti\Pi^\vee=\Pi^\vee_1/\mu_2=\Bigl(\ti\al_1^\vee=\al_6\,,~\ti\al_2^\vee=\al_3\,,~
\ti\al_3^\vee=\al_2+ \al_4\,,~\ti\al_4^\vee=\al_1 +\al_5\Bigr)=
$$
\beq{cbe7}
=\Bigl(e_4+e_3\,,e_2-e_3\,,e_1-e_2+e_3-e_4\,,\oh(e_4-e_3-e_2-3e_1)+\f1{\sqrt{2}}(e_5-e_7)
\Bigr)\,.
\eq
It is the coroot basis in the invariant subalgebra $\ti\gH_0$.
The dual root system
\beq{srf4}
\ti\al_1=e_3+e_4\,,~\ti\al_2=e_2-e_3\,,~\ti\al_3=\oh(e_1-e_2+e_3-e_4)\,,
\eq
$$
\ti\al_4=\frac{1}4(-3e_1-e_2-e_3+e_4)+\f1{2\sqrt{2}}(e_5-e_7)\,.
$$
defines simple roots of type ${\bf f_4}$.  Then $\ti\gH_0=\gH(\baf)$.


\subsubsection*{GS basis in $\bae$ subalgebra}
 Under the $\ti\la_7$-action $\bae$ is decomposed as
\beq{e6f4}
\bae=\baf\oplus\underline{26}\,,
\eq
where $\underline{26}$ is a fundamental representation of $\baf$. In terms of (\ref{gse7})
 $\ti\gg_0=\baf$ and $\underline{26}\subset\gg_1$.

Let us describe (\ref{e6f4}) in terms of the familiar decomposition (\ref{e6d})
\beq{povt}
\bae={\bf so(8)}\oplus\clJ=\underline{28}\oplus\underline{50}
\eq
 taking into account the $\ti\la_7$-action.

 \bigskip
 \noindent
 \textbf{GS basis in ${\bf so(8)}$-component.}\\
The  ${\bf so(8)}$ subalgebra  is generated by the simple roots $(\al_2,\al_3,\al_4,\al_6)$. The $\ti\la_7$ invariant subsystem $(\ti\al_1,\ti\al_2,\ti\al_3)$
(\ref{srf4}) forms $B_3$ root subsystem. It implies that
\beq{so8ex}
{\bf so(8)}={\bf so(7)}\oplus\underline{7}\,,~~\underline{28}=\underline{21}\oplus\underline{7}\,,
\eq
$$
\ti\la_7({\bf so(7)})={\bf so(7)}\,,~\ti\la_7(\underline{7})=-\underline{7}\,,
$$
where  $\underline{7}$, is the fundamental representations of ${\bf so(7)}$.

The 24 root subspaces of ${\bf so(8)}$ contains 18 root subspaces of ${\bf so(7)}$.
Then according with the definition of ${\bf so(8)}$ root subspaces (\ref{rso8}) we
define the GS-basis in ${\bf so(8)}$.

\vspace{3mm}
\begin{center}
\begin{tabular}{|l|l|}
  \hline
  ${\bf so(7)}$ & $\underline{7}$ \\
  \hline
  $E_{12}^0=E_{12}+E_{34}$ & $\gt_{12}^1=\f1{\sqrt{2}}(E_{12} - E_{34})  $\\
  $E_{13}^0=E_{13}+E_{24}$&$ \gt_{13}^1=\f1{\sqrt{2}}(E_{13}- E_{24})$  \\
$E_{15}^0=E_{15}+E_{26}$&$ \gt_{15}^1=\f1{\sqrt{2}}(E_{15}- E_{26} )   $\\
  $E_{23}^0=E_{23}\,,~E_{14}^0=E_{14}\,,$ & $\gt_{j,1}^1\,,~j=2,3,5$    \\
  $E^0_{35}=E_{35}\,,~E_{25}^0=E_{25}$ &  $\gh^1_{\al_2}$ (\ref{ghf4})  \\
  $E^0_{16}=E_{16}\,,~E_{17}^0=E_{17}$ &    \\
  \hline
\end{tabular}
\\
\vspace{5mm}

\textbf{Table 2.} GS basis in ${\bf so(8)}$.
\end{center}
The left column represents 9 positive root subspaces of ${\bf so(7)}$, and the right
column represents the root subspaces and a Cartan element in $\underline{7}$.
The ${\bf so(8)}$ roots that parameterized the GS-basis are
\beq{re7gs}
\begin{array}{ll}
  \al_{12}=(e_1-e_2)\to \gt_{12}^1\,, & -\al_{12}=(e_2-e_1)\to \gt_{21}^1\,, \\
  \al_{13}= (e_1-e_3)\to \gt_{13}^1\,, &  -\al_{13}= (e_3-e_1)\to \gt_{31}^1\,, \\
   \al_{15}=(e_1+e_4)\to \gt_{15}^1\,, & -\al_{15}=(-e_1-e_4)\to \gt_{51}^1\,, \\
\end{array}
\eq

Consider the embedding of $\gH({\bf so(7)})$ in $\gH({\bf so(8)})$. Remember that
the basis in $\gH({\bf so(7)})$ is $(\ti\al_k^\vee\,,~k=1,2,3)$.
In addition we have the anti-invariant generator
\beq{ghf4}
\gh^1_{\al_2}=\f1{\sqrt{2}}(\al_2-\al_4)=\f1{\sqrt{2}}(-e_1+e_2+e_3-e_4)\,,
\eq
 $$
 \gH({\bf so(8)})=\gH({\bf so(7)})\oplus\mC\gh^1_{\al_2}\,.
  $$
  In this way we have defined the $\ti\la_7$-action on ${\bf so(8)}$ component in (\ref{povt}).
In correspondence with (\ref{e6f4}) we have
$$
{\bf so(7)}\subset\baf\,,~~   \underline{7}\subset\underline{26}\,.
$$

 \bigskip
 \noindent
  \textbf{GS basis in $\clJ$-component}\\
Now consider the $\la$-action on the space $\clJ$ (\ref{povt}). It is represented by
$48$ root subspaces  (\ref{are6r}) and two elements from $\gH(\bae)$. Let us define the
latter generators. They are
\beq{ce6a}
\gh^1_{\al_1}=\f1{\sqrt{2}}(\al_1-\al_5)=\f1{\sqrt{2}}\Bigl(\oh(e_1-e_2-e_3+e_4)+
\f1{\sqrt{2}}(e_5-2e_6+e_7)\Bigr)
\eq
and the already defined invariant generator $\ti\al_4^\vee=2\ti\al_4$ (\ref{srf4})
that completes $\gH({\bf so(7)})$ to $\gH(\baf)$.

The $\ti\la_7$-action on the
the root subspaces takes the form
$$
\ti\la_7~:~\left\{
\begin{array}{c}
  (P^R\pm\f1{\sqrt{2}}(e_5-e_6))\leftrightarrow (P^V\pm\f1{\sqrt{2}}(e_6-e_7))\,, \\
  (P^L\pm\f1{\sqrt{2}}(e_5-e_7))\leftrightarrow(P^L\pm\f1{\sqrt{2}}(e_5-e_7))\,.
\end{array}
\right.
$$
Since
$\ti\la_7\,:\,\varpi_4^{L}\to -\varpi_4^{L}=\varpi_8^{L}$,
all  roots $P^L\pm\f1{\sqrt{2}}(e_5-e_7)$
are fixed under the $\ti\la_7$-action, except
  $(\varpi_4^{L}=\oh(e_1-e_2-e_3+e_4))\pm\f1{\sqrt{2}}(e_5-e_7)$.
Then from the $\ti\la_7$ action on the Chevalley basis in $\bae$ (\ref{sroe6})
we obtain generators of the GS basis
\beq{te67}
\begin{array}{cc}
  \gt^{R,k}_{a,\pm}=\f1{\sqrt{2}}(E^R_{a,\pm}+(-1)^kE^V_{a,\pm})\,, & k=0,1\,, \\
  \gt^{L,k}_{4,\pm}=\f1{\sqrt{2}}(E^L_{4,\pm}+(-1)^kE^L_{8,\pm})\,, & k=0,1\,, \\
  \gt^{L,0}_{a,\pm}=E^L_{a,\pm}\,, & ~a\neq 4,8\,.
\end{array}
\eq
The $\mu_2$-gradation  $\clJ=\clJ_0\oplus\clJ_1$ takes the form
\beq{clj0}
\clJ_0=\{\sqrt{2}\gt^{R,0}_{a,\pm}=E_{\ti\al^R_{a,\pm}}\,,~a=1,\ldots,8\,,~~
\sqrt{2}\gt^{L,0}_{a,\pm}=E_{\ti\al^L_{a,\pm}}\,,~a\neq 4,8\,,~~
\sqrt{2}\gt^{L,0}_{4,\pm}=E_{\ti\al^L_{4,\pm}}\,,~\ti\al_4^\vee\}\,,
\eq
\beq{clj1}
\clJ_1=\{\gt^{R,1}_{a,\pm}\,,~a=1,\ldots,8\,,~~\gt^{L,1}_{4,\pm}\,,~\gh^1_{\al_1}\}\,,
\eq
$$
\dim\,\clJ_0=31\,,~~\dim\,\clJ_1=19\,.
$$
Here $E_{\ti\al^R_{a,\pm}}$ are invariant root subspaces  constructed from the
root subspaces of $\bae$ (\ref{sroe6}).
Finally, by comparing (\ref{e6f4}) and (\ref{povt}) we come to the decompositions
\beq{f4d}
\baf={\bf so(7)}\oplus\clJ_0\,,~~(52=21+31)\,,
\eq
\beq{26d}
\underline{26}=\clJ_1\oplus\underline{7}\,,~~(26=19+7)\,,
\eq
where $\underline{7}$ is represented by the right column in Table 2.


\subsubsection*{Chevalley basis in $\baf$}

In what follows we need the Chevalley basis in $\baf$ (\ref{f4d})
in terms of the Chevalley basis in $\bae$. We pass to the canonical basis in $\gH(\baf)$. From (\ref{cbe7}) we find
\beq{cf4}
\gH(\baf)=\{\ti\bfu=u_1e_1+u_2e_2+u_3e_3+u_4e_4+
u_5e_5-u_5e_7 \}\,,
\eq
where
\beq{fres}
u_1-u_2-u_3+u_4=0\,.
\eq
In other terms it can be written in the form (see (\ref{f4d}))
\beq{f4e7}
\gH(\baf)=\gH({\bf so(7)})\oplus\mC(\frac{1}2(-3e_1-e_2-e_3+e_4)+\f1{\sqrt{2}}(e_5-e_7))\,.
\eq

It follows from (\ref{clj0}), (\ref{f4d}) that the $\baf$ roots are
\beq{rf4}
R(\baf)=R({\bf so(7)})\cup R(\clJ_0)\,,
\eq
$$
R^+(\clJ_0)=\{\ti\al^R_{a,+}=\f1{2}(\varpi_a^R+\varpi_a^V+\f1{\sqrt{2}}(e_5-e_7))\,,~
\ti\al^L_{a,+}=\oh(\varpi_a^L+\f1{\sqrt{2}}(e_5-e_7))\,,~(a\neq 4,8)\,,~\}\,,
$$
$$
\ti\al_4\,,~(\ref{srf4})\,,~~R^+({\bf so(7)})=\Bigl\{\ti\al_j\,,~j=1,2,3,~(\ref{srf4})\,,\Bigr.
$$
$$
\left.
\oh(e_1+e_2-e_3-e_4)\,,~\oh(e_1+e_2+e_3+e_4)\,,~(e_1-e_4)\,,~(e_2+e_4)\,,~
(e_1+e_3)\,,~(e_1+e_2)\right\}\,.
$$
The ${\bf so(7)}$  root subspaces are read of from Table 2, while the
root subspaces in $\clJ_0$ are defined in (\ref{clj0}).

On $\gH(\baf)$ with the basis $H_{\ti\al_j}=\ti\al_j^\vee$, $j=1,\ldots,4$ (\ref{cbe7}) the Killing form is defined by the $\baf$ Cartan matrix (A.24.I).
On $\gL(\baf)$ the Killing form is
is
$$
(E^0_{23},E^0_{32})=(E^0_{14},E^0_{41})=(E^0_{35},E^0_{53})=
(E^0_{25},E^0_{52})=(E^0_{16},E^0_{61})=(E^0_{17},E^0_{71})=1\,,
$$
\beq{kff4}
(E^0_{12},E^0_{12})=(E^0_{13},E^0_{31})=(E^0_{15},E^0_{51})=2\,,
\eq
$$
(\gt^{R,0}_{a,+},\gt^{R,0}_{b,-})=2\de_{a,-b}\,,~~
(\gt^{L,0}_{a,+},\gt^{L,0}_{b,-})=2\de_{a,-b}\,.
$$


\subsubsection*{GS basis in $\clI$}

Consider the $\ti\la_7$-action on the subspace $\clI$ in (\ref{e7e6}).
 55 Chevalley generators in  $\clI$ (\ref{sroe7}) form the GS basis similarly to (\ref{te67}).
Then we come to the GS basis
\beq{1te67}
\begin{array}{l}
\gt^{R,+,k}_{a,+}=\f1{\sqrt{2}}(E^{R,+}_{a,+}+(-1)^kE^{V,+}_{a,-})\,,~~k=0,1\,,
\\
\gt^{R,+,k}_{a,-}=\f1{\sqrt{2}}(E^{R,+}_{a,-}+(-1)^kE^{V,+}_{a,+})\,,~~k=0,1\,,
\\
\gt^{L,+,k}_{4,+}=\f1{\sqrt{2}}(E^{L,+}_{4,+}+(-1)^kE^{L,+}_{8,-})\,,~~k=0,1\,,
\\
\gt^{L,+,k}_{4,-}=\f1{\sqrt{2}}(E^{L,+}_{4,-}+(-1)^kE^{L,+}_{8,+})\,,~~k=0,1\,,
\\
\gt^{L,+,k}_{a,+}=\f1{\sqrt{2}}(E^{L,+}_{a,+}+(-1)^kE^{L,+}_{a,-})\,,~~a\neq 4,8\,,~~k=0,1\,,
\\
\gt^{k}_{5,\pm}=\f1{\sqrt{2}}(E_{5,\pm}+(-1)^kE_{7,\mp})\,,~~k=0,1\,,
\\
\gt^{k}_{6,+}=\f1{\sqrt{2}}(E_{6,+}+(-1)^kE_{6,-})~~k=0,1\,,
\\
\gh^1_{e_5}=\f1{\sqrt{3}}(e_5+e_6+e_7)\,.
\end{array}
\eq
The generators with $k=0$ form the GS basis in $V$ and with $k=1$ along with
the GS basis in $\underline{26}$ (\ref{26d}) form basis in $\gg_1$.

Thus, we come to the following GS basis
$$
\ble=\baf\oplus V\oplus\gg_1\,, ~~
(\underline{133}=\underline{52}\oplus\underline{27}\oplus\underline{54})\,,
$$
\beq{gsbe7}
V=\{\gt^{R,+,0}_{a,\pm}\,,~\gt^{L,+,0}_{a,+}\,,\,(a\neq 4,8)\,,~~
\gt^{L,+,0}_{4,\pm}\,,~\gt^{0}_{5,\pm}\,,~\gt^{0}_{6,+}\}\,,
\eq
$$
\gg_1=\Bigl\{\gt^{R,1}_{a,\pm}\,,~\gt^{L,1}_{4,\pm}\,,~\gt^{R,+,1}_{a,\pm}\,,~\gt^{L,+,1}_{a,+}\,,~
(a\neq 4,8)\,,~\gt^{L,+,1}_{4,\pm}\,,\Bigr.
$$
$$
\gt^1_{1j}\,,~\gt^1_{j1}\,,\,(j=2,3,5)\,,~\gt^{1}_{5,\pm}\,,~\gt^{1}_{6,+}\,,~
\gh_{\al_1}^1\,,~\gh_{\al_2}^1\,,~\gh^1_{e_5}
\Bigr\}\,.
$$

The Killing form on the GS generators takes the form
$$
(\gt^{R,+,k_1}_{a,+},\gt^{R,+,k_2}_{b,-})=\de_{a,-b}\de^{k_1+k_2,0\,,~mod(2)}\,,
~~(\gt^{L,+,k_1}_{4,\pm},\gt^{L,+,k_2}_{4,\pm})=\de^{k_1+k_2,0\,,~mod(2)}\,,
$$
$$
(\gt^{L,k_1}_{4,\pm},\gt^{L,k_2}_{4,\mp})=\de^{k_1+k_2,0\,,~mod(2)}\,,~~
(\gt^{R,k_1}_{a,\pm},\gt^{R,k_2}_{b,\mp})=\de_{a,-b}\de^{k_1+k_2,0\,,~mod(2)}\,,
$$
\beq{kfgse7}
(\gt^{L,+,k_1}_{a,+},\gt^{L,+,k_2}_{b,+})=\de_{a,-b}\de^{k_1+k_2,0\,,~mod(2)}\,,~
(a\neq 4,8)\,,~~(t^1_{1,j},t^1_{k,1})=\de_{jk}\,,
\eq
$$
(\gh^1_{e_6},\gh^1_{e_6})=1\,,~~~
(\gt^{k_1}_{5,+},\gt^{k_2}_{5,-})=\de^{k_1+k_2,0\,,~mod(2)}\,,~~~
(\gt^{k_1}_{6,+},\gt^{k_2}_{6,-})=\de^{k_1+k_2,0\,,~mod(2)}\,.
$$

It allows us to define the dual basis (5.9.I)
$$
\gT^{R,+,k}_{a,+}=\gt^{R,+,k}_{-a,-}\,,~~\gT^{R,+,k}_{a,-}=\gt^{R,+,k}_{-a,+}\,,~~
\gT^{L,+,k}_{4,\pm}=\gt^{L,+,k}_{4,\pm}\,,
$$
$$
\gT^{R,k}_{a,+}=\gt^{R,k}_{-a,-}\,,~~
\gT^{L,+,k}_{4,\pm}=\gt^{L,+,k}_{4,\mp}\,,
$$
\beq{dbe7}
\gT^{L,+,k}_{a,\pm}=\gt^{L,+,k}_{-a,\mp}\,,~(a\neq 4,8)\,,
~~\gH^1_{e_5}=\gh^1_{e_5}=\f1{\sqrt{3}}(e_5+e_6+e_7)\,,
\eq
$$
\gT^{k}_{5,+}=\gt^{k}_{5,-}\,,~~~\gT^{k}_{6,+}=\gt^{k}_{6,-}\,,
~~\gT^1_{1,j}= t^1_{j,1}\,,~~ \gT^1_{j,1}=t^1_{1,j}\,.
$$
It follows from (\ref{ghf4}) and (\ref{ce6a})  that the scalar product matrix
$a_{j,k}=(\gh_{\al_j}^1,\gh_{\al_k}^1)$ for $(j,k=1,2)$ is the Cartan matrix $A_2$.
We need the inverse matrix to define  the dual basis (5.15.I)
 \beq{scpr7}
 \clA=\f1{3}\left(
        \begin{array}{cc}
          2 & 1 \\
          1 & 2 \\
        \end{array}
      \right)\,.
\eq
Thus,
\beq{dbce7}
\gH_{\al_1}^1=\frac{2}3\gh_{\al_1}^1+\f1{3}\gh_{\al_2}^1=
\f1{3}(e_5-2e_6+e_7)\,,
\eq
\beq{dbce7a}
\gH_{\al_2}^1=\frac{1}3\gh_{\al_1}^1+\frac{2}{3}\gh_{\al_2}^1=
\f1{2\sqrt{2}}(-e_1+e_2-e_3-e_4)+\f1{6}(e_5-2e_6+e_7)\,.
\eq
The scalar product $(\gH_{\al_j}^1,\gH_{\al_k}^1)$ is defined by $\clA$ (\ref{scpr7}), while
$\gH^1_{e_5}$ is orthogonal to them.


\subsection*{Lax operators and Hamiltonians}

\subsubsection*{Trivial bundles}

The moduli space of trivial $\bG=E_7$ bundles is the quotient
$$
\gH(\ble)/W_{\ble}\ltimes(\tau(Q(\ble)+Q(\ble))\,.
$$

The moduli space of trivial $G^{ad}=E_7/\mu_2$ bundles is the quotient
$$
\gH(\ble)/W_{\ble}\ltimes(\tau(P(\ble)+P(\ble))\,,
$$
where $P(\ble)$ are defined by the basis (\ref{wle7}).

For trivial bundles Lax operator takes the form
$$
L^{CM}_{\ble}(z)=L^{CM}_{\bae}(z)+
\sum_{a=1}^8\sum_{A=L,R,V}S_{a,\pm}^{A,+}\phi((\bfu,\varpi_a^A)\pm\f1{\sqrt{2}}(u_j+u_k),z) E_{a,\pm}^{A,+}
+\sum_{j=5,6,7}
 S_{j,\pm}\phi( \pm\sqrt{2}u_j,z) E_{j,\pm}\,.
$$
Here we have the correspondence $L\to (j=5,7)$, $R\to (j=5,6)$, $V\to (j=6,7)$.
In contrast with $\bae$ there is no restrictions  $\sum_{k=1}^7 v_k=\sum_{k=1}^7 u_k=0$.

The Hamiltonian after the symplectic reduction with respect to the $\clH(\ble)$ action
assumes the form
$$
H^{CM}_{\ble}=H^{CM}_{\bae}+\sum_{a=1}^4\sum_{A=L,R,V}S_{a,+}^{A,+}S_{a,-}^{A,+}
E_2((\bfu,\varpi_a^A)+\f1{\sqrt{2}}(u_j+u_k))+
\sum_{j=5,6,7} S_{j,+}S_{j,-}E_2(\sqrt{2}u_j,z)\,.
$$


\subsubsection*{Nontrivial bundles}

\textbf{The moduli space.}\\
The moduli space are described by elements $\ti\bfu\in\gH(\baf)$
(\ref{cf4}).
The Weyl group $W(\baf)$ is generated by reflections with respect the planes, orthogonal
to simple roots (\ref{srf4}).

The coroot lattice $\ti Q^\vee(\baf)$ is generated by the simple coroots (\ref{cbe7}).
The moduli space of nontrivial $E_7$ bundles is defined as
$$
\gH(\baf)/(W(\baf)\ltimes(\tau\ti Q^\vee(\baf)+\ti Q^\vee(\baf)))\,.
$$

\bigskip
\textbf{$F_4$ Calogero-Moser system.}\\
Represent the Lax operator in the form (6.7.I)
$$
L(z)=\ti L_0(z)+L'(z)+L_1(z)\,,
$$
where $\ti L_0(z)=L_{\baf}^{CM}(z)$. In defined below expression $\ti\bfv=(v_1,v_2,v_3,v_4,v_5)$ the momentum vector satisfies the same restriction as the vector $\ti\bfu$ (\ref{fres}).
It follows from (\ref{f4d}) it takes the form
$$
L_{\baf}^{CM}(z)=L_{\bf so(7)}^{CM}(z)+
\frac{1}4\Bigl(-3(v_1+S_{0,1}E_1(z))e_1-(v_2+S_{0,2}E_1(z))e_2-(v_3+S_{0,3}E_1(z))e_3
$$
$$
+(v_4+S_{0,4}E_1(z))e_4)+
\f1{2\sqrt{2}}((v_5+S_{0,5}E_1(z))e_5-(v_5+S_{0,5}E_1(z))e_7)\Bigr)+
$$
$$
+\sum_{a=1}^8\Bigl(S_{a,+}^{R,0}\phi((\ti\bfu,\varpi_a^R)+\f1{\sqrt{2}}u_5,z)\gt^{R,0}_{-a,-}
+S_{-a,-}^{R,0}\phi(-(\ti\bfu,\varpi_a^R)-\f1{\sqrt{2}}u_5,z)\gt^{R,0}_{a,+}\Bigr)
$$
$$
S_{4,+}^{L,0}\phi((\ti\bfu,\varpi_4^L)+\sqrt{2}u_5,z)\gt^{L,0}_{4,-}+
S_{4,-}^{L,0}\phi(-(\ti\bfu,\varpi_4^L)-\sqrt{2}u_5,z)\gt^{L,0}_{4,+}+
$$
$$
\sum_{a\neq 4}^8\Bigl(S_{a,+}^{L,0}\phi((\ti\bfu,\varpi_a^L)+\sqrt{2}u_5,z)\gt^{L,0}_{-a,-}
+S_{-a,-}^{L,0}\phi(-(\ti\bfu,\varpi_a^L)-\sqrt{2}u_5,z)\gt^{R,0}_{a,+}\Bigr)\,.
$$
We find the corresponding quadratic Hamiltonian
$$
H_{\baf}^{CM}=H_{\bf so(7)}^{CM}+\oh\Bigl(v_1^2+v_2^2+v_3^2+v_4^2+2v_5^2\Bigr)
$$
$$
+\sum_{a=1}^8S_{a,+}^{R,0}S_{-a,-}^{R,0}E_2((\ti\bfu,\varpi_a^R)+\f1{\sqrt{2}}u_5)
+S_{4,+}^{L,0}S_{4,-}^{L,0}E_2((\ti\bfu,\varpi_4^L)+\sqrt{2}u_5)
$$
$$
+\sum_{a\neq 4}^8S_{a,+}^{L,0}S_{-a,-}^{L,0}E_2((\ti\bfu,\varpi_a^L)+\sqrt{2}u_5)\,.
$$

\bigskip
\textbf{The Lax operators and Hamiltonians}\\
Define as in the general case (6.14.I) the function
$$
\varphi^k_\be(\ti\bfu,z)=\bfe\,\Bigl(\lan\ka_{\bf e_7},\be\ran z\Bigr)
\phi(\lan\ka_{\bf e_7}-\ti\bfu,\be\ran+k/2, z)\,,~~~(k=0,1)\,,
$$
where $\ka_{\bf e_7}$ (\ref{ka7}) and $\be\in R({\bf e_7})$ generating the GS basis.
The following  $R({\bf e_7})$ roots define GS generators  (\ref{gsbe7})
\beq{be7}
\be=
\left\{
  \al_{(a,\pm)}^{(L,+)}\,,~~
  \al_{(a,\pm)}^{(R,+)}\,, ~ ~\al_{1j}=e_1-e_j\,,~(j=2,3,5)\,,~~(\ref{sroe7})\,,  \right.
\eq
$$
\left.
\al_{(a,\pm)}^{L}\,,~
\al_{(a,\pm)}^{R}~(\ref{sroe6})\,, ~~  \pm \al^{(+)}_j,~(j=5,6,7)\,,~(\ref{re7gs})\,,~~
\right\}\,.
$$
Then following (6.15.I) and  (6.17.I) from (\ref{gsbe7})  we find
$$
L_0'(z)=\sum_{a=1}^8S_{a,\pm}^{R,+,0}\varphi^0_{ \al_{(a,\pm)}^{(L,+)}}(\ti\bfu,z)
\gt^{R,+,0}_{-a,\mp}+
\sum_{a\neq 4,8}S_{a,\pm}^{L,+,0}\varphi^0_{ \al_{(a,\pm)}^{(L,+)}}(\ti\bfu,z)\gt^{L,+,0}_{-a,\mp}+
$$
$$
S^{L,+,0}_{4,\pm}\varphi^0_{ \al_{(4,\pm)}^{(L,+)}}(\ti\bfu,z)\gt^{L,+,0}_{4,\pm}+
S^{0}_{5,\pm}\varphi^0_{ \pm\al_{(5)}^{(+)}}(\ti\bfu,z)\gt^{0}_{5,\mp}+
S^{0}_{6,\pm}\varphi^0_{ \pm\al_{(6)}^{(+)}}(\ti\bfu,z)\gt^{0}_{6,\mp}\,.
$$
$$
L_1(z)=\Bigl(S^1_{\al_1}\gH_{\al_1}^1+S^1_{\al_2}\gH_{\al_2}^1
+S^1_{e_5}\gh^1_{e_5}\Bigr)\phi(\oh,z)
$$
$$
\sum_{a=1}^8S^{R,1}_{a,\pm}\varphi^1_{ \al_{(a,\pm)}^{(R)}}(\ti\bfu,z)\gt^{R,1}_{a,\mp}+
S^{L,1}_{4,\pm}\varphi^1_{ \al_{(4,\pm)}^{(L)}}(\ti\bfu,z)\gt^{L,1}_{4,\mp}
$$
$$
+\sum_{a=1}^8S_{a,\pm}^{R,+,1}
\varphi^1_{\al_{(a,\pm)}^{(R,+)}}(\ti\bfu,z) \gt^{R,+,1}_{-a,\mp}+ \sum_{a\neq
4,8}S_{a,\pm}^{L,+,1}\varphi^1_{
\al_{(a,\pm)}^{(L,+)}}(\ti\bfu,z)\gt^{L,+,1}_{-a,\mp}+
$$
$$
\sum_{j=2,3,4,5}\Bigl(S^1_{1,j}\varphi^1_{ \al_{(1,j)}}(\ti\bfu,z)t^1_{j,1}+
S^1_{j,1}\varphi^1_{ \al_{(j,1)}}(\ti\bfu,z)t^1_{1,j}\Bigr)+
$$
$$
S^{L,+,1}_{4,\pm}\varphi^1_{ \al_{(4,\pm)}^{(L,+)}}(\ti\bfu,z)\gt^{L,+1}_{4,\mp}+
S^{1}_{5,\pm}\varphi^1_{\pm\al_{(5)}^{(+)}}(\ti\bfu,z)\gt^{1}_{5,\mp}+
 S^{1}_{6,\pm}\varphi^1_{\pm\al_{(6)}^{(+)}}(\ti\bfu,z)\gt^{1}_{6,\mp}\,.
$$
where for
$\gH_{\al_1}^1$ and $\gH_{\al_2}^1$ see (\ref{dbce7}) and (\ref{dbce7a}).

For the $E_7$ quadratic Hamiltonians we have
$$
H_{\ble}=H^{CM}_{\baf}+H_0'+H_1\,,
$$
where $H_0'$ comes from $\oh(L_0^{'2})$ and $H_1$ from $\oh(L_1^{2})$.
To calculate the Hamiltonians we use (\ref{kfgse7}) and (\ref{scpr7}) for scalar products of the
dual Cartan generators $\gH_{\al_j}^1$. Then
$$
-H_0'=\sum_{a=1}^8S_{a,+}^{R,+,0}S_{-a,-}^{R,+,0}E_2(\lan \al_{(a,+)}^{(L,+)},\ti\bfu\ran)+
\sum_{a\neq 4,8}S_{a,+}^{L,+,0}S_{-a,-}^{L,+,0} E_2(\lan \al_{(a,+)}^{(L,+)},\ti\bfu\ran)+
$$
$$
S^{L,+,0}_{4,+}S^{L,+,0}_{4,-}E_2( \lan\al_{(4,+)}^{(L,+)},\ti\bfu\ran)+
S^{0}_{5,+}S^{0}_{5,-}E_2(\lan\al_{(5)}^{(+)},\ti\bfu\ran)+
S^{0}_{6,+}S^{0}_{6,-}E_2(\lan\al_{(6)}^{(+)},\ti\bfu\ran)\,.
$$

$$
H_1=\frac{2}3\Bigl((S^1_{\al_1})^2+(S^1_{\al_2})^2+S^1_{\al_1}S^1_{\al_2}\Bigr)E_2\Bigl(\oh\Bigr)
+(S^1_{e_5})^2E_2\Bigl(\oh\Bigr)
$$
$$
+\sum_{a=1}^8S^{R,1}_{a,+}S^{R,1}_{-a,-}E_2(\lan \al_{(a,+)}^{(R)},\ti\bfu+\oh\ran)+
S^{L,1}_{4,+}S^{L,1}_{4,-}E_2(\lan\al_{(4,+)}^{(L)},\ti\bfu\ran+\oh)
$$
$$
+\sum_{a=1}^8S_{a,+}^{R,+,1}S_{-a,-}^{R,+,1}E_2(\lan\al_{(a,+}^{(R,+)},\ti\bfu\ran+\oh)
+ \sum_{a\neq 4,8}S_{a,+}^{L,+,1}S_{-a,-}^{L,+,1}E_2(\lan\al_{(a,+)}^{(L,+)},\ti\bfu\ran+\oh)
$$
$$
+\sum_{j=2,3,4,5}S^1_{1,j}S^1_{j,1}E_2(\lan \al_{(1,j)},\ti\bfu\ran+\oh)
+S^{L,+,1}_{4,+}S^{L,+,1}_{4,-}E_2(\lan \al_{(4,+)}^{(L,+)},\ti\bfu\ran+\oh)
$$
$$
+
S^{1}_{5,+}S^{1}_{5,-}E_2(\lan\al_{(5)}^{(+)},\ti\bfu\ran+\oh)+
S^{1}_{6,+}S^{1}_{6,-}E_2(\lan\al_{(6)}^{(+)},\ti\bfu\ran+\oh) \,,
$$
where $\ti\bfu$ is defined by (\ref{cf4}), (\ref{fres}).


\small{



\begin{thebibliography}{60}


\bibitem{Bou}
N.~Bourbaki,
 \emph{Lie Groups and Lie Algebras: Chapters 4-6},
  Springer-Verlag, Berlin-Heidelberg-New York, (2002).

\bibitem{FFZ}
D.~Fairlie, P.~Fletcher and C.~Zachos,
\emph{Infinite Dimensional Algebras and a Trigonometric Basis for the Classical Lie Algebras,} Journal
of Mathematical Physics, {\bf 31} (1990), 1088-1094.

\bibitem{GH}
J.Gibbons, and T.Hermsen,
\emph{A generalization of the Calogero-Moser systems,} Physica \textbf{11D}
(1984), 337-348.


\bibitem{Jac}\
N.~Jacobson,
\emph{Exceptional Lie algebras,}
Lecture Notes in Pure and Applied Mathematics, (1971) NY.

\bibitem{Ka} V.~Kac,
\emph{Automorphisms of  finite order of semisimple Lie algebras},
Funct.Anal. and Applic., {\bf 3}, (1969), 94-96.


\bibitem{KLO}
B.~Khesin, A.~Levin, M.~Olshanetsky,
{\em Bihamiltonian structures and
quadratic algebras in hydrodynamics and on non-commutative torus},
 Comm.Math.Phys., {\bf 250} (2004) 581-612.


\bibitem{LOSZ}
   A.~Levin, M.~Olshanetsky, A.~Smirnov,  A.~Zotov,
\emph{Integrable systems and Characteristic Classes.  General construction.}

\bibitem{LZ}
   A.~Levin, and A.~Zotov,
   {\em Integrable systems of interacting elliptic tops},
   Theor. Math.Phys., {\bf 146:1}, (2006),  55-64.

\bibitem{LX}
Luen-Chau Li, Ping Xu
\emph{Integrable spin Calogero-Moser systems}
 Commun.Math.Phys. {\bf 231} (2002), 257-286.


\bibitem{OP}  M.~Olshanetsky,  A.~Perelomov,
 {\em Classical integrable
finite-dimensional systems related to Lie algebras}, Physics Reports, v.71
(1981), 313-400.

\bibitem{RSTS}
A.~Reyman and M.~Semenov-Tian-Schansky,
{\em Lie algebras and Lax equations with spectral parameter on elliptic curve},
(Russian)  Zap. Nauchn. Sem. Leningrad. Otdel. Mat. Inst. Steklov. (LOMI)
{\bf 150}  (1986),  Voprosy Kvant. Teor. Polya i Statist. Fiz. 6, 104--118, 221;
translation in  J. Soviet Math.,  {\bf 46}, no. 1, (1989),   1631--1640.


\bibitem{Wo}
S.Wojciechowski,
 \emph{An integrable marriage of the Euler equations with the Calogero-Moser
systems,}
Phys. Lett. A, \textbf{111} (1985), 101-103.

\end{thebibliography}
\end{document}